\documentclass[a4paper,11pt]{article}

\usepackage{jcappub}

\usepackage{epsfig,amsmath,amssymb,verbatim,mathrsfs,array,layout,textcomp,latexsym, slashed,graphicx,bbm,physics,tensor}

\usepackage[normalem]{ulem}
\usepackage{appendix}

\usepackage{rotating}
\usepackage{hyperref}
\usepackage[usenames,dvipsnames]{color}

\usepackage{soul}

\newcommand{\be}{\begin{equation}}
\newcommand{\ee}{\end{equation}} 
\newcommand{\bea}{\begin{eqnarray}}
\newcommand{\eea}{\end{eqnarray}}

\definecolor{gbcolor}{rgb}{.1,.1,.8}

\usepackage{dsfont}

\newcommand{\eq}[1]{(\ref{#1})}

\begin{document}

\begin{flushleft}                          
\hspace{12cm} \footnotesize IFT UAM-CSIC 24-60
\end{flushleft} 

\title{One-loop power spectrum in ultra slow-roll inflation and implications for primordial black hole dark matter}

\author[a,b]{Guillermo Ballesteros,}
\author[b]{Jes\'us Gamb\'in Egea}

\affiliation[a]{Departamento de F\'isica Te\'orica, Universidad Aut\'onoma de Madrid (UAM), Campus de Cantoblanco, 28049 Madrid, Spain}
\affiliation[b]{instituto de F\'isica Te\'orica UAM-CSIC, Campus de Cantoblanco, 28049 Madrid, Spain}

\emailAdd{guillermo.ballesteros@uam.es}
\emailAdd{j.gambin@csic.es}

\abstract{We apply the in-in formalism to address the question of whether the size of the one-loop spectrum of curvature fluctuations in ultra-slow-roll inflation models designed for producing a large population of primordial black holes implies a breakdown of perturbation theory. We consider a simplified piece-wise description of inflation, in which the ultra-slow-roll phase is preceded and followed by slow-roll phases linked by transitional periods. We work in the $\delta\phi$-gauge, including all relevant cubic and quartic interactions and the necessary counterterms to renormalize the ultraviolet divergences, regularized by a cutoff. The ratio of the one-loop to the tree-level contributions to the spectrum of curvature perturbations is controlled by the duration of the ultra-slow-roll phase and of the transitions. Our results indicate that perturbation theory does not necessarily break in well-known models proposed to account for all the dark matter in the form of primordial black holes.}

\maketitle

\flushbottom

\newpage

\section{Introduction}

Primordial black holes (PBHs) in the range that goes from $10^{-16}$ to $10^{-12}$ Solar masses might account for all the dark matter (DM) of the Universe \cite{Carr:2009jm, Niikura:2017zjd, Katz:2018zrn, Montero-Camacho:2019jte, Ballesteros:2019exr, Carr:2020gox, Green:2020jor}. Although the existence of PBHs has not been demonstrated, that possibility alone makes them worth studying. Indeed, they have become serious contenders for the solution to the dark matter problem, on the same footing as particle candidates such as e.g.\ WIMPS and warmer candidates, the QCD axion and axion-like particles. Moreover, whether they are discovered or ruled out in the future, PBH physics can help us learn about the very early Universe. 

A popular mechanism to form PBHs consists in the collapse of Hubble-sized density fluctuations in the radiation epoch \cite{Carr:1974nx}. These collapsing regions, with densities above a certain threshold \cite{Nakama:2013ica,Musco:2018rwt}, could have originated from specific dynamics during a preceding inflationary period. The most studied model of that kind posits a phase of so-called ultra slow-roll (USR) \cite{Tsamis:2003px,Kinney:2005vj} during inflation. In terms of a single scalar field --the inflaton, $\phi$-- driving inflation as it descends a potential $V(\phi)$, this corresponds to $\phi$ undergoing a significant deceleration, such that $\ddot\phi \simeq - 3 H \dot \phi$, where dots indicate derivatives with respect to cosmic time, $t$, and $H$ is the Hubble function measuring the (accelerated) expansion of the Universe. This dynamics implies a rapid growth of curvature fluctuations, which in turn leads to large density perturbations during the subsequent radiation epoch. 

A simple estimate assuming Gaussian primordial curvature fluctuations indicates that their spectrum $\mathcal{P}_\zeta$ (which we will define later on) must be $\sim 10^{-2}$ at comoving wavenumbers of the order of $10^{12}$ Mpc$^{-1}$ -- $10^{14}$ Mpc$^{-1}$ if all the dark matter is made of PBHs formed from that mechanism in the aforementioned mass range. This spectral value is much larger than the inferred one from the cosmic microwave background (CMB), which is $\mathcal{P_\zeta}\simeq 2 \cdot 10^{-9}$ \cite{Fixsen:1996nj}. Since $\mathcal{P_\zeta}$ scales as the square of the curvature fluctuation, $\zeta$, values of $\mathcal{P_\zeta}$ of order $10 ^{-2}$ almost beg the question of the validity of perturbation theory. A paper by J.\ Kristiano and J.\ Yokoyama from November 2022 \cite{Kristiano:2022maq} studied this question considering the one-loop correction to $\mathcal{P_\zeta}$ in USR inflation, using the in-in formalism \cite{Schwinger:1960qe,Keldysh:1964ud,Weinberg:2005vy}. They concluded that such large values of $\zeta$ imply the breakdown of perturbation theory at CMB scales, which, according to their work, would severely threaten USR as a possibility for PBH formation. Other papers on the same issue have been appearing since then, with contradicting conclusions that have kept the question open, see \cite{Riotto:2023hoz, Firouzjahi:2023bkt, Franciolini:2023lgy, Tada:2023rgp, Iacconi:2023ggt} for a non-exhaustive but representative list of references. 

We weigh in this debate improving the treatment of the problem in several respects. As in most earlier works on the topic, we use the in-in formalism to compute $\mathcal{P}_\zeta$ at one-loop and compare the result to its tree-level counterpart. However, we include the complete set of relevant interactions (cubic, quartic and boundary terms) for fluctuations, 
several of which were ignored in previous analyses.\footnote{Although there are no boundary terms in the set of interactions we consider, these interactions are related to boundary terms in the $\zeta$-gauge, on which the problem has been studied in previous works.} We manage to do this by working on the $\delta\phi$-gauge, choosing spacetime coordinates for which $\zeta =0$ and the physical content of the primordial fluctuations is contained in the perturbation of the inflaton, $\delta\phi$. {This choice of gauge, and a convenient model definition, allow us to unveil a key divergence of the two-point function arising at loop-level in the limit of instantaneous transitions between SR and USR, which had been missed in the previous studies on this subject.} In addition, we implement a regularization procedure of ultraviolet divergences at the level of the 
action for fluctuations.
We use a cutoff to regularize the ultraviolet divergences, similarly to what has been done in previous literature on the topic, see e.g.\ \cite{Kristiano:2022maq,Franciolini:2023lgy}. However, we also absorb the divergences by introducing adequate counterterms, whereas in previous works the cutoff was given a numerical value motivated by physical intuition (which made the result depend on its value). Although we do not implement a full renormalization procedure, we are able to analyze the validity of perturbation theory, providing an answer to the issue raised in  \cite{Kristiano:2022maq}.

We find that whether perturbation theory breaks down depends on the duration of the transition between slow-roll (SR) inflation and USR inflation. Our results indicate that for $\mathcal{P_\zeta}\sim 10^{-2}$ and well-motivated inflationary models considered in the literature \cite{Ballesteros:2017fsr,Ballesteros:2020qam}, cosmological perturbation theory is valid, in the sense that the one-loop spectrum is significantly smaller than the tree-level one. 

\section{Model of USR inflation}

We consider a two-parameter, piece-wise, description of the inflationary dynamics leading to a large tree-level $\mathcal{P_\zeta}$. We assume a phase of SR followed by USR and then, again, SR. The transitions between these phases are characterized by a single parameter, which controls the duration of both transitions. The other parameter of the model is the duration of the USR phase. Imposing a value for $\mathcal{P_\zeta}$ establishes a one-to-one correspondence between both parameters. The dynamics of the SR phases is controlled by the small quantity $0<\epsilon=- \dot{H}/H^2\ll 1$, whose actual value is largely irrelevant in what follows. The quantity $\eta = \dot{\epsilon}/(H\epsilon)$ is, by definition, exactly equal to $-6$ in the USR phase and, will be assumed to vanish during both SR phases. The transitions between phases are described in terms of a third quantity,
$\nu$, as we explain next. 

\begin{figure}[t]
	\begin{center}
		\includegraphics[width=0.6\textwidth]{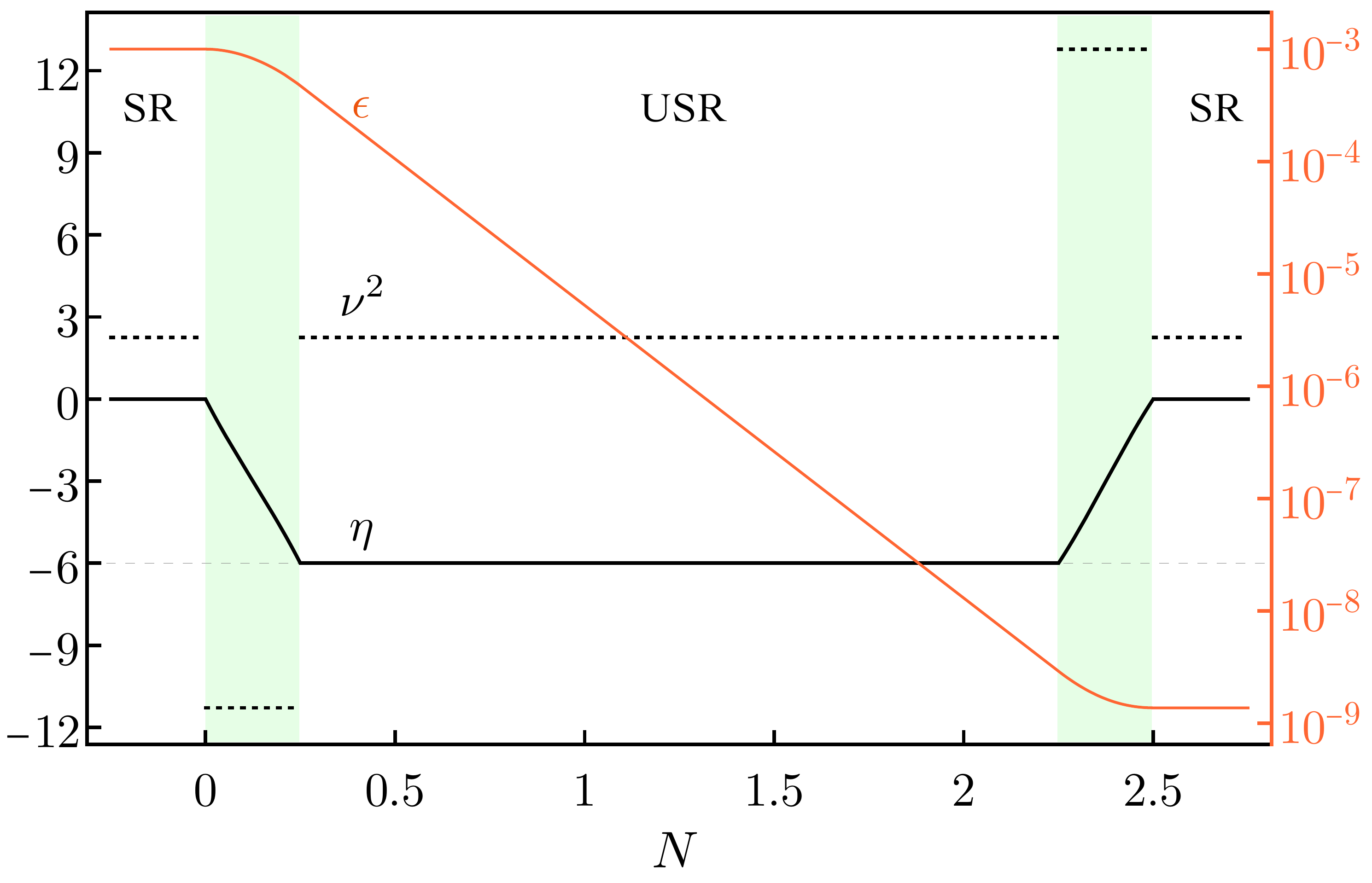}
		\caption{\label{fig:model} \it     \small Evolution of $\nu^2$ (black dotted), $\eta$ (black continuous) and $\epsilon$ (orange continuous, right axis) as functions of $N$ for $\delta N = 0.25$, $\Delta N = 2$ and 
	$\epsilon(N<0) = 10^{-3}$. The transitions (of duration $\delta N$) between SR ($\eta = 0$) and USR ($\eta = -6$) are shown as shaded bands.}
	\end{center}
\end{figure}

Assuming a single inflaton $\phi$, with a canonical kinetic term and a potential $V(\phi)$, working in conformal time, $\tau$, and neglecting terms suppressed by powers of $\epsilon$, the action for fluctuations in the $\delta\phi$-gauge is
\begin{align}	\label{eq:Expanded_Action}
	S = {\int}\dd \tau\, \dd^3 \vb{x}\, \left[\frac{a^2}{2}\left(\left(\partial_\tau\delta\phi\right)^2 - \left(\partial_i\delta\phi\right)^2  \right) -a^4 \sum_{n\geq 2} \frac{V_n\, \delta\phi\,^n}{n!}\right]
\end{align}
and $V_n = \dd^n V / \dd \phi^n$. The interactions that arise from the 
metric fluctuations are suppressed (see Appendix \ref{sec:calculation_action}), and only the interactions coming from the potential survive at lowest order in $\epsilon$:
\begin{align} \label{derivs}
	a^2 V_2 = -(aH)^2 (\nu^2-9/4)\,, \quad a^2V_3 = -\dfrac{aH (\nu^2)'}{\sqrt{2\epsilon}M_P}\,,\\
\label{derivs2}	a^2V_4 = -\dfrac{1}{2\epsilon M_P^2} \left( (\nu^2)'' - aH (\nu^2)'\left(1+\dfrac{\eta}{2} \right)  \right)\,,
\end{align}
where primes denote derivatives with respect to conformal time, 
\begin{align}
\nu^2 \equiv \frac{9}{4} + \frac{1}{2}\left(3\,\eta + \frac{\eta^2}{2} + \frac{\eta'}{aH}\right) \in \mathbb{R}
\end{align}
and $M_P^2 = 1/(8\pi\, G)$ is the reduced Planck mass squared.
In both SR ($\epsilon,\,|\eta|\ll1$) and USR, $\nu = 3/2$, see also \cite{Wands:1998yp,Leach:2000yw,Leach:2001zf}. 
We impose that $\nu^2$ is piece-wise constant. The function $\nu^2$ in the transition from SR to USR and, also in the subsequent transition from USR to SR, is then set by their (equal) duration. This in turn determines $\eta$ and $\epsilon$ completely, which are found integrating their respective definitions. Figure \ref{fig:model} shows in an example $\nu^2$, $\epsilon$ and $\eta$ as functions of the number of e-folds of inflation ($N=\int H dt$) elapsed from the beginning of the first transition. In terms of this variable, the duration of the USR phase is denoted $\Delta N$ and that of the transitions is $\delta N$. For $\delta N = 0$, we recover the model used in \cite{Kristiano:2022maq}. For the (well-motivated) known potentials that lead to transient USR compatible with $\mathcal{P_\zeta}\sim 10^{-2}$ \cite{Ballesteros:2017fsr,Ballesteros:2020qam}, the function $\nu^2$ is indeed approximately constant during the transitions, which satisfy $\delta N\lesssim 1 < \Delta N$. 

Since $\nu$ is discontinuous at the beginning and at the end of each transition, the self-interactions of $\delta\phi$ (proportional to $V_{3	,4}$) are Dirac deltas centered on those instants.
In the limit $\delta N \to 0$, $\nu^2$ satisfies {$\abs{\nu^2} \rightarrow 3/\delta N$ during the transitions}; i.e.\ the interactions diverge 
in the limit of instantaneous transitions between SR and USR.
It is therefore important to consider smooth transitions. 
This effect is not as transparent in the $\zeta$-gauge
because the dependence on $\nu^2$ does not arise so naturally.

Although we use the $\delta\phi$-gauge, we are interested in $\mathcal{P}_\zeta$, defined, at any order in perturbation theory, through the two-point correlation function:
\begin{align}
	\langle \zeta(x) \zeta(y) \rangle = \int \dfrac{\dd^3 \vb{k}}{4\pi k^3} e^{i\vb{k}(\vb{x}-\vb{y})} \mathcal{P}_\zeta(\tau,k)\,.
	\label{eq:Def_PowerSpectrum}
\end{align}
For modes satisfying $k \equiv\abs{\vb{k}}\ll aH$ in the last SR phase ($\eta =0$) we have
$ \left\langle \delta\phi(x) \delta\phi(y) \right\rangle =	2\epsilon M_P^2\left\langle \zeta(x) \zeta(y) \right\rangle $ (see Appendix \ref{sec:App_Correlation_Z_DP}).\footnote{In a general model where in the last phase of inflation $\eta\neq 0$, this expression has corrections proportional to the value of $\eta$ in that phase.}

\section{In-In formalism, regularization and counterterms}

In the in-in formalism \cite{Weinberg:2005vy,Chen:2010xka}, at second order in the interaction Hamiltonian, $H_I$, the vacuum expectation value of an operator $Q(\tau)$ can be obtained as {(see Appendix \ref{sec:in_in_form})}
\begin{align} \nonumber
	\expval{Q(\tau)} &= \bra{0} Q_I(\tau) \ket{0} + 2\Im{\int_{-\infty}^\tau \dd \tau' \bra{0} Q_I(\tau) H_I(\tau'_-) \ket{0}} \\&\quad +2\Re{\int^\tau_{-\infty} \dd \tau' \int ^\tau_{\tau'} \dd \tau'' \bra{0} {\left( H_I(\tau''_+)Q_I(\tau) - Q_I(\tau) H_I(\tau''_-) \right) }H_I(\tau'_-) \ket{0}}\,.
	\label{eq:In-In_General}
\end{align}
On the right hand side of this expression, the fields are in the interaction picture ($_I$), i.e.\ they are described by the free Hamiltonian and satisfy canonical commutation rules. $\ket{0}$ is the vacuum in the interaction picture. 
We use $\tau_\pm \equiv \tau(1\pm i \omega)$ (with $\omega>0$) to guarantee the correct projection onto the interaction vacuum in $\tau\to-\infty$. 

\begin{figure}[t]
	\begin{center}
		\includegraphics[width=1\textwidth]{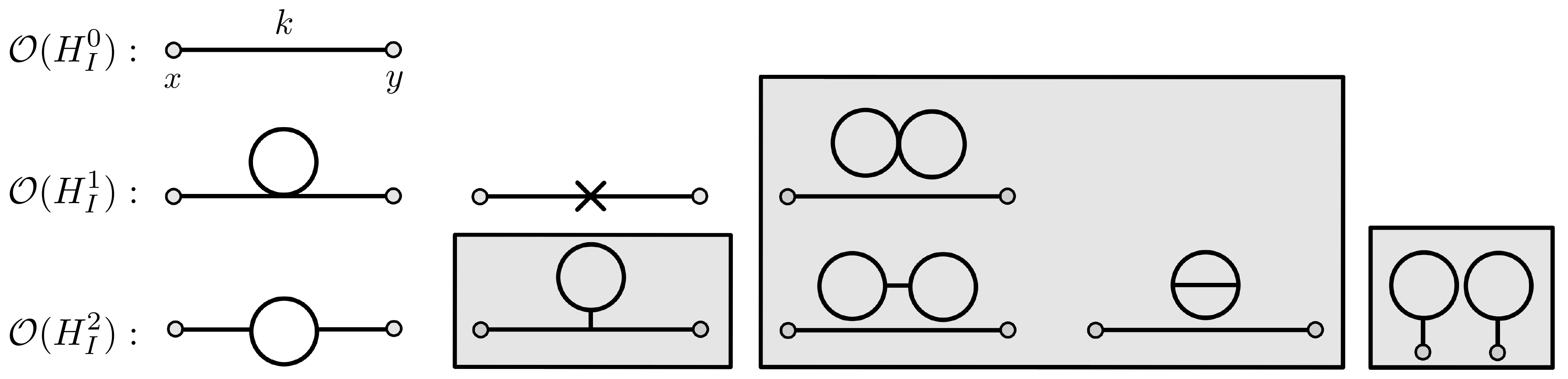}
		\caption{\label{fig:one-loops}  \small \it Diagrams arising from eq.\ (\ref{eq:In-In_General}) when calculating $\expval{\delta\phi^2}$. We distinguish the tree-level (first row), the one-loop (first diagrams of rows second and third) and the counterterms (second diagram of the second row). Additionally, there is a term proportional to the tadpole, bubble diagrams and a term composed of two tadpoles. None of the latter diagrams (in shaded boxes) contribute to $\mathcal{P}_\zeta$.}
	\end{center}
\end{figure}

{In a model with cubic and quartic interactions, the topology of the diagrams}
generated from equation (\ref{eq:In-In_General}) for the two-point function of $\delta\phi$  are shown in Figure~\ref{fig:one-loops}.\footnote{{Diagrams that are suppressed  by powers of $M_P$
are not shown, e.g.\ the diagrams induced at second order by the quartic interaction.}}
It can be checked that we do not have to include the bubble diagrams to the actual computation we perform.
The calculation at one-loop of the two-point function gives 
an ultraviolet (UV) divergence
that needs to be renormalized, see \cite{Senatore:2009cf,Weinberg:2008nf}. For the regularization of the divergences, we use a UV cutoff $\Lambda_{\rm UV}$. Whereas the time integrals can be done explicitly, for the spatial part we work in Fourier space. The cutoff goes into the momentum loops as usual, but its effect on the time integrals needs to be worked out. In practice, we remove from the integral a domain where the time intervals are smaller than those allowed by $\Lambda_{\rm UV}$ \cite{Senatore:2009cf}, as follows:
\begin{equation}
	\int^\tau_{-\infty} \dd \tau' \int ^\tau_{\tau'} \dd \tau'' \to \int^{\tau + {1/(a(\tau)\Lambda_{\rm UV})}}_{-\infty} \dd \tau' \int ^\tau_{\tau' + 1/(a(\tau')\Lambda_{\rm UV})} \dd \tau'' \,.
	\label{eq:reg_time_integrals}
\end{equation}
Once the divergences have been regularized, we introduce counterterms to absorb the dependence of the loops on the regulator $\Lambda_{\rm UV}$ \cite{Weinberg:2010wq}. 
To obtain the counterterms we can think in terms of the action for the inflaton, $\phi$. General covariance implies that the wave function renormalization, $\phi \to Z_\phi\,  \phi_R$ (where the subscript $_R$ stands for {\it renormalized}) requires the quantity $Z_\phi \equiv 1 + \delta_\phi$ to be constant.\footnote{This renormalization also changes the background action. However, by introducing a renormalization of the vacuum expectation value (VEV) of $\phi$, $\phi_0(t)$, one can impose that the background dynamics remains unchanged, equivalent to what happens with the Higgs VEV (see e.g.\ \cite{Hollik:1988ii}).} Similar arguments may in principle be used for the counterterms coming from the potential. 
However, in the model under consideration, $V(\phi)$ is given by the background evolution described in the previous section. Therefore, just as $V_n$ are functions of time,  
by renormalizing $V_n\to V_{R,n} + \delta_{V_n}$, the counterterms $\delta_{V_n}$ are functions of time. The counterterms that will affect the renormalization of the two-point correlation will be $\delta_\phi$ and $\delta_{V_2} \equiv \delta_V$. Including these counterterms
in the action (\ref{eq:Expanded_Action}) we extract the complete interaction Hamiltonian in the interaction picture:\footnote{For convenience, we will omit the subscript $_R$, always keeping in mind that we are working with renormalized quantities.}
\begin{align}
	H_I(\tau)   = \int \dd^3\vb{x}\,\bigg{[}a^4\left(\dfrac{V_3 \delta\phi^3}{3!}+\dfrac{V_4 \delta\phi^4}{4!} \right) +    (a^2\delta_\phi\, \delta\phi\, \delta\phi')' +
	2 a^2\, \delta\phi\, \left(a^2\tilde{\delta}_V-\delta_\phi \nabla^2 \right) \delta\phi\bigg{]}\,,
	\label{eq:Int_Hamiltonian} 
\end{align}
where $\tilde{\delta}_V  = \delta_\phi V_2 + \delta_V /4$. This Hamiltonian includes all possible interaction terms and counterterms that affect $\mathcal{P}_\zeta$ at one-loop at leading order in $\epsilon$. 

We quantize $\delta\phi$ --in the interaction picture-- with creation and annihilation operators satisfying standard commutation relations,
\begin{equation}
	\delta\phi(x) = \int \dfrac{\dd^3 \vb{k}}{(2\pi)^{3/2}} e^{i\vb{k}\vb{x}} \left( \delta\phi_k(\tau)a_{\vb{k}} + \delta\phi_k^*(\tau)a^\dagger_{-\vb{k}} \right) \,,
\end{equation}
where $^*$ indicates complex conjugation.
The modes $\delta\phi_k$ obey
$\delta\phi''_k + 2aH\delta\phi_k' + (k^2+a^2V_2)\delta\phi_k = 0$ with Bunch-Davies initial conditions $\delta\phi_k(\tau\to-\infty) = e^{-i k \tau} / \sqrt{2ka^2}$, which guarantee canonical commutation rules for $\delta\phi$ \cite{Bunch:1978yq}. Their time evolution is
\begin{equation}
	\delta\phi_k(\tau) = (-k\tau)^{3/2} \left(\alpha_k J_\nu(-k\tau) + \beta_k Y_\nu(-k\tau) \right)\,,
\end{equation}
where $J_\nu$ and $Y_\nu$ are the first and second kind Bessel functions, respectively. We stress that during the first transition $\nu^2$ 	can be negative and therefore $\nu$ can be imaginary. The coefficients $\alpha_k$ and $\beta_k$ are obtained in each phase imposing that $\delta\phi_k$ and $\delta\phi_k'$ are continuous across the boundaries. 

\section{Structure of $\mathcal{P}_\zeta$}\label{structureP} 

\begin{figure}[t]
	\begin{center}
		\includegraphics[width=0.45\textwidth]{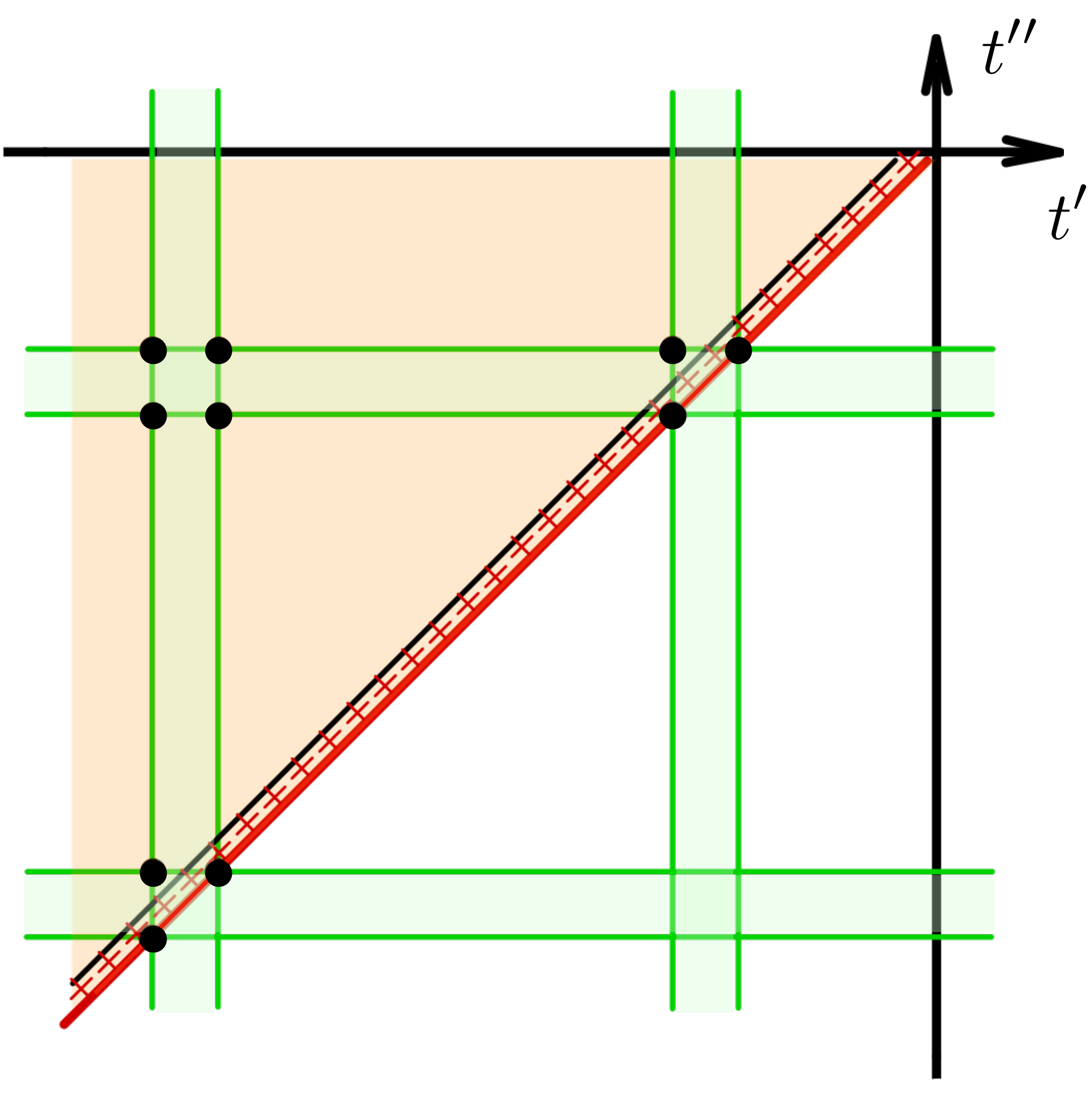}
		\caption{\label{fig:ttplott}  \small \it  The {orange} shaded region represents the integration domain of the (cosmic) time integrals that arises calculating {$\mathcal{P_\zeta}$} at second-order in $H_I$. The red diagonal line represents $t' = t''$. The (thinner) black line parallel to it restricts the integration regime due to the cutoff, see eq.\ (\ref{eq:reg_time_integrals}). The separation between the two lines is given by {$\Lambda^{-1}_{\rm UV}$}. The green lines mark the times of the transitions between different phases. Their intersections within the shaded area correspond to the points (black dots) contributing to the time integrals in our model. At each point, the interaction is a Dirac delta in time, which simplifies the integration greatly.}
	\end{center}
\end{figure}

Due to momentum conservation,
\begin{equation}
	\expval{\delta\phi(\tau,\vb{x}) \delta\phi(\tau,\vb{y})} \equiv \int \dfrac{\dd^3\vb{k}}{(2\pi)^3} e^{i\vb{k}(\vb{x} - \vb{y})} D(\tau,k)\,,
\end{equation}
at all loop orders, for some function
$ D(\tau,k)=D(\tau,k)_{\rm tl}+D(\tau,k)_{\rm 1l}+D(\tau,k)_{\rm ct}\,,$
where we have split the contributions at tree-level (tl), one-loop (1l) and those coming from the counterterms (ct) (see Figure \ref{fig:one-loops}), so that $\mathcal{P}_\zeta (\tau,k) = k^3 D(\tau,k)/(4\pi^2M_P^2\,\epsilon)$.
The tree-level contribution is just
\begin{equation}
	\mathcal{P}_\zeta ^{\rm tl}(\tau,k) = \dfrac{k^3}{4\pi^2M_P^2\epsilon(\tau)} \abs{\delta\phi_k(\tau)}^2 \,.
\end{equation}
The one-loop can be separated into terms coming from cubic and quartic interactions. The quartic part is
\begin{align}\label{eq:D_V4}
\mathcal{P}_\zeta ^{\rm 1l, V_4}(\tau,k) =\dfrac{k^3}{4\pi^2M_P^2\epsilon(\tau)} \Im \left\lbrace  \delta\phi^2_k(\tau) \int ^\tau_{-\infty}  \dd\tau' a^4(\tau') V_4(\tau') \delta\phi^{*2}_k(\tau')  \int \dfrac{\dd^3 \vb{p}}{(2\pi)^3} \abs{\delta\phi_p(\tau')}^2  \right\rbrace\,,
\end{align}
{Since the transitions are instantaneous, the prescription $\tau'_- = \tau'(1-i\omega)$ has no effect on {$\mathcal{P}_\zeta ^{\rm 1l, V_4}$} and we omit it.}
The loop integral has no dependence on the external momentum $k$, so {$\mathcal{P}_\zeta ^{\rm 1l, V_4}$} can be fully absorbed by the counterterms, as we shall see below. The cubic part is
\begin{align}
	\nonumber
	\mathcal{P}_\zeta ^{\rm 1l, V_3}(\tau,k)  = \dfrac{k^3}{\pi^2M_P^2\epsilon(\tau)} & \Im \bigg \lbrace \int_{-\infty}^{\tau+ \frac{1}{a(\tau)\Lambda}}\dd \tau' a^4(\tau'_-)\, V_3(\tau'_-)\, \delta\phi_k(\tau) \delta\phi_k^*(\tau'_-) \times \\ \nonumber 
	& \quad\quad   \times \int_{\tau' + \frac{1}{a(\tau')\Lambda}}^\tau\dd \tau'' a^4(\tau'') V_3(\tau'') \Im{\delta\phi_k(\tau) \delta\phi_k^*(\tau'') } \times \\ 
	 & \quad\quad  \times \int \dfrac{\dd^3 \vb{p}}{(2\pi)^3} \delta\phi_p(\tau'')\, \delta\phi_p^*( \tau'_-) \,\delta\phi_{|\vb{k} - \vb{p}|}(\tau'') \,\delta\phi_{|\vb{k} - \vb{p}|}^*( \tau'_-) \bigg \rbrace\,.
	\label{eq:D_V3}
\end{align}
In this case, the $i\omega$ prescription is important for the convergence of the loop integral, and therefore we must keep it.\footnote{{In principle, we should also keep this prescription in $\tau''$. However, it has no effect and convergence is guaranteed simply by keeping $\tau'_-$.}}
Since the interactions $V_3$ are Dirac deltas at the edges of the transitions, only a few points contribute to the integrals of (\ref{eq:D_V3}). Let us determine which ones are relevant. We can denote by $\tau_1$ the conformal time corresponding to $N=0$ in Figure \ref{fig:model}. Then, we denote as $\tau_2$ and $\tau_3$ the conformal times at which the USR phase starts and ends, respectively. The beginning of the final SR phase is identified with $\tau_4$. Considering the time integrals in \eq{eq:D_V3}, points falling along the diagonal ($\tau' = \tau''$) in the plane $\{\tau',\tau''\}$ are removed by the regulator (see eq.\ \eq{eq:reg_time_integrals}).\footnote{If we used a regulator that does not exclude the points on the diagonal (red line in Figure \ref{fig:ttplott}), such as dimensional regularization, the result of the finite part of $\mathcal{P}_\zeta$ at one-loop would be different. This difference would be compensated by
the finite part of the counterterms. See Appendix \ref{sec:Estimation 1l}, where the equivalence between this type of regulators and a cutoff is shown in the limit $\delta N \to 0$.}
The pairs $(\tau_i, \tau_j)$ that contribute to the one-loop spectrum are those that satisfy $\tau' = \tau_i < \tau'' = \tau_j $, see Figure \ref{fig:ttplott}.
Taking this into account,
\begin{align} \label{cubic1l} \nonumber
	\mathcal{P}_\zeta ^{\rm 1l, V_3}(\tau,k) & = \dfrac{k^3}{2\pi^2M_P^2\epsilon(\tau)} \dfrac{H^2}{M_P^2} \sum_{i = 1}^4 \sum_{j=i + 1}^4 \dfrac{a^3\Delta\nu^2}{\sqrt{\epsilon}}\eval_{\tau_i} \dfrac{a^3\Delta\nu^2}{\sqrt{\epsilon}}\eval_{\tau_j} \times  \\ &  \times \Im{\delta\phi_k(\tau) \delta\phi_{k}^*(\tau_j)} \Im{\delta\phi_k(\tau) \delta\phi_{k}^*(\tau_i)\, l(k;j,i)}\,,
\end{align}
where $\Delta\nu^2(\tau_i)$ is the change in $\nu^2$ at $\tau_i$ and $l(k;j,i)$ comes from the loop integral in the second line of (\ref{eq:D_V3}). We can write  $l(k;j,i)$ conveniently with a change of variables \cite{Espinosa:2018eve},\footnote{We define $p = k(s-d)/2$ and $|\vb{k} - \vb{p}| = k(s+d)/2$ and we use the fact that eq.\ \eq{eq:D_V3} is symmetric under the exchange of $\vb{p}$ and $\vb{k} - \vb{p}$ to integrate over half of the integration domain. }
\begin{align} \nonumber
	 l(k;j,i)  = \quad\quad\quad \quad\quad\quad  \quad\quad\quad  \quad\quad\quad  \quad\quad\quad \quad\quad\quad \quad\quad\quad \quad\quad\quad \quad\quad\quad \quad\quad\quad \quad\quad\quad \quad\quad\quad \\ \dfrac{k^3}{16\pi^2}\int_{1+\Lambda_{\rm IR}/k}^\infty \dd s\int_0^1\dd d\,(s^2-d^2)\,\delta\phi_{k(s-d)/2}(\tau_j)  \delta\phi^*_{k(s-d)/2}(\tau_{i-}) \delta\phi_{k(s+d)/2}(\tau_j) \delta\phi^*_{k(s+d)/2}(\tau_{i-})\,.
	\label{eq:loop_integral}
\end{align}
We note that $l(k;i,j)$ has no UV divergence, thanks to the $\tau_{i-}$ prescription. Without this prescription, $l$ would be oscillatory in the UV
{(see Appendix \ref{sec:limits}). It is also crucial that the contributions from the diagonal of Figure~\ref{fig:ttplott} are excluded by the regulator, as these would introduce a divergent contribution that cannot be absorbed by the counterterms because of the different momentum dependence of eq.\ \eq{cubic1l} and the contribution from the counterterms (see eq.\ \eq{eq:D_ct} below).\footnote{If the points on the diagonal are included, $l(k;i,i)$ would be real and linearly divergent in the UV. The structure of the divergence would be $\mathcal{P}_\zeta^{\rm 1l, V_3} \sim k^3 \Im{\delta\phi_k(\tau) \delta\phi_{k}^*(\tau_i)}^2$, while the structure of the counterterms would be $\mathcal{P}_\zeta^{\rm ct} \sim k^3 \Im{\delta\phi_k^2(\tau) \delta\phi_{k}^{*2}(\tau_i)}$.} Previous works (see e.g.\ \cite{Kristiano:2022maq}) obtained UV divergences from a cubic interaction in the $\zeta$-gauge, whereas we obtain a finite result thanks to the regulator we use (and the appropriate use of the $i\omega$ prescription).}
However, $l$ shows an infrared (IR) divergent part ($l^{\rm IR}$), which we regularize with an IR cutoff $\Lambda_{\rm IR}$. This type of IR divergence arises from eq.\ (\ref{eq:In-In_General}) in perturbation theory for massless free fields. 
In this work we are only concerned with UV divergences, and assume that IR divergences are either unphysical and disappear when calculating physical observables \cite{Gerstenlauer:2011ti,Giddings:2011zd,Senatore:2012nq} or can be addressed beyond perturbation theory \cite{Gorbenko:2019rza}.\footnote{In the latter case, we will assume that the finite contribution of the IR effect will not be larger than the finite part of the rest of the calculation we make.}
Therefore, in practice, we redefine $l$ as follows:	 $l\to l-l^{\rm IR}$.
Finally, the contribution of the counterterms (ct) will be
\begin{align}
	\mathcal{P}_\zeta ^{\rm ct} (\tau,k) = \dfrac{k^3}{4\pi^2M_P^2\epsilon(\tau)}\left( 2\,\delta_\phi\, \abs{\delta\phi_k(\tau)}^2 +
	8 \Im{\delta\phi_k^2(\tau) \int_{-\infty}^\tau \dd \tau' a^2(\delta_\phi k^2 + \tilde{\delta}_V) \delta\phi_k^{*2}\eval_{\tau'}}\right) \,.
	\label{eq:D_ct}
\end{align}
After making the replacement
\begin{equation} \label{eq:D_ct_2}
	\tilde{\delta}_V \to \tilde{\delta}_V -\dfrac{1}{8} a^2V_4 \int \dfrac{\dd^3 \vb{p}}{(2\pi)^3} \abs{\delta\phi_p}^2
\end{equation}
the (divergent and finite) contributions of {$\mathcal{P}_\zeta ^{\rm 1l, V_4}$} can be completely absorbed by the counterterms 
{--as it happens in $\lambda \phi^4$ in Minkowski--. We stress that we can make this redefinition thanks to the arbitrary time dependence of $\tilde{\delta}_V$.}
The situation is different for $\mathcal{P}_\zeta ^{\rm 1l, V_3}$, where the counterterms cannot absorb its finite part for all $k$. Therefore, the only relevant contribution at one-loop --and the only one we include-- comes from {$\mathcal{P}_\zeta ^{\rm 1l, V_3} \equiv \mathcal{P}_\zeta ^{\rm 1l} $}.
Since the only UV divergence appearing in the one-loop calculation is reabsorbed with the counterterms, the two-point correlation is, after this procedure, completely finite.

We note that in Figure \ref{fig:one-loops} there are diagrams that are not relevant in our calculation at one-loop. These diagrams fall in one of the following three categories: 1) bubble diagrams (whose contribution vanishes), 2) diagrams that do not contribute to $\mathcal{P}_\zeta$,\footnote{We have only one diagram in this category, the disconnected one formed by two tadpoles (last diagram of the last row of Figure \ref{fig:one-loops}). Although this diagram does affect the two-point correlation, its momentum structure is different from that of the power spectrum, so it does not modify $\mathcal{P}_\zeta$.} and 3) diagrams proportional to the tadpole (which vanishes
{imposing $\expval{\delta\phi} = 0$ at loop level, using counterterms, as shown in Appendix \ref{sec:Tadpoles}}).

\begin{figure}[t]
	\begin{center}
		\includegraphics[width=0.65\textwidth]{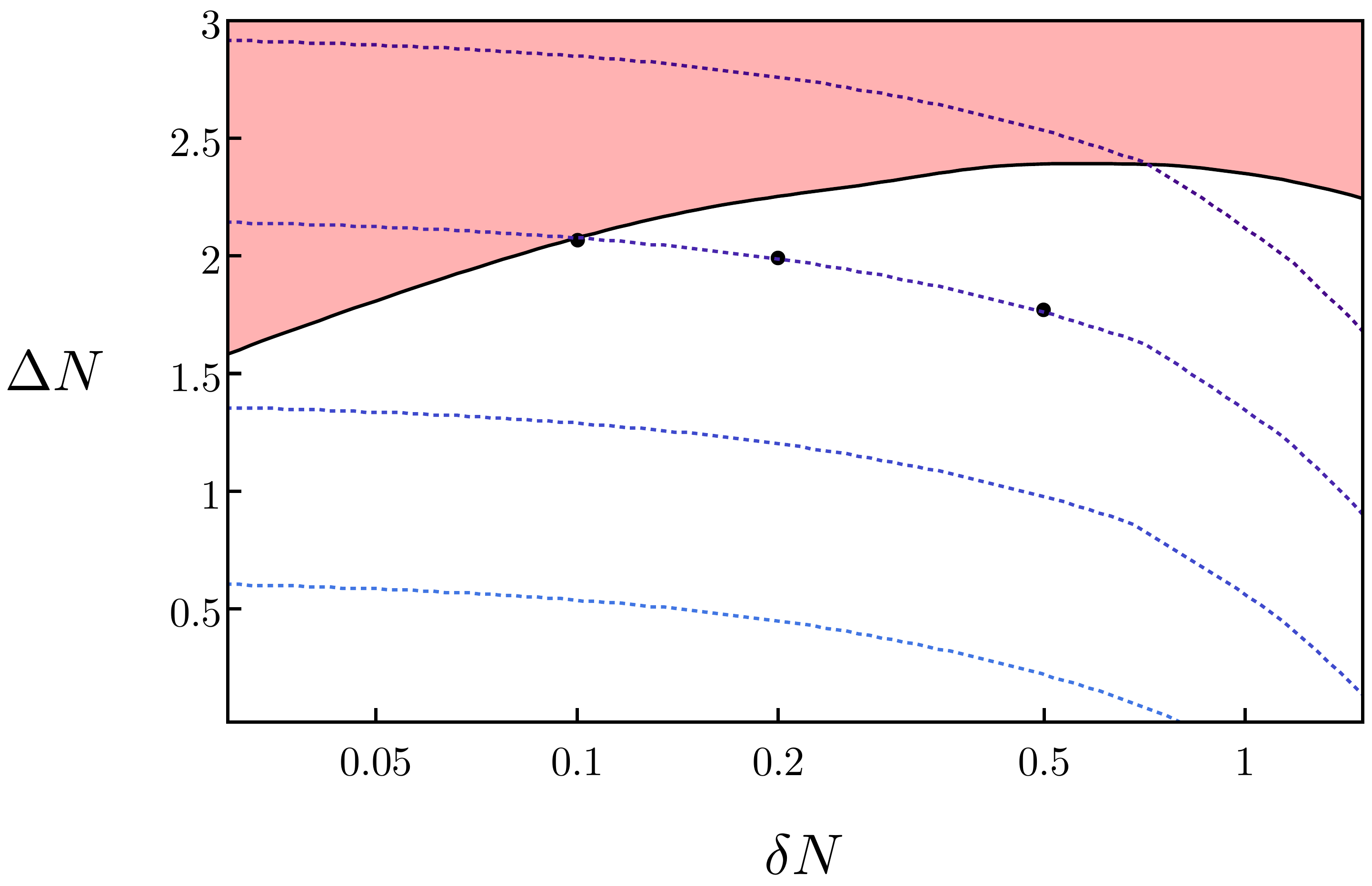}
		\caption{\label{fig:PTbreak}  \small \it The dotted curves represent lines of constant $\mathcal P_{\zeta}$ (with values $1$, $10^{-2}$, $10^{-4}$ and $10^{-6}$ from top to bottom), computed at tree-level. In the shaded region, the one-loop contribution to $\mathcal P_{\zeta}$ is larger than its tree-level counterpart at the scales near the maximum of the latter. The three black dots correspond to the three examples shown in Figure \ref{fig:Set_Of_1loop}. {The region where $\mathcal{P}_\zeta^{\rm 1l} > \mathcal{P}_\zeta^{\rm tl}$ is approximately scale invariant, as can be seen in	 Figure \ref{fig:PTbreak3x3}.}}
	\end{center}
\end{figure}

\section{Discussion}

A full renormalization procedure would require imposing a set of conditions on $\mathcal{P}_\zeta$ allowing us to extract the finite part of the counterterms that make the theoretical prediction of the total power spectrum coincide with the one inferred from observations, which is currently unconstrained at small enough scales. 
Although we have not performed a full renormalization, which would have required working with a tractable functional form for the potential, we can use our results to draw conclusions about the validity of perturbation theory.
The validity of eq.\ (\ref{eq:In-In_General}), which is the basis of our analysis, assumes $\mathcal{P}_\zeta^{\rm 1l}\ll \mathcal{P}_\zeta^{\rm tl}$. For consistency, eq.\ (\ref{eq:In-In_General}) also requires $\mathcal{P}_\zeta^{\rm ct}\sim\mathcal{P}_\zeta^{\rm 1l}\ll \mathcal{P}_\zeta^{\rm tl}$. The contributions to $\mathcal{P}_\zeta$ coming from the counterterms must be of the same order as those coming from the loops because otherwise the divergences cannot be absorbed. 
Taking the condition $\mathcal{P}_\zeta^{\rm 1l}\sim \mathcal{P}_\zeta^{\rm tl}$ as a proxy for perturbation theory breakdown, we find that whether perturbation theory breaks or not depends on the values chosen for the two parameters, $\delta N$ and $\Delta N$, of the model. 

The contribution to $\mathcal{P}_\zeta$ at one-loop depends strongly on the width $\delta N$ of the transitions between SR and USR. Moreover, we find that when $\delta N$ is made arbitrarily small, the one-loop contribution diverges (see Appendix \ref{sec:Estimation 1l} {for details),} invalidating any prediction based on perturbation theory through the in-in formalism. {Identifying this divergence is straightforward in the $\delta\phi$-gauge, but it had been missed in the previous works on the subject, most of which used the $\zeta$-gauge. We find that in the latter gauge, the divergence becomes manifest once the interaction Hamiltonian in the interaction picture, $H_I$, is correctly computed from the single cubic interaction that is more important in that gauge ($S\propto \int d^4x\, \epsilon\,\eta'\,a^2\, \zeta'\,\zeta^2$). Specifically, the piece of $H_I$ that carries the divergence is $\propto \int \dd^3\vb{x}\, a^2 \epsilon\, {\left( \eta'\right) ^2} \zeta^4$.}

\begin{figure}[t!]
	\begin{center}
		\includegraphics[width=1\textwidth]{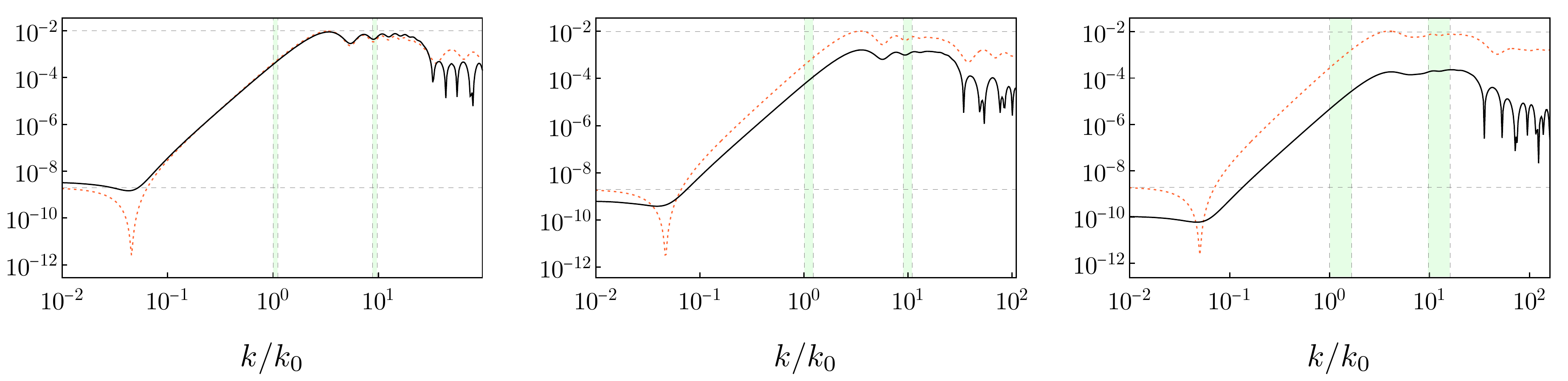}
		\caption{\label{fig:Set_Of_1loop}  \small \it Tree-level (red dashed) and {absolute value} of renormalized one-loop (black continuous) contributions to $\mathcal{P}_{\zeta}(k)$ for three different choices of $\delta N$ and $\Delta N$. From left to right: $\{\delta N =0.1\,,\Delta N=2.07\}$, $\{\delta N =0.2\,,\Delta N=1.99\}$, $\{\delta N = 0.5\,,\Delta N=1.77\}$. The vertical green bands indicate the ranges of momenta becoming super-Hubble during the equally-shaded transitions of Figure \ref{fig:model}. The value $k = k_0$ corresponds to the Fourier mode becoming super-Hubble at the beginning of the first transition (between SR and USR).}
	\end{center}
\end{figure}

In models that can account for all DM in the form of PBHs --which requires {a primordial spectrum} $\mathcal{P}_{\zeta} \sim \order{0.01}$ at the peak in the Gaussian approximation, see e.g.\ \cite{Ballesteros:2017fsr}--, $\delta N$ needs to be  $\gtrsim 0.1 $ for perturbation theory to hold, as shown in Figures \ref{fig:PTbreak} and \ref{fig:Set_Of_1loop}. In realistic USR inflationary models \cite{Ballesteros:2017fsr,Ballesteros:2020qam} this parameter is $\delta N \sim 0.4 - 0.5$. For such values of $\delta N$, perturbation theory does indeed appear to hold in the model we have considered for our analysis. Even though inflation does not end in our model and $\mathcal{P}_\zeta$ does not feature a peak but rather a plateau-like feature for $k\to \infty$, we think that our results are evidence that models in which all of the DM is comprised of PBHs originating from USR inflation are not necessarily hampered by perturbation theory breaking, contrary to what was argued in \cite{Kristiano:2022maq}. 
In future work, our analysis can be extended to include $|\eta| \sim\mathcal{O}(1)$ in the last phase of Figure~\ref{fig:model}, as well as using dimensional regularization to deal with UV divergences in $\mathcal{P_\zeta^{\rm 1l}}$. Another possible direction consists in exploring what happens assuming a given functional form for the potential $V(\phi)$, such as the one in \cite{Ballesteros:2020qam}.

We note that the Gaussian approximation, and hence the benchmark $\mathcal{P}_{\zeta} \sim \order{0.01}$, may not be enough to compute the PBH abundance for realistic USR models accurately. We expect that the effect of small non-Gaussian corrections around the maximum of the distribution of $\zeta$ will not change our results significantly \cite{Taoso:2021uvl}. However, there exist indications that the tail of the distribution in USR models may fall exponentially, differing substantially from a Gaussian one, see e.g.\ \cite{Ezquiaga:2019ftu, Figueroa:2020jkf}. If this were the case, a smaller maximum $\mathcal{P}_{\zeta}$ would be sufficient to account for all DM, which would make attaining the conditions needed for perturbation theory to hold even easier. 

The debate has been intense since \cite{Kristiano:2022maq} appeared. An early reaction already suggested the need of smoothing the transition between the SR and USR phases, see \cite{Riotto:2023hoz}. More detailed calculations implementing smooth transitions gave contradictory results, in some cases supporting the thesis of \cite{Kristiano:2022maq} (see e.g.\ \cite{Firouzjahi:2023aum}) and in others indicating that the tension could be relaxed, see e.g.\ \cite{Franciolini:2023lgy}. Before our work, previous analyses have primarily focused on the specific cubic interaction in the $\zeta$-gauge that was first considered in \cite{Kristiano:2022maq} 
($S\propto \int d^4x\, \epsilon\,\eta'\,a^2\, \zeta'\,\zeta^2$).  {As we have mentioned above, it} turns out that the quartic interaction induced by this cubic term in the action for $\zeta$, which has been ignored in previous works, is the origin of the divergence in the limit of instantaneous transitions ($\delta N \to 0$) in this gauge (as shown in Appendix~\ref{sec:Estimation 1l}).
However, the quartic action \cite{Firouzjahi:2023aum} and some boundary terms \cite{Fumagalli:2023hpa,Tada:2023rgp,Kristiano:2023scm,Firouzjahi:2023bkt} were also partially considered before our work.\footnote{In the cubic action in the $\zeta$-gauge \cite{Maldacena:2002vr}, there are several boundary terms arising from integrating by parts, see also \cite{Wang:2013zva}. It is necessary to include all of them in $H_i$ to compute correlation functions correctly. However, the boundary terms without time derivatives of the fields cannot contribute to any correlator, as it is argued e.g.\ in \cite{Burrage:2011hd} using path integral arguments. See \cite{Weinberg:2005vy} for more details on path integral methods in the in-in formalism {and \cite{Braglia:2024zsl} for recent arguments about the validity of this statement.} Although several boundary terms disappear from the cubic action for $\zeta$ after the field redefinition of \cite{Maldacena:2002vr}, care must be taken to always include all relevant contributions.}
It is worth mentioning the work presented in \cite{Tada:2023rgp}, where it was claimed that Maldacena's consistency relation \cite{Maldacena:2002vr} and the $i\omega$ prescription guarantee that the one-loop power spectrum has to be suppressed with respect to the tree-level one. We have however verified that the consistency relation is satisfied in our analysis (see Appendix~\ref{sec:Bispectrum}). 

In previous works, the loop integrals (mainly involving the cubic interaction in the $\zeta$-gauge mentioned above) have been cut between two values of the momentum (in particular, between the wavenumbers becoming super-Hubble at the beginning and at the end of USR). That procedure makes the resulting computation of the one-loop contribution to $\mathcal{P}_\zeta$ dependent on those cutoffs, casting a doubt on any conclusions extracted from it. In our analysis, we have dealt with this problem regularizing and absorbing the relevant divergences with appropriate counterterms. We have found that, after regularization, the cubic interactions in the $\delta\phi$-gauge give a finite one-loop contribution to $\mathcal{P}_\zeta$. However, the quartic interactions (on the same gauge) required appropriate counterterms to absorb divergent contributions.

{A fraction of the earlier literature on the topic of our work has questioned whether $\mathcal{P}_\mathcal{\zeta}$ at long wavelengths can be affected (substantially or even at all) by linear Fourier modes of $\zeta$ that are orders of magnitude shorter \cite{Riotto:2023hoz,Firouzjahi:2023ahg,Tada:2023rgp,Inomata:2024lud}, as it indeed occurs in the problem we have studied. This question raises from the general notion of decoupling, according to which very small distance scales should not affect phenomena at much larger distances. This idea is at the core of any effective field theory, many of which have proven to be very successful to describe nature. 
To address this question (decoupling) we note that one-loop and counterterm contributions to $\mathcal{P}_\mathcal{\zeta}$ are scale invariant in the limit of vanishing wavenumber (see Appendix \ref{sec:limits}), so the former can be entirely absorbed by the latter. In addition, the tree-level contribution is also scale invariant, which makes it indistinguishable from the other two.
}

\vspace{0.3cm}

\section*{Acknowledgments}

This work has been funded by the following grants: 1) Contrato de Atracci\'on de Talento (Modalidad 1) de la Comunidad de Madrid (Spain), 2017-T1/TIC-5520 and 2021-5A/TIC-20957, 2)  PID2021-124704NB-I00 funded by MCIN/AEI/10.13039/501100011033 and by ERDF A way of making Europe, 3) CNS2022-135613 MICIU/AEI/10.13039/501100011033 and by the European Union NextGenerationEU/PRTR, 4) the IFT Centro de Excelencia Severo
Ochoa Grant CEX2020-001007-S, funded by MCIN/AEI/10.13039/501100011033. JGE has been supported by a PhD contract {\it contrato predoctoral para formaci\'on de doctores} PRE2021-100714 associated to the aforementioned Severo Ochoa program, CEX2020-001007-S-21-3. 
We thank Gabriele Franciolini and Thomas Konstadin for comments and  Marco Taoso for questions. We thank Jaime Fern\'andez Tejedor for comments and discussions. We thank Jos\'e Ram\'on Espinosa for discussions.

\appendix

\section{In-in formalism in more detail} \label{sec:in_in_form}

Let $Q(t)$ be a generic Hermitian operator whose vacuum expectation value (VEV) we want to calculate
\begin{equation}
	\expval{Q(t)} \equiv \sideset{_H}{}{\mathop{\bra{\Omega}}}  Q_H(t) \ket{\Omega}_H,
\end{equation}
where $\ket{\Omega}_H$ 
{is the ground state}, i.e.\ 
the vacuum of the system. This is customarily done in the so-called in-in formalism; see \cite{Schwinger:1960qe,Keldysh:1964ud} and also \cite{Weinberg:2005vy,Chen:2010xka,Wang:2013zva} for later descriptions. We work in the Heisenberg ($_H$) picture, in which $\ket{\Omega}_H$ does not evolve in time, and instead all the time evolution falls on the operators, such as $Q_H(t)$. The time evolution of the operators in this picture is given by the equation
\begin{equation}
	Q_H(t) = U^{-1} (t,t_0)Q_S\, U(t,t_0)\,,
\end{equation}
where $U^{-1} (t,t_0)U(t,t_0)=1$\,.
Here we are introducing an auxiliary time $t_0$ in which all objects in the Heisenberg and Schr\"odinger ($_S$) picture coincide. $U(t,t_0)$ is the time evolution operator that satisfies
\begin{equation}
	\partial_t U(t,t_0) = -i H(t) U(t,t_0)\,,\quad U(t_0,t_0) = 1\,.
\end{equation}
The Hamiltonian $H(t)$ can be separated into a free part and an interaction part, $H = H_0 + H_{\rm i}$. Thus, we can define the operators in the interaction picture ($_I$) 
\begin{align}
	Q_I(t) = U_0^{-1}(t,t_0) Q_S\, U_0(t,t_0)\,,
\end{align}
where
\begin{align}
\partial_t U_0(t,t_0) = -i H_0(t) U_0(t,t_0)\,,\quad U_0(t_0,t_0) = 1\,.
\end{align}
The operators in the interaction picture follow the dynamics governed by the free (quadratic) Hamiltonian, without interactions. 
That is, one can understand the operators in the interaction picture as operators in the Heisenberg picture, but with the Hamiltonian of the system being simply the free Hamiltonian. Therefore, the fields in the interaction picture can be 
expressed in terms of the creation and annihilation operators {--which will allow us to use Wick's theorem when computing correlators--}.
We can now write the operators in the Heisenberg picture in terms of the operators in the interaction picture,
\begin{align}
	Q_H(t) = F^{-1}(t,t_0) Q_I(t) F(t,t_0),\quad F(t,t_0) \equiv U_0^{-1}(t,t_0) U(t,t_0)\,.
\end{align}
Differentiating, we obtain the equation that is satisfied by the operator $F(t,t_0)$:
\begin{align} \label{eqf}
\partial_t F(t,t_0) = -i H_I(t) F(t,t_0)\,,
\end{align}
where the Hermitian operator
\begin{align}
H_I(t) \equiv U_0^{-1}(t,t_0) H_{\rm i}(t) U_0(t,t_0)
\end{align}
represents the interaction Hamiltonian in the interaction picture, and then the fields that compose $H_I$ are in the interaction picture. The formal solution of eq.\ \eq{eqf}  and its inverse are, respectively:
\begin{align} \label{Fdef}
F(t,t_0) = T \exp \left( -i \int_{t_0}^t \dd t' H_I(t') \right)\,,\quad F^{-1}(t,t_0) = \bar{T} \exp \left(i \int_{t_0}^t \dd t' H_I(t') \right)\,,
\end{align}
where $T$ and $\bar T$ indicate time and anti-time ordering.
The VEV of $Q(t)$ is then
\begin{equation}
	\expval{Q(t)} = \sideset{_H}{}{\mathop{\bra{\Omega}}}  F^{-1}(t,t_0) Q_I(t)F(t,t_0) \ket{\Omega}_H\,.
\end{equation}
We note that all the elements of the latter expression are in the interaction picture except for $\sideset{_H}{}{\mathop{\bra{\Omega}}} $ and $\ket{\Omega}_H$.  Therefore, we need to rewrite this expression using the vacuum in the interaction picture $\ket{0}_I$ (i.e.\ the vacuum of the Hamiltonian without interactions). Like $\ket{\Omega}_H$, $\ket{0}_I$ does not evolve over time. Let us consider
\begin{align}
	\nonumber
	&\exp \left(-i \int_{t_0}^t \dd t' H(t') \right) \ket{0}_I = \exp \left(-i \int_{t_0}^t \dd t' H(t') \right) \sum_n  \ket{n}_H \sideset{_H}{_I}{\mathop{\braket{n}{0}}}\\ \nonumber
	&=\exp \left(-i \int_{t_0}^t \dd t' E_\Omega(t') \right) \bigg\{\sideset{_H}{_I}{\mathop{\braket{\Omega}{0}}} \ket{\Omega}_H \\ 
	& \quad\quad\quad\quad\quad\quad\quad\quad\quad\quad\quad + \sum_{n\neq \Omega}  \exp \left(-i \int_{t_0}^t \dd t' \left(E_n(t')-E_\Omega(t')\right) \right) \:\sideset{_H}{_I}{\mathop{\braket{n}{0}}}\ket{n}_H\bigg\}\,.
\end{align} 
Since $E_n(t) > E_\Omega(t)$ for all $n$ and $t$, if we make the replacement $t_0 \to  -\infty^- \equiv -\infty(1-i\omega)$ with $\omega>0$, {all excited states are suppressed and we get}
\begin{equation}
	\exp \left(-i \int_{-\infty^-}^t \dd t' H(t') \right) \ket{0}_I = \sideset{_H}{_I}{\mathop{\braket{\Omega}{0}}}  \exp \left(-i \int_{-\infty^-}^t \dd t' H(t') \right) \ket{\Omega}_H  .
\end{equation}
{Applying time ordering,}
\begin{equation}
	U(t,-\infty^-) \ket{\Omega}_H = \dfrac{U(t,-\infty^-)\ket{0}_I}{\sideset{_H}{_I}{\mathop{\braket{\Omega}{0}}}}\,,
\end{equation}	
	{and we obtain the following useful relation between the vacuum in the interaction and Heisenberg picture:\footnote{We interpret the prescription $i\omega$ as a way to turn off the interactions in the limit $t\to -\infty$, well motivated by the Bunch-Davies initial conditions.}}
\begin{equation}
F(t,-\infty^-) \ket{\Omega}_H = \dfrac{F(t,-\infty^-)\ket{0}_I}{\sideset{_H}{_I}{\mathop{\braket{\Omega}{0}}} } \equiv \mathcal{N}\, F(t,-\infty^-)\ket{0}_I\quad\end{equation}
and
\begin{equation}
\sideset{_H}{}{\mathop{\bra{\Omega}}}  F^{-1}(t,-\infty^+) 
= \mathcal{N}^* \sideset{_I}{}{\mathop{\bra{0}}} F^{-1}(t,-\infty^+) \,,
\end{equation}
where $\infty^+ \equiv \infty(1+i\omega)$. This allows us to write all the elements of the VEV in the interaction picture:
\begin{equation}
 \expval{Q(t)} = \abs{\mathcal{N}}^2\sideset{_I}{}{\mathop{\bra{0}}} F^{-1} (t,-\infty^+)Q_I(t)F(t,-\infty^-) \ket{0}_I \,,
\end{equation}
where  $F^{-1} (t,-\infty^+)$ can be read from \eq{Fdef}.
The last piece we need to understand is what is $\abs{\mathcal{N}}^2$. Assuming that $\ket{\Omega}_H$ is properly normalized, i.e.\ $\sideset{_H}{_H}{\mathop{\braket{\Omega}{\Omega}}}  = 1$, if we calculate the VEV of the identity operator ($Q = 1$), we get that 
\begin{equation}
	\frac{1}{\abs{\mathcal{N}}^2} =  \sideset{_I}{}{\mathop{\bra{0}}} F^{-1}(t,-\infty^+) F(t,-\infty^-) \ket{0}_I\,.
\end{equation}
Making a perturbative expansion, it becomes clear that  $\abs{\mathcal{N}}^{-2}$ is the sum over the bubble-type diagrams, defined as the set of diagrams with at least one part disconnected from any external leg.
Thus, the final formula to be used for the calculation of the VEV of any operator $Q(t)$ is (see also \cite{Senatore:2016aui})
\begin{align} \nonumber
	\expval{Q(t)}  & = \dfrac{\sideset{_I}{}{\mathop{\bra{0}}} F^{-1}(t,-\infty^+) Q_I(t)F(t,-\infty^-) \ket{0}_I}{\sideset{_I}{}{\mathop{\bra{0}}} F^{-1}(t,-\infty^+) F(t,-\infty^-) \ket{0}_I}\\ & = \sideset{_I}{}{\mathop{\bra{0}}} F^{-1}(t,-\infty^+) Q_I(t)F(t,-\infty^-) \ket{0}_I \eval_{\textrm{no bubbles}}\,.
\end{align}
For brevity, we write $\ket{0}_I$ as $\ket{0}$ in what follows (and in the main body of this work) and, even if we will not write it explicitly, bubble diagrams will be assumed to be removed.

The main expression we use in this paper, eq.\ \eq{eq:In-In_General}, is a second order expansion in the interaction Hamiltonian of the in-in formalism for the calculation of a VEV. To derive it we proceed by first expanding $F(t,-\infty^-) $ from eq.\ \eq{Fdef}:	
\begin{align}
	F(t,-\infty^-) = 1 -i \int_{-\infty^-}^t \dd t' H_I(t') - \dfrac{1}{2} T \int_{-\infty^-}^t \dd t' \int_{-\infty^-}^t \dd t'' H_I(t') H_I(t'')\,.
\end{align}
Now, we make a redefinition of time such that the prescription $i\omega$ passes from the lower integration limit to the time of the Hamiltonian, {$t\rightarrow t_{-}=t(1-i\omega)$,} such that
\begin{align}
	F(t,-\infty^-) = 1 -i \int_{-\infty}^t \dd t' H_I(t'_-) - \dfrac{1}{2} T \int_{-\infty}^t \dd t' \int_{-\infty}^t \dd t'' H_I(t'_-) H_I(t''_-)\,.
\end{align}
In principle, the upper integration limit should also be modified, 
{but this change has no effect when, at the end of the calculation, we take $\omega\to 0$. The only effect of the prescription $i \omega$ is observed in the limit $t',t''\to -\infty$, introducing a damping in the time integrals, 
which is interpreted as a projection onto the interaction vacuum.} Now we use the definition of time ordering to rewrite
\begin{align} \nonumber
	\dfrac{1}{2} T \int_{-\infty}^t \dd t' \int_{-\infty}^t \dd t'' H_I(t'_-) H_I(t''_-) & = \int_{-\infty}^t \dd t' \int_{-\infty}^{t'} \dd t'' H_I(t'_-) H_I(t''_-) \\ & = \int_{-\infty}^t \dd t' \int_{t'}^t \dd t'' H_I(t''_-) H_I(t'_-)\,.
\end{align}
Therefore, $F$ and its inverse are
\begin{align}
	F(t,-\infty^-) & = 1 -i \int_{-\infty}^t \dd t' H_I(t'_-) - \int_{-\infty}^t \dd t' \int_{t'}^t \dd t'' H_I(t''_-) H_I(t'_-)\,,\\
	F^{-1}(t,-\infty^+	) & = 1 + i \int_{-\infty}^t \dd t'\left(  H_I(t'_-)\right) ^\dagger - \int_{-\infty}^t \dd t' \int_{t'}^t \dd t'' \left( H_I(t''_-) H_I(t'_-)\right) ^\dagger\,,
\end{align}
so that the VEV of $Q$ can be written as
\begin{align}
	\nonumber
	\expval{Q(t)}  = \bra{0} Q_I(t)\ket{0} & +  2\Im{\int_{-\infty}^t\dd t'\bra{0} Q_I(t) H_I(t'_-) \ket{0}} \\ \nonumber & 
	+\int_{-\infty}^t\dd t' \int_{-\infty}^t\dd t'' \bra{0} H_I(t''_+) Q_I(t) H_I(t'_-)\ket{0}\\ & - 2 \Re{ \int_{-\infty}^t \dd t' \int_{t'}^t \dd t'' \bra{0} Q_I(t) H_I(t''_-) H_I(t'_-)\ket{0}}\,.
	\label{eq:App_InIn_ExpVa}
\end{align}
Using that
\begin{equation}
	\bra{0} H_I(t'_+) Q_I(t) H_I(t''_-)\ket{0} = \bra{0} H_I(t''_+) Q_I(t) H_I(t'_-)\ket{0}^*
\end{equation}
we can rewrite the last term of the first line of (\ref{eq:App_InIn_ExpVa}) as
\begin{equation}
	\int_{-\infty}^t\dd t' \int_{-\infty}^t\dd t'' \bra{0} H_I(t''_+) Q_I(t) H_I(t'_-)\ket{0} = 2\Re{\int_{-\infty}^t \dd t' \int_{t'}^t \dd t'' \bra{0} H_I(t''_+) Q_I(t)  H_I(t'_-)\ket{0}}\,,
\end{equation}
so that the final expression is
\begin{align} \nonumber
	\expval{Q(t)} &=  \bra{0} Q_I(t)\ket{0} + 2\Im{\int_{-\infty}^t\dd t'\bra{0} Q_I(t) H_I(t'_-) \ket{0}} \\ &\quad + 2\Re{\int_{-\infty}^t \dd t' \int_{t'}^t \dd t'' \bra{0} \left( H_I(t''_+)Q_I(t) - Q_I(t) H_I(t''_-) \right)  H_I(t'_-)\ket{0}}\,.
\end{align}
We emphasize that this expression does not include the bubble diagrams and we stress the importance of the prescription $i \omega$ (represented by $t_\pm$) for the convergence of the {time} integrals.

\section{Action for fluctuations in the $\delta\phi$-gauge} \label{sec:calculation_action}

In the ADM formalism \cite{Arnowitt:1962hi,Maldacena:2002vr,Wang:2013zva} the spacetime metric is written as follows:
\begin{align}
	ds^2 = -N^2 \dd t^2 + \gamma_{ij} \left(N^i\dd t + \dd x^i \right) \left(N^j\dd t + \dd x^j \right)\,,
	\label{eq:ADM_Decomposition}
\end{align}
where the {\it lapse} $N$ and the {\it shift $N^i$} are non-dynamical quantities acting as Lagrange multipliers in the action, and are therefore determined in terms of the dynamical quantities of the system. We use the scalar-vector-tensor decomposition of the metric
\begin{align}
	\gamma_{ij} = a^2 \left[ e^\Gamma\right]_{ij}\,,\quad  \Gamma_{ij} = 2\zeta \delta_{ij} + \partial_{ij} E + \partial_{(i} E_{j)} + h_{ij}\,,\quad
	N = 1 + \alpha\,,\quad N_i = \partial_i \beta + \beta_i\,,
	\label{eq:SVT_Decomposition}
\end{align}
where $\partial^iE_i =\partial^i\beta_i = \partial^ih_{ij} = \tensor{h}{^i_i} = 0$. We also decompose the inflaton field $\phi$ into a homogeneous background $\phi_0$ and fluctuations:
\begin{align} \label{decphi}
\phi(t,x) = \phi_0(t) + \delta\phi(t,x)\,.
\end{align}
Doing an infinitesimal coordinate (gauge) transformation we can eliminate two scalar and two vector degrees of freedom, leaving only one scalar and two tensor physical degrees of freedom. 
{Although we are primarily interested in scalar modes, for completeness we will also include tensor modes.\footnote{By including the tensor modes, we can obtain the action describing the gravitational waves induced at second order by scalar fluctuations.}}
One possible choice of gauge is: $\delta\phi = E = 0$, which is known as the $\zeta$-gauge. The variable $\zeta$, defined via eq.\ \eq{eq:SVT_Decomposition}, {eventually} becomes constant for $k \ll H$ \cite{Maldacena:2002vr,Weinberg:2003sw}, which allows us to relate it to observable quantities in a simple way, see e.g.\ \cite{Baumann:2022mni}.
However, the one-loop calculation of $\mathcal{P}_\zeta$ in the model we are interested in is simplified if we work in the $\delta\phi$-gauge, where $\zeta = E = 0$. This is because in the inflationary model we consider, the scalar interactions coming from the Ricci scalar $R$ in the action
\begin{equation}
	S = \dfrac{1}{2}\int \sqrt{-g} \:\dd^4x \bigg\{ M_P^2 R - g^{\mu\nu}\partial_\mu \phi \partial_\nu \phi - 2V(\phi)\bigg\}\,,
	\label{eq:General_Action}
\end{equation}
are suppressed with respect to the interactions coming from the potential $V(\phi)$. The $\delta\phi$-gauge allows us to study simultaneously cubic and quartic interactions, avoiding writing interactions in the form of boundary terms that may become relevant in the loop calculation \cite{Arroja:2011yj,Fumagalli:2023hpa}.\footnote{In ref.\ \cite{Arroja:2011yj}, the role of the boundary terms in the tree-level calculation of the bispectrum of $\zeta$ is analyzed, concluding that there is only one relevant boundary term and that it is cancelled by the field redefinition introduced by Maldacena in \cite{Maldacena:2002vr}. However, in the calculation of other observable quantities, and specifically at loop level, there is a larger set of relevant boundary terms that needs to be included in the calculations. This difficulty is circumvented using the $\delta\phi${-}gauge.} 

Let us see which are the most relevant interactions of the action (\ref{eq:General_Action}) in the $\delta\phi$-gauge under the assumption that $\epsilon \equiv  -\dot{H}/H^2\ll 1$ and $\eta\equiv \dot{\epsilon}/(H\epsilon) \sim 1$ at most. We write the components of the metric and its inverse as
\begin{align}
	g_{\mu \nu}=\left(\begin{array}{cc}
		-N^{2} + N^{k} N_{k} & N_{i} \\
		N_{j} & \gamma_{i j}
	\end{array}\right) 	 , \quad g^{\mu \nu}=\left(\begin{array}{cc}
		-1 / N^{2} & N^{i} / N^{2} \\
		N^{j} / N^{2} & \gamma^{i j} - N^{i} N^{j} / N^{2}
	\end{array}\right) .
\end{align}
The Ricci scalar decomposes according to
\begin{align}
	R = R^{(3)} - \dfrac{1}{N^2}(K^2 -K^{ij} K_{ij}) + \textrm{{boundary term}}\,,\quad
	K_{ij} = \dfrac{1}{2}\left(\dot\gamma_{ij} - \nabla_iN_j-\nabla_j N_i \right)\,,\quad K = \tensor{K}{^i_i}\,.
\end{align}
We use that the spatial indices (which we have been denoting with Latin letters) are risen and lowered with $ \gamma_{ij} $, so that $ N^i = \gamma^{ij}N_j $, where $ \gamma^{ij} $ is the inverse of $ \gamma_{ij} $. In our notation, $R^{(3)} $ is the Ricci {scalar} for $ \gamma_{ij} $, $ \nabla_i $ is the covariant derivative associated to $ \gamma_{ij} $ and $ N^{-1} K_{ij} $ is the extrinsic curvature. In addition, we now have $ \sqrt{-g} = N \sqrt{\gamma} $. Therefore, the action (\ref{eq:General_Action}) is rewritten as
\begin{align} \nonumber
	S = \int \dd^4 x N\sqrt{\gamma} \bigg\lbrace & \dfrac{M_P^2}{2} \left( R^{(3)} - \dfrac{1}{N^2}\left(K^2 -K^{ij} K_{ij}\right)\right)  \\ &+ \dfrac{1}{2N^2}(\dot{\phi}-N^i\partial_i\phi)^2-\dfrac{1}{2}\gamma^{ij}\partial_i\phi\partial_j\phi -V(\phi)\bigg\rbrace\,,
	\label{eq:Action_ADM}
\end{align}
where we have eliminated the boundary terms so that the variational principle is well defined \cite{Gibbons:1976ue,York:1972sj,Dyer:2008hb}. This action is valid at any order in perturbation theory. As anticipated, the lapse $ N $ and the shift $ N_i $ have no time derivatives, so these fields behave as Lagrange multipliers.  Varying the action with respect to them, we obtain that
\begin{align}
	M_P^2\left( \partial_j \dfrac{K}{N} - \gamma_{jk} \nabla_i \dfrac{K^{ik}}{N} \right) +\dfrac{1}{N} (\dot{\phi}-N^i\partial_i\phi) \partial_j \phi & = 0\,, \label{eq:eom_Ni}\\
	\dfrac{M_P^2}{2}\left(R^{(3)} +\dfrac{1}{N^2}\left(K^2 -K^{ij} K_{ij}\right) \right) -\dfrac{1}{2N^2} (\dot{\phi}-N^i\partial_i\phi)^2-\dfrac{1}{2}\gamma^{ij}\partial_i\phi\partial_j\phi - V(\phi) & = 0\,, \label{eq:eom_N}
\end{align}
where
\begin{align}
	\nabla_i \dfrac{K^{ik}}{N} \equiv \partial_i \dfrac{K^{ik}}{N} + \dfrac{K^{ik}}{N} \Gamma^j_{ij} + \dfrac{K^{ij}}{N}\Gamma^k_{ij}\,.
\end{align}
By solving these algebraic equations, we can reintroduce $N$ and $N_i$ into the action, eliminating the Lagrange multipliers.
{We are interested in the action in the $\delta\phi$-gauge up to order four in powers of $\delta\phi$ and $h_{ij}$.} In order to get there, we start by studying next the properties of the potential, the lapse and the shift.

\subsection{Expansion of the potential}
The interaction terms coming from the potential are obtained by making the expansion
\begin{equation}
	V(\phi) = V(\phi_0 + \delta\phi) = \sum_{n=0}^\infty V_n(\phi_0) \dfrac{\delta\phi^n}{n!} \;\; \textrm{where} \;\;
	V_n(\phi_0) =  \dfrac{\partial^n V(\phi_0)}{ \partial \phi_0^n}  = \dfrac{1}{\dot{\phi_0}}  \dfrac{\dd V_{n-1}(\phi_0)}{\dd t}\,.
\end{equation}
Using that $\dot{\phi_0}^2=2H^2M_P^2 \epsilon$ and $ V_0=H^2M_P^2(3-\epsilon)$ and defining $\epsilon = -\dot{H}/H^2$, $\eta = \dot{\epsilon}/(H\epsilon)=\epsilon_2$ and, in general,	 $\epsilon_{n}=\dot{\epsilon}_{n-1}/(H\epsilon_{n-1})$ for $n\geq 2$, the derivatives of the potential are
\begin{align}
	V_1 & = \frac{H^2 M_P \sqrt{\epsilon }  }{\sqrt{2}} (-6+2 \epsilon-\eta  )\,,\\
	V_2 & = -\frac{1}{4} H^2 \left(8 \epsilon ^2 -2 \epsilon(12 + 5 \eta )+ \eta  (6+\eta +2 \epsilon_3 )  \right) \simeq -H^2 \left(\nu ^2 - \dfrac{9}{4} \right)\,,\\ \nonumber
	V_3 & = \frac{H^2}{2 \sqrt{2} \sqrt{\epsilon } M_P} \left(8 \epsilon ^3-6 \epsilon ^2 (4 + 3 \eta) +\epsilon \eta (18 + 6 \eta +7 \epsilon_3)-\eta  \epsilon_3  (3 + \epsilon_4 +\eta +\epsilon_3) \right)\\ &  \simeq - \dfrac{H}{\sqrt{2\epsilon}M_P} \dot{\left(\nu^2 \right)}\,,\\
	\nonumber
	V_4 &  = -\frac{H^2}{8\epsilon  M_P^2} \bigg(32 \epsilon ^4-16 \epsilon ^3 (6+7 \eta ) +2 \epsilon ^2\,\eta\, (72 + 39 \eta +32 \epsilon_3) \\\nonumber & \quad\quad\quad\quad\quad \quad -\epsilon\, \eta \left(6 \eta ^2+\eta  (18 + 35 \epsilon_3) + 6 \epsilon_3  (8 + 3 \epsilon_4 +3 \epsilon_3 )\right)\\ \nonumber
	& \quad\quad\quad\quad\quad\quad+\eta\, \epsilon_3  \left(-\eta ^2 + \eta  (-3 + \epsilon_4 +3 \epsilon_3)+2 \epsilon_4  (3 + \epsilon_4 +\epsilon_5 )+6 \epsilon_3 (1 + \epsilon_4)  +2\epsilon_3 ^2\right) \bigg) \\ & \simeq - \dfrac{1}{2\epsilon M_P^2} \left(\ddot{\left(\nu^2 \right)} - \dfrac{H\eta}{2}\dot{\left(\nu^2 \right)} \right)\,,
\end{align}
where
\begin{align}
\nu^2 \equiv \frac{9}{4} + \frac{1}{2}\left(3\,\eta + \frac{\eta^2}{2} + \frac{\eta'}{a\,H	}\right) \in \mathbb{R}\,,
\end{align}
and we recall that primes denote derivatives with respect to conformal time {and dots to cosmic time}. 
In the approximations leading to the above expressions, we have applied that $\epsilon\ll1$ and 
{$\epsilon_n \sim 1$ for $n\geq 2$}. By changing to conformal time we recover the expressions given in eqs.\ \eq{derivs} and \eq{derivs2}.
We note that $V_n \sim \epsilon^{1-n/2}$ for $n\geq 1$. This implies that in the limit $\epsilon\ll1$ the interactions coming from the potential  dominate over those coming from the purely gravitational part of the action, since these are not accompanied by this enhancement, {as we shall see below}. 

\subsection{Expansions of $N$ and $N_i$}
Since we want to calculate the action up to fourth order, we must obtain the fluctuations of $N$ and $N_i$ up to second order. This is because when calculating the action at order $n$, the fluctuations at order $n$ and $n-1$ of both the lapse and the shift cancel out \cite{Wang:2013zva}. We define
\begin{equation} \label{defalphabeta}
	N = 1 + \alpha = 1 + \alpha_1 + \alpha_2 + \dots\,,\quad N_i = \partial_i \beta + \beta_i = \partial_i \beta_1 + \partial_i \beta_2 + \beta_{1,i} + \beta_{2,i} + \dots,
\end{equation}
{where the subscript $_n$ (e.g. in $\alpha_n$)} refers to the perturbative order. 
In this way, we seek to obtain $\alpha$, $\beta$ and $\beta_i$, for which we only have to use the eqs.\ (\ref{eq:eom_Ni}) and (\ref{eq:eom_N}) and expand them perturbatively.
Starting with the first order, we obtain
\begin{equation}
	\alpha_1 = \sqrt{\dfrac{\epsilon}{2}} \dfrac{\delta\phi}{M_P}\,,\quad  \dfrac{1}{a^2}\partial^2 \beta_1 = \dfrac{\sqrt{\epsilon}}{2\sqrt{2}M_P} \left( H\eta\delta\phi - 2\dot{\delta\phi} \right)\,,\quad \beta_{1,i} = 0.
	\label{eq:N_Ni_order1}
\end{equation}
We stress that $\alpha_1$ and $ \beta_1$ are order $\sqrt{\epsilon}$.  We now study the lapse and the shift at second order,
\begin{align} \nonumber
	0 =\,& 2H^2 M_P^2(\epsilon-3)\alpha_2 - \dfrac{2H M_P^2}{a^2}\partial^2\beta_2  - \dfrac{2-\epsilon}{4} \dot{\delta\phi}^2 - \dfrac{H\epsilon\eta}{4} \delta\phi \dot{\delta\phi}\\
	\nonumber
	& + \dfrac{1}{2a^2} \left( 2\sqrt{2\epsilon}H M_P \partial^i \beta_1  -\partial^i\delta\phi \right) \partial_i\delta\phi + \dfrac{H^2}{16}\left( -8\epsilon^2 +\epsilon\left(24-4\eta +\eta^2 \right) +8\left( \nu^2-\dfrac{9}{4} \right)  \right) \delta\phi^2   \\
	&- \dfrac{M_P^2}{8} \left( \dot{ h}^{ij}\dot{ h}_{ij} + \dfrac{1}{a^2}\partial^k h^{ij} \partial_k h_{ij} \right) + \dfrac{M_P^2}{2a^2} \left( 4Hh^{ij} +\dot{ h}^{ij} - \dfrac{1}{a^2} \partial^{ij} \beta_1 \right) \partial_{ij} \beta_1\,,
\end{align}
\begin{align} \nonumber
	0 =\,& \dfrac{M_P^2}{2a^2}\partial^2\beta_{2,j} - 2H M_P^2 \partial_j \alpha_2 + \dfrac{M_P^2}{4} \left(\dot{h}^{ik} \left(\partial_j h_{ik} - \partial_k h_{ij} \right) +h^{ik} \partial_k \dot{h}_{ij}  \right) - \dfrac{M_P^2}{2a^2} \partial^2h_{ij} \partial^i\beta_1  \\
	& + \dfrac{\sqrt{\epsilon}M_P}{2\sqrt{2}} \dot{h}_{ij} \partial^i\delta\phi - \dfrac{\sqrt{\epsilon}M_P}{\sqrt{2}a^2} \partial^i\delta\phi \partial_{ij}\beta_1 + \dfrac{1}{4} \left( H\epsilon\left( 4+\eta \right) \delta\phi -2\left(\epsilon-2 \right) \dot{\delta\phi} \right) \partial_j \delta\phi \,. 
\end{align}
Even though the expressions for the lapse and the shift fluctuations at second order are cumbersome, since our goal is to obtain the action in the limit $\epsilon\ll 1 $ and $\eta \sim \mathcal{O}(1)$ (at most), it is enough to note that $\alpha_2 \sim \beta_2 \sim \beta_{2,i} \sim \delta\phi^2 + \sqrt{\epsilon}\, h\, \varphi + h^2$. 

\subsection{Quadratic, cubic and quartic actions for fluctuations}
To obtain the quadratic, cubic and quartic actions, we have to introduce the 
{elements} studied above in the action (\ref{eq:Action_ADM}) and make a perturbative expansion in powers of the fields. In addition, we will systematically apply integration by parts with spatial derivatives, but not with time derivatives, due to known subtleties regarding boundary terms \cite{Arroja:2011yj} (which affect the n-point correlation functions). The quadratic action is then
\begin{align}
	\nonumber
	S_2 & = \int \dd^4x\,  \bigg\{  \dfrac{M_P^2 a^3}{8} \left( \dot{h}^{ij} \dot{h}_{ij} - \dfrac{1}{a^2} \partial^k h^{ij} \partial_k h_{ij} \right) -\dfrac{1}{2} \partial_t\left(Ha^3\epsilon\delta\phi^2 \right)\bigg\} \\ & \quad\quad\quad\quad\quad+\dfrac{a^3}{2} \left(\dot{\delta\phi}^2-\dfrac{1}{a^2} \partial^i\delta\phi \partial_i \delta\phi  + \dfrac{2}{a^3}\partial_t\left(Ha^3\epsilon \right) \delta\phi^2 -V_2 \delta\phi^2  \right) 	\\ \nonumber
	&\simeq \int \dd^4x\, \bigg\{ \dfrac{M_P^2 a^3}{8} \left( \dot{h}^{ij} \dot{h}_{ij} - \dfrac{1}{a^2} \partial^k h^{ij} \partial_k h_{ij} \right) +\dfrac{a^3}{2} \left(\dot{\delta\phi}^2-\dfrac{1}{a^2} \partial^i\delta\phi \partial_i \delta\phi  -V_2 \delta\phi^2  \right)\bigg\}\,. \quad\quad\quad
\end{align}
In the second line we have taken the limit $\epsilon\ll 1 $ to keep only the terms we are interested in. As we have commented above, the self-interactions {of $\delta \phi$} coming from the {purely gravitational part of the action} are suppressed with respect to those coming from the potential. Therefore, at this level, the action for the scalar field is simply the action of a free scalar field in a FLRW universe.

The complete cubic action (without any approximations) is
\begin{align}
	\nonumber
	S_3 &= \int \dd^4x\, \Bigg\{  \dfrac{M_P^2 a}{24} h^{ij} \left(4h^{kl} \partial_{lj} h_{ik} - 3h^{kl} \partial_{kl} h_{ij} +2\partial_kh_{jl}\partial^l \tensor{h}{_i^k} \right) \\ \nonumber  & \quad\quad\quad\quad- \dfrac{M_P a^3 \sqrt{\epsilon}}{8\sqrt{2}}\delta\phi \left(\dot{h}^{ij} \dot{h}_{ij} + \dfrac{1}{a^2} \partial^k h^{ij} \partial_k h_{ij} \right) - \dfrac{1}{4}M_P^2a\dot{h}^{jk}\partial_ih_{jk} \partial^i\beta_1  \\
	\nonumber
	& \quad \quad \quad \quad  +\dfrac{M_P a \sqrt{\epsilon}}{2\sqrt{2}} \dot{h}^{ij} \delta\phi  \partial_{ij}\beta_1 + \dfrac{M_P^2}{2a} \partial_i h_{jk} \partial^i\beta_1\partial^{jk} \beta_1 + \dfrac{a}{2} h^{ij} \partial_i\delta\phi \partial_j\delta\phi \\
	\nonumber
	& \quad \quad \quad \quad  +\dfrac{H^2a^3\epsilon^{3/2}\left(24-8\epsilon+4\eta+\eta^2 \right) }{16\sqrt{2}M_P} \delta\phi^3 - \dfrac{a^3\sqrt{\epsilon} V_2}{2\sqrt{2}M_P} \delta\phi^3 -\dfrac{a^3}{6}V_3 \delta\phi^3  \\ \nonumber
	& \quad \quad \quad \quad - \dfrac{Ha^3\epsilon^{3/2}(\eta-2)}{4\sqrt{2}M_P} \delta\phi^2\dot{\delta\phi}  + \dfrac{a^3\sqrt{\epsilon}(\epsilon-2)}{4\sqrt{2}M_P}\delta\phi \dot{\delta\phi}^2 -a\dot{\delta\phi}\partial_i\delta\phi \partial^i\beta_1 \\  & \quad\quad\quad \quad - \dfrac{M_P \sqrt{\epsilon}}{2\sqrt{2}a} \delta\phi \partial_{ij}\beta_1 \partial^{ij} \beta_1 - \dfrac{a\sqrt{\epsilon}}{2\sqrt{2}M_P} \delta\phi \partial_i \delta\phi \partial^i \delta\phi\Bigg\}\,.
\end{align}
Considering that $\beta_1 \sim \sqrt{\epsilon}\, \delta\phi$ and taking the limit $\epsilon\ll 1$, we obtain
\begin{align}
	\nonumber
	S_3 &\simeq\int \dd^4x\, \bigg\{ \dfrac{M_P^2 a}{24} h^{ij} \left(4h^{kl} \partial_{lj} h_{ik} - 3h^{kl} \partial_{kl} h_{ij} +2\partial_kh_{jl}\partial^l \tensor{h}{_i^k} \right) \\\nonumber & \quad \quad\quad \quad - \dfrac{M_P a^3 \sqrt{\epsilon}}{8\sqrt{2}} \left(\dot{h}^{ij} \dot{h}_{ij} + \dfrac{1}{a^2} \partial^k h^{ij} \partial_k h_{ij} \right) \delta\phi  \\
	& \quad \quad \quad \quad - \dfrac{1}{4}M_P^2a\dot{h}^{jk}\partial_ih_{jk} \partial^i\beta_1 + \dfrac{a}{2} h^{ij} \partial_i\delta\phi \partial_j\delta\phi -\dfrac{a^3}{6}V_3 \delta\phi^3\bigg\}\,.
\end{align}
Again, since $V_3 \sim \epsilon^{-1/2}$, the term $\sim V_3 \delta\phi^3$ dominates over any other self-interaction of the scalar field. 

We now turn to study the quartic action. 
{Due to the large number of terms that appear at this order, we will keep only those that are of our interest. For the calculation of the scalar power spectrum, we need the terms that scale as $\sim \delta\phi^4$. We will also keep the terms $\sim h^2 \delta\phi^2$, which contribute to the tensor power spectrum induced by scalar modes. We obtain:}
\begin{align}
	\nonumber
	S_4^{\delta\phi^4} &+ S_4^{h^2\delta\phi^2}\simeq \int\dd^4 x\, \Bigg\{ -\dfrac{a}{4} \tensor{h}{_i^k} \tensor{h}{_j_k} \partial^i \delta\phi \partial^j \delta\phi -\dfrac{a^3V_4}{24}\delta\phi^4 \\
	\nonumber
	&- a^3\alpha_2\bigg[\dfrac{M_P^2}{8} \left( \dot{h}^{ij} \dot{h}_{ij} + \dfrac{1}{a^2} \partial^k h^{ij} \partial_k h_{ij} \right) +\dfrac{1}{2} \bigg(\dot{\delta\phi}^2+\dfrac{1}{a^2} \partial^i\delta\phi \partial_i \delta\phi  +V_2 \delta\phi^2  \bigg) \\\nonumber  & + 3H^2M_P^2\alpha_2 + \dfrac{2HM_P^2}{a^2}\partial^2\beta_2 \bigg] - a \left(\dfrac{M_P^2}{4} \dot{h}^{jk} \partial^ih_{jk} + \dot{\delta\phi}\partial^i\delta\phi  \right) \left( \partial_i \beta_2 + \beta_{2,i} \right) \\
	& + \dfrac{M_P^2a}{2} \tensor{h}{^k^{[i}} \tensor{\dot{h}}{^{j]}_k} \partial_i \beta_{2,j} -\dfrac{M_P^2}{4a} \beta_{2,i} \partial^2 \beta_2^i\Bigg\}\,.
\end{align} 
The second and third lines are simplified using the equations of motion of $\alpha_2$, $\beta_2$ and $\beta_{2,i}$ at lower order in $\epsilon$, 
\begin{align} \nonumber
6 H^2 M_P^2\alpha_2 & + \dfrac{2H M_P^2}{a^2}\partial^2\beta_2  + \dfrac{M_P^2}{8} \left(\dot{ h}^{ij}\dot{ h}_{ij} + \dfrac{1}{a^2}\partial^k h^{ij} \partial_k h_{ij} \right) \\ & + \dfrac{1}{2}\left( \dot{\delta\phi}^2 + \dfrac{1}{a^2} \partial^i\delta\phi \partial_i\delta\phi  + V_2\delta\phi^2  \right)=0\,,
\end{align}
\begin{align}
\dfrac{M_P^2}{2a^2}\partial^2\beta_{2,j} - 2H M_P^2 \partial_j \alpha_2 + \dfrac{M_P^2}{4} \left(\dot{h}^{ik} \left(\partial_j h_{ik} - \partial_k h_{ij} \right) +h^{ik} \partial_k \dot{h}_{ij}  \right) + \dot{\delta\phi}  \partial_j \delta\phi= 0 \,,
\end{align}
obtaining
\begin{align}
	\nonumber
	S_4^{\delta\phi^4} + S_4^{h^2\delta\phi^2} &\simeq \int\dd^4 x\, \Bigg\{  -\dfrac{a}{4} \tensor{h}{_i^k} \tensor{h}{_j_k} \partial^i \delta\phi \partial^j \delta\phi -\dfrac{a^3V_4}{24}\delta\phi^4 + 3a^3H^2M_P^2\alpha_2^2\Bigg\} \\
	&\simeq \int\dd^4 x\, \Bigg\{ 	-\dfrac{a}{4} \tensor{h}{_i^k} \tensor{h}{_j_k} \partial^i \delta\phi \partial^j \delta\phi -\dfrac{a^3V_4}{24}\delta\phi^4 \nonumber \\  &\quad\quad\quad\quad + \dfrac{3a^3}{8} \partial^{-2} \partial^i \left( \dot{h}^{jk} \partial_i h_{jk} \right) \partial^{-2} \partial^l \left( \dot{\delta\phi} \partial_l \delta\phi \right)\Bigg\}\, .
\end{align}
We conclude that the set of important self-interactions of $\delta\phi$ up to fourth order are those coming from the potential. This conclusion can be generalized to any perturbative order.
{The reason is that the interactions coming from the potential at order $n$ depend on $V_n$, while the interactions coming from {the metric} depend on $N_{(n-2)}$ and $N_{(n-2),i}$ {at most}, and therefore on $V_{(n-2)}$. As we have seen above $V_{(n-2)}/ V_n \sim \epsilon$. It is important to stress that this only takes into account that $\epsilon\ll 1$, assuming that the rest of SR parameters are $\sim 1$ --at least in some phase of inflation that we are interested in--. Hence, this result does not apply to standard SR but does apply to USR.}\\

\section{Classical dynamics of super-Hubble modes}
We study now the classical dynamics of the fields $\delta\phi$ and $h_{ij}$. 
First, let us define a variable with respect to which all calculations will be much simpler, 
\begin{equation}
	\varphi \equiv -H \dfrac{\delta\phi}{\dot{\phi}_0}\,.
	\label{eq:def_varphi}
\end{equation}
We work in the super-Hubble limit, i.e.\ $k\ll aH$ (and not in the $\epsilon\ll 1$ limit, which we will not use again), in which {we neglect gradients}. We are also interested in {taking a late time limit,}
which is formally the limit $a\to \infty$. Moreover, we will assume that $1/(a^3\epsilon) \to 0$. 
Under these simplifications, and working with the variable $\varphi$, the quadratic action simplifies to 
\begin{align}
	S_2 = \int \dd^4x \Bigg\{ \dfrac{M_P^2 a^3}{8}  \dot{h}^{ij} \dot{h}_{ij} + M_P^2 a^3 \epsilon \dot{\varphi}^2 + \partial_t\left( H M_P^2 a^3 \epsilon \left(-\epsilon+\dfrac{\eta}{2} \right)\varphi^2  \right)\Bigg\}\,.
\end{align}
The equations of motion are
\begin{equation}
	\partial_t\left(a^3\dot{h}_{ij} \right) = \order{2}\,,\quad \partial_t\left(a^3\epsilon \dot{\varphi} \right) = \order{2},  
\end{equation}
where we write $\order{2}$ because we have not included the effects of the cubic action. That is, these equations of motion have corrections that are quadratic in the fields. Neglecting them, 
for $k\ll aH$ and at {late times}, both $\varphi$ and $h_{ij}$ are constant, i.e.\ $\dot{\varphi} =  \order{2}$ and $\dot{h}_{ij} = \order{2}$. 

We will now consider what those quadratic corrections are, for which we have to study the cubic action under the same approximations. When neglecting gradients, special care must be taken with terms of the form $\sim \partial^2\beta_1$, since $\beta_1$ is defined with the inverse of the Laplacian, see eq.\ \eq{eq:N_Ni_order1}. Terms with two time derivatives contribute to the equations of motion as a third order correction, so we can neglect them.\footnote{For instance, for $\mathcal{L} \supset f(t)\varphi\dot{\varphi}^2$, we obtain $\sim\partial_t(f(t)\varphi\dot{\varphi}) + \order{4} \sim \order{3}$.} With these considerations, the cubic action simplifies to
\begin{align}\nonumber
	S_3  =  \int \dd^4x \,  \frac{H M_P^2 a^3 \epsilon}{6}  \bigg\{  & \left(2 \epsilon^3-3 \epsilon^2 (2 + 3 \eta)+\epsilon\, \eta\, (9 + 6 \eta+4 \epsilon_3)-\eta\, \epsilon_3\, (3 + \eta+\epsilon_4+\epsilon_3)\right)H\,\varphi^3 \\  \quad\quad\quad\quad\quad - &  3\epsilon\,(2 \epsilon -3 \eta)\,\varphi^2\, \dot{\varphi} \bigg\}\,.
\end{align}
Taking into account simultaneously $S_2$ and $S_3$, the updated equations of motion are
\begin{equation}
	\partial_t\left(a^3\dot{h}_{ij} \right) = \order{3},\; \partial_t\left(a^3\epsilon \partial_t\left(\varphi + \dfrac{\eta}{4}\varphi^2 \right)  \right) = \order{3}\,,  \label{eq:din_varphi_out_horizon}
\end{equation}
where we have used the fact that $\varphi^2$ enters and leaves the time derivative as if it were a constant, since the corrections left by this operation are of cubic order. Now, neglecting gradients and at {late times} we have that 
\begin{align}\label{eq:eom varphi second order}
\partial_t \left( \varphi + \eta\,\varphi^2/4 \right) =  \order{3}\quad \text{and }\quad \dot{h}_{ij} = \order{3}\,.
\end{align} 
As we will see in
{Appendix \ref{sec:App_Correlation_Z_DP}}
, the variable $\zeta$ of Maldacena is $\zeta = \varphi + \eta/4 \varphi^2$ at second order in fluctuations. That is, obtaining that $\varphi + \eta\, \varphi^2/4$ is constant for super-Hubble modes is consistent with the result that $\dot \zeta = 0$ at all orders in the limit $k\ll aH$ \cite{Maldacena:2002vr,Weinberg:2003sw}.

Finally, we analyze the lapse and the shift in this limit in terms of $\varphi$. At first order,
\begin{equation}
	\alpha_1 = -\epsilon\varphi\,,\quad \dfrac{1}{a^2}\partial^2 \beta_1 = \epsilon \dot{\varphi}\,,\quad \beta_{1,i} = 0\,.
\end{equation}
By using the equation of motion of $\varphi$, we obtain that $\partial^2 \beta_1/a^2 =  - H \, \epsilon\, \eta \, \epsilon_3\,\varphi^2/4  =  \order{2}$. At second order, neglecting gradients and, again, using the equations of motion of $\varphi$ and $h$ (i.e.\ imposing that $\dot{\varphi} \sim \dot{h} \sim \order{2}$), we obtain that
\begin{equation}
	\alpha_2 = \dfrac{\epsilon (2\epsilon +\eta)}{4}\varphi^2\,,\quad \dfrac{1}{a^2}\partial^2 \beta_2 = \dfrac{H \, \epsilon\, \eta \, \epsilon_3}{4} \varphi^2\,,\quad \beta_{2,i} = 0\,. \label{eq:beta2_out_horizon}
\end{equation}

\section{Change of gauge} \label{sec:App_Correlation_Z_DP}

Starting from the decomposition of the metric and the scalar field of the eqs.\ (\ref{eq:SVT_Decomposition}) and \eq{decphi}, we can choose the coordinate system so that we are in the $\zeta$-gauge ($\delta\phi = E = E_{i}= 0$) or the $\delta\phi$-gauge ($\zeta = E =E_{i}= 0$). The question we are going to solve is what is the relation between the variables in one gauge and the other at third order, see also e.g.\ \cite{Maldacena:2002vr,Malik_2008im,Malik_2009fp}.  In particular, starting from the $\delta\phi$-gauge, we are going to make a coordinate transformation that leads to the $\zeta$-gauge. 
At the end of this appendix we will explain why we must go up to third order to be consistent with the calculation we do of the power spectrum of $\zeta$.

In what follows we use a tilde to distinguish among the two gauges. Concretely,
the transformation we are looking for is such that
\begin{equation}
	\delta\phi \to \tilde{\delta\phi} = 0\,,\quad \zeta = 0 \to \tilde{\zeta}\,,\quad E = 0\to \tilde{E} = 0\,,\quad E_i = 0\to \tilde{E}_i = 0\,,\quad h_{ij} \to \tilde{h}_{ij}\,.
\end{equation}
{Both the lapse and the shift will also transform under the change of gauge, but they will still be algebraic variables and therefore we do not need to worry about their transformation properties.}
The fields in the new gauge will be a series expansion of the fields in the old gauge, {e.g.\ $\tilde{\zeta} \equiv \sum_n \tilde{\zeta}_n $, where the subscript $_n$ indicates the order of the expansion.} The coordinate change that sends one gauge to the other is
\begin{equation}
	x^\mu \to \tilde{x}^\mu = x^\mu - \xi^\mu(x)\,,\quad \textrm{where}\quad\xi^\mu(x) \equiv \left( \xi^0(x) ,\xi^i(x) = \partial^i\vartheta(x) + \vartheta^i(x) \right) = {\sum _n \xi^\mu_n} \,. 
\end{equation}
We also need the inverse transformation $\tilde{x}^\mu \to x^\mu = \tilde{x}^\mu + \tilde{\xi}^\mu(\tilde{x})$, where $\tilde{\xi}^\mu(\tilde{x})$ is related to $\xi^\mu(x)$ by the expression
\begin{align} \nonumber
	\tilde{\xi}^\mu(\tilde{x}) & = \xi^\mu(x) = \xi^\mu(\tilde{x} + \tilde{\xi}(\tilde{x})) \\ & = \xi^\mu (\tilde{x}) + \left(\xi^{\rho}(\tilde{x}) + \xi^{\sigma}(\tilde{x})\partial_{\sigma}\xi^{\rho}(\tilde{x}) \right) \partial_{\rho}\xi^{\mu}(\tilde{x}) +\dfrac{1}{2}\xi^{\rho}(\tilde{x})\xi^{\sigma}(\tilde{x}) \partial_{\rho\sigma}\xi^{\mu}(\tilde{x}) + \order{4}\,.
\end{align}

\subsection{Scalar field transformation}
Being the inflaton field $\phi$ a scalar, {it transforms as} $\phi(x) \to \tilde{\phi}(\tilde{x}) = \phi(x)$, which implies that $\tilde{\phi}_0(\tilde{t}) = \phi_0(t) + \delta\phi(x)$. From here, we obtain
\begin{align}
		\delta\phi(x) = \tilde{\phi}_0(t) - \phi_0(t) - \xi^0(x) \dot{\tilde{\phi}}_0(t) + \dfrac{1}{2} \left( \xi^0(x)\right) ^2 \ddot{\tilde{\phi}}_0(t) -\dfrac{1}{3!} \left( \xi^0(x)\right) ^3 \dddot{\tilde{\phi}}_0(t)\,.
\end{align}
At zeroth order, $\phi_0(t) = \tilde{\phi}_0(t)$. To study higher orders, we have to expand $\xi^0(x) = {\sum_n \xi^0_n(x)}$ in the expression above:
\begin{equation}
	\delta\phi = -\xi^0_1 \dot{\phi}_0 + \left(-\xi^0_2 \dot{\phi}_0 + \dfrac{1}{2} \left(\xi^0_1 \right)^2 \ddot{\phi}_0  \right) + \left(-\xi^0_3\dot{\phi}_0 + \xi^0_1\xi^0_2 \ddot{\phi}_0 -\dfrac{1}{3!} \left(\xi^0_1 \right)^3 \dddot{\phi}_0  \right) 
\end{equation}
so that, up to order three we obtain
\begin{align}
	\xi^0_1(x) = -\dfrac{\delta\phi(x)}{\dot{\phi}_0(t)} = \dfrac{\varphi(x)}{H}\,,\quad
	\xi^0_2(x) = \dfrac{1}{2}\left(\xi^0_1(x) \right)^2 \dfrac{\ddot{\phi}_0}{\dot{\phi}_0}\,,\\
	\xi^0_3(x) = \dfrac{\left(\xi^0_1(x) \right)^3}{2}\left( \left( \dfrac{\ddot{\phi}_0}{\dot{\phi}_0}\right) ^2-\dfrac{1}{3}\dfrac{\dddot{\phi}_0}{\dot{\phi}_0} \right)\,,
\end{align}
where the variable $\varphi$ was defined in (\ref{eq:def_varphi}) and, in terms of slow-roll parameters:
\begin{align}
	\dot{\phi}_0 = H M_P \sqrt{2\epsilon}\,,\quad \dfrac{\ddot{\phi}_0} {\dot{\phi}_0} = H\left(-\epsilon + \dfrac{\eta}{2} \right)\,,\quad \dfrac{\dddot{\phi}_0} {\dot{\phi}_0} = H^2\left(2\epsilon^2 + \dfrac{\eta^2}{4} -\dfrac{5}{2} \epsilon \eta + \dfrac{\dot{\eta}}{2H} \right)\,.
\end{align}
We note that the scalar field transformation only constrains $\xi^0$ but leaves $\xi^i$ completely free, which allows us to impose the rest of the conditions of the gauge transformation.

\subsection{Metric transformation}
The metric transforms as
\begin{align}
	g_{\mu\nu}(x) \to \tilde{g}_{\mu \nu}(\tilde{x}) = \dfrac{\partial x^\rho}{\partial \tilde{x}^\mu}\dfrac{\partial x^\sigma}{\partial \tilde{x}^\nu} g_{\rho \sigma}(x) = \left(\delta_\mu^\rho + \partial_\mu \tilde{\xi}^\rho(\tilde{x}) \right) \left(\delta_\nu^\sigma + \partial_\nu \tilde{\xi}^\sigma(\tilde{x})  \right) g_{\rho \sigma}(\tilde{x} + \tilde{\xi}(\tilde{x}))\,, \label{eq:App_TransGauge_TransMetrica}
\end{align}
where $ \partial_\mu \tilde{\xi}^\rho(\tilde{x}) = \partial \tilde{\xi}^\rho(\tilde{x}) / \partial \tilde{x}^\mu $ and $g_{\rho \sigma}(\tilde{x} + \tilde{\xi}(\tilde{x})) = \sum_n \frac{1}{n!}\tilde{\xi}^{\delta_1}(\tilde{x}) \dots \tilde{\xi}^{\delta_n}(\tilde{x}) \partial_{\delta_1 \dots \delta_n} g_{\rho \sigma}(\tilde{x})$ .
Since we have written all the fields at the same point $ \tilde{x} $, for simplicity we will no longer write the spatial dependence.
By expanding eq.\ (\ref{eq:App_TransGauge_TransMetrica}), we will be able to impose the rest of the conditions of our gauge transformation, obtaining $\tilde{\xi}^i$, $\tilde{\zeta}$ {and $\tilde{h}_{ij}$}. 

Let us consider an object $M_{ij}=M_{ji}$ with two spatial indices, which we can decompose as follows: 
\begin{align}
	M_{ij} = \delta_{ij} A + 2 \partial_{ij} B + 2 \partial_{(i} C_{j)} + D_{ij}\,,
\end{align}
where $\partial^iC_i = 0$ and $\delta^{ij}D_{ij} = \partial^iD_{ij} = 0$.
To extract each of these variables from $M_{ij}$, we define ad-hoc projectors 
\begin{align} \label{p1}
	\hat{P}^{kl}\,M_{kl} & \equiv \dfrac{1}{2} \left( \delta^{kl} - \partial^{-2} \partial^{kl}\right) M_{kl} = A  \,,\\ \label{p2}
	\hat{Q}^{kl}\,M_{kl} & \equiv \dfrac{1}{4} \partial^{-2}\left(3 \partial^{-2}\partial^{kl} - \delta^{kl} \right) M_{kl} =  B  \,,\\ \label{p3}
	\hat{W}_{i}^{kl}\,M_{kl} & \equiv \partial^{-2} \left(\tensor{\delta}{_i^l}\partial^k - \partial^{-2} \tensor{\partial}{_i^k^l} \right)\,M_{kl} =  C_i  \,,\\ \label{p4}
	\hat{F}_{ij}^{kl}\,M_{kl} & \equiv  \left(\tensor{\delta}{_i^k} \tensor{\delta}{_j^l}-\dfrac{1}{2} \delta_{ij} \delta^{kl} +\dfrac{1}{2} \partial^{-2}\left( \delta_{ij} \partial^{kl} - 4 \tensor{\delta}{_{\left(i\right.} ^l} \tensor{\partial}{_{\left. j\right)}^k}  + \partial_{ij}\delta^{kl} + \partial^{-2} \tensor{\partial}{_i_j^k^l} \right) \right) M_{kl} = D_{ij} \,.
\end{align}
These projectors will allow us to analyze the gauge transformations of the fields.
Before expanding eq.\ (\ref{eq:App_TransGauge_TransMetrica}) in fluctuations, let us write down the metric fluctuations we will need, in both gauges. Starting from $g_{ij} = \gamma_{ij}$, we have
\begin{equation}
	g_{0,ij} = a^2 \delta_{ij}\,,\quad
	g_{1,ij} = a^2 h_{ij}\,,\quad
	g_{2,ij} = \dfrac{1}{2} a^2 \tensor{h}{_i^k} \tensor{h}{_k_j}\,,\quad
	g_{3,ij} = \dfrac{1}{6} a^2 \tensor{h}{_i^k} \tensor{h}{_k^l}\tensor{h}{_l_j}\,,
\end{equation}
{where, again, the subscript $_n$ denotes the perturbative order of $g_{n,ij}$ and we use a comma just to separate it from the spacetime indices.}
The expressions of $ \tilde{g}_{ij} $ are:
\begin{align}
\tilde{g}_{0,ij} & = a^2 \delta_{ij}\,,\\
\tilde{g}_{1,ij} & = a^2 \left(2\tilde{\zeta}_1\delta_{ij} + \tilde{h}_{1,ij} \right)\,,\\
\tilde{g}_{2,ij} & = a^2 \left( 2\left( \tilde{\zeta}_2+ \tilde{\zeta}_1^2 \right)\delta_{ij} + 2\tilde{\zeta}_1\tilde{h}_{1,ij}+  \tilde{h}_{2,ij} + \dfrac{1}{2}\tensor{\tilde{h}}{_{1,i}^k} \tilde{h}_{1,kj} \right),\\
\nonumber
\tilde{g}_{3,ij}  & = a^2 \bigg\{ 2 \left(\tilde{\zeta}_3 + 2\tilde{\zeta}_2\tilde{\zeta}_1 + \dfrac{2}{3} \tilde{\zeta}_1^3 \right)\delta_{ij}+ \left( 2\tilde{h}_{2,ij}   + \tensor{\tilde{h}}{_{1,i}^k} \tilde{h}_{1,kj}\right)\tilde{\zeta}_1 \\ &\quad  \quad \quad +2\left( \tilde{\zeta}_2+ \tilde{\zeta}_1^2 \right)\tilde{h}_{1,ij} + \tilde{h}_{3,ij} + \tensor{\tilde{h}}{_{2,\left( i\right.}^k} \tilde{h}_{1,k\left. j\right)} + \dfrac{1}{6} \tensor{\tilde{h}}{_{1,i}^k}\tensor{\tilde{h}}{_{1,k}^l} \tilde{h}_{1,lj} \bigg\} \,.
\end{align}
We stress that it is here that we have made use of the gauge freedom we had left ($ \xi_i $) to impose that $ \tilde{E} = \tilde{E}_i = 0 $. In addition, we will also need to know $ g_{00} = -N^2 + \gamma^{ij}N_iN_j $ and $ g_{0i} = N_i $,
\begin{align}
	&g_{0,00} = -1\,,\quad  g_{1,00} = -2\alpha_1\,,\\
	&g_{0,0i} = 0\,,\quad g_{1,0i} = \partial_i \beta_1\,,\quad g_{2,0i} = \partial_i \beta_2 + \beta_{2,i}\,,
\end{align}
where $\alpha$ and $\beta$ are introduced in eq.\ \eq{defalphabeta}.
Now we can expand eq.\ (\ref{eq:App_TransGauge_TransMetrica}), so that at zeroth order  $ \tilde{g}_{0,\mu\nu} = g_{0,\mu\nu} $. At first order, 
\begin{align}
	\tilde{g}_{1,\mu\nu} = g_{1,\mu\nu} + \xi_1^\delta\partial_\delta g_{0,\mu\nu} + 2\partial_{\left( \mu \right.}\xi^\rho_1\,  g_{0,\left. \nu \right) \rho}\,.
\end{align}
Since we are only interested in obtaining $ \tilde{\zeta} $ and $ \tilde{h} $, we can focus on the component $ ij $,
\begin{align}
	\tilde{g}_{1,ij} = g_{1,ij} + 2a^2\varphi \delta_{ij} + 2a^2 \partial_{(i} \xi_{1,j)}\,.
\end{align}
To obtain $\xi^i$, $\tilde{\zeta}$ and $\tilde{h}_{ij}$, we use the projectors introduced in eqs.\ \eq{p1}--\eq{p4},
\begin{align}
	\tilde{\zeta}_1 = \varphi\,,\quad \tilde{h}_{1,ij} = h_{ij}\,,\quad\xi^i_1 = 0\,.
\end{align}
At second order,
\begin{align}
	&\tilde{g}_{2,ij} = g_{2,ij} + 2a^2\left( \left(1+\dfrac{\eta}{4} \right) \varphi^2 + \dfrac{\varphi\dot{\varphi}}{H} \right) \delta_{ij} + 2a^2 \partial_{(i} \xi_{2,j)} + 2a^2\varphi h_{ij} - \chi_{2,ij}\,,\\ 
	&\chi_{2,ij} \equiv \partial_i \xi^0_1 \partial_j \xi^0_1 - 2 \partial_{\left( i \right.} \beta_1 \partial_{\left. j \right)} \xi^0_1 - \dfrac{a^2}{H} \varphi \dot{h}_{ij}\,.
\end{align}
Projecting, we obtain 
\begin{align}
	\tilde{\zeta}_2 = \dfrac{\eta}{4} \varphi^2 + \dfrac{\varphi \dot{\varphi}}{H}   - \dfrac{1}{2a^2}\hat{P}^{kl}\, \chi_{2,kl}\,,\quad \tilde{h}_{2,ij} = -\dfrac{1}{a^2} \hat{F}^{kl}_{ij} \, \chi_{2,kl}\,,\\
	\vartheta_2 = \dfrac{1}{a^2} \hat{Q}^{kl} \chi_{2,kl}\,, \quad \vartheta_{2,i} = \dfrac{1}{a^2} \hat{W}^{kl}_{i}\, \chi_{2,kl} \,.
\end{align}
At third order, the number of terms {involved in the gauge transformation} becomes very large. Therefore, we use the super-Hubble limit, where $ \dot{\varphi} = \dot{h} = \order{2} $ and gradients are neglected.
Then, the transformation at third order is
\begin{align} \nonumber
	\tilde{g}_{3,ij} = g_{3,ij} & + 2a^2  \left( \dfrac{2}{3}  +\dfrac{\eta}{2} + \dfrac{\eta^2}{12}+ \dfrac{\eta\epsilon_3}{6} \right) \varphi^3 \delta_{ij} + 2a^2 \left(1+\dfrac{\eta}{4} \right) \varphi^2 h_{ij} \\ & + a^2 \varphi \tensor{h}{_i^k} \tensor{h}{_k_j} + 2a^2 \partial_{(i} \xi_{3,j)} - \chi_{3,ij}\,,
\end{align}
where $\chi_{3,ij} = -2 \partial_{(i} \beta_2 \partial_{j)} \xi^0_1 $, which is the only non-zero term for {$k\ll aH$}, 
{because $ \beta_2 $ depends on the inverse of the Laplacian (eq.\ (\ref{eq:beta2_out_horizon})).}
Again, we project to obtain that
\begin{align} 
	\tilde{\zeta}_3 = \dfrac{\eta^2 + 2\eta\epsilon_3}{12} \varphi^3 - \dfrac{1}{2a^2}\hat{P}^{kl} \, \chi_{3,kl} \,,\quad \tilde{h}_{3,ij} = -\dfrac{1}{a^2} \hat{F}^{kl}_{ij} \, \chi_{3,kl}\,,\\
	\vartheta_3 = \dfrac{1}{a^2} \hat{Q}^{kl} \, \chi_{3,kl}\,,\quad \vartheta_{3,i} = \dfrac{1}{a^2} \hat{W}^{kl}_{i} \, \chi_{3,kl} \,.
\end{align}
Although it seems that $\chi_{2,ij}$ and $\chi_{3,ij}$ have an effect on the gauge transformation, their contribution vanishes for super-Hubble modes, as we will demonstrate. Let us consider $\tilde{\zeta}$,
\begin{align}
	\tilde{\zeta} = \tilde{\zeta}_1 + \tilde{\zeta}_2 + \tilde{\zeta}_3 = \varphi + \dfrac{\eta}{4} \varphi^2 + \dfrac{\varphi \dot{\varphi}}{H} + \dfrac{\eta^2 + 2\eta\epsilon_3}{12} \varphi^3  - \dfrac{1}{2a^2}\hat{P}^{kl} \left( \chi_{2,kl}  + \chi_{3,kl}\right) \,.
\end{align}
Because we are working at third order, and because $\varphi \dot{\varphi} = \order{3}$, we cannot neglect this term.
In the above expression, we get the combination 
\begin{align} \label{subtle1}
	\chi_{2,kl}  + \chi_{3,kl} = -2 \partial_{(i} \xi^0_1 \partial_{j)}\left(\beta_1 + \beta_2 \right) -\dfrac{a^2}{H} \varphi \dot{h}_{ij}\,.
\end{align}
Since $ \dot{h}_{ij} = \order{3} $, we can neglect the term $ \sim \varphi\, \dot{h}_{ij} $. On the other hand, 
\begin{align} \label{subtle2}
	\beta_1 + \beta_2 = a^2 \epsilon \partial^{-2}\left( \dot{\varphi} + \dfrac{H \eta \epsilon_3}{4}\varphi^2 \right) =  a^2 \epsilon \partial^{-2} \partial_t\left(\varphi + \dfrac{\eta}{4}\varphi^2\right) + \order{3} = \order{3}\,,
\end{align}
where we have used  eq.\ (\ref{eq:eom varphi second order}).
We conclude that $ \chi_2 + \chi_3 = \order{4} $ and therefore we can neglect its effect at third order,
\begin{align}
	\tilde{\zeta} = \varphi + \dfrac{\eta}{4} \varphi^2 + \dfrac{\eta^2 - \eta\,\epsilon_3}{12} \varphi^3 \,.
\end{align}
The same applies to $ \tilde{h}_{ij} $ and $ \xi_i $, so that finally
\begin{align}
	\tilde{h}_{ij} &= \tilde{h}_{1.ij}+ \tilde{h}_{2,ij} + \tilde{h}_{3,ij} = h_{ij},\\ \label{cancelc}
	\xi_i &= \xi_{1,i}+\xi_{2,i}+\xi_{3,i} = 0\,.
\end{align}
An important conclusion is that at third order {$h_{ij}$ is invariant under a change of gauge from the $\zeta$-gauge to the $\delta\phi$-gauge. Although we do not prove it, this result suggests it can be generalized at higher orders in perturbation theory.}

Let us assume that eq.\ \eq{cancelc} can be extended to all orders in perturbation theory. We stress that this is not a given. As we have just seen eq.\ \eq{cancelc} holds thanks to a subtle cancellation involving the inverse of the Laplacian, see eqs.\ \eq{subtle1} and \eq{subtle2}. Therefore in the transformation of the metric (\ref{eq:App_TransGauge_TransMetrica}) we are going to neglect $\partial_i \tilde{\xi}^\rho$: 	
\begin{align} \nonumber
	\tilde{g}_{ij}(\tilde{x}) =  g_{ij}(\tilde{x} + \tilde{\xi}(\tilde{x})) & =\sum_n \dfrac{1}{n!}\tilde{\xi}^{\delta_1}(\tilde{x}) \dots \tilde{\xi}^{\delta_n}(\tilde{x}) \partial_{\delta_1 \dots \delta_n} g_{ij}(\tilde{x}) \\ 	& \simeq \sum_n \dfrac{1}{n!} \left( \tilde{\xi}^{0}(\tilde{x}) \right) ^n \dfrac{\partial^n}{\partial t^n} g_{ij}(\tilde{x}) = g_{ij}(\tilde{t} + \tilde{\xi}^0(\tilde{x}),\tilde{\vb{x}})\,,
\end{align}
where we have also neglected the gradients of the metric, $\partial_k g_{ij}$. Then, from $\tilde{\gamma}_{ij}(\tilde{x}) = \gamma_{ij}(\tilde{t} + \tilde{\xi}^0(\tilde{x}),\tilde{\vb{x}})$ we get
\begin{equation}
2\left(\log a(\tilde{t}) +\tilde{\zeta}(\tilde{x}) \right) \delta_{ij} +\tilde{h} _{ij} (\tilde{x}) = 2 \log a(\tilde{t} + \tilde{\xi}^0(\tilde{x})) \delta_{ij} + h_{ij}(\tilde{t} + \tilde{\xi}^0(\tilde{x}),\tilde{\vb{x}}) \,.
\end{equation}
Taking into account that $h_{ij}$ is constant for $k\ll a H$ ($h_{ij}(\tilde{t} + \tilde{\xi}^0(\tilde{x}),\tilde{\vb{x}}) \simeq h_{ij}(\tilde{t},\tilde{\vb{x}})$), we have finally that\footnote{See e.g.\ \cite{Sugiyama:2012tj} for the connection with the $\delta N$-formalism for large scale modes of $\zeta$.}
\begin{equation}
	\tilde{\zeta}(\tilde{x}) = \log \dfrac{a(t)}{a(\tilde{t})}\,,\quad \tilde{h}_{ij}(\tilde{x}) = h_{ij}(\tilde{x}) \,.
\end{equation}
We stress that the decomposition (\ref{eq:SVT_Decomposition}) has been instrumental in the derivation of this result. A different decomposition would mix $h_{ij}$ with lower helicity variables. 

\subsection{Two-point correlation function of $\zeta$}
At late times, the relation between the variables $\zeta$ and $\delta\phi$ in their respective gauges is
\begin{equation}
	\zeta = -\dfrac{\delta\phi}{\sqrt{2\epsilon} M_P} + \dfrac{\eta}{4} \left( \dfrac{\delta\phi}{\sqrt{2\epsilon} M_P}\right) ^2 - \dfrac{\eta^2 - \eta\epsilon_3}{12} \left(\dfrac{\delta\phi}{\sqrt{2\epsilon} M_P} \right) ^3 + \order{\dfrac{\delta\phi}{M_P}}^4 \,.
	\label{eq:App_Expanion_Zeta}
\end{equation}
This relation is exact in the sense that it does not make any SR approximation. We compute $\mathcal{P}_\zeta$ at one-loop using correlations of $\delta\phi$. Knowing the expression above, we can explain why it is necessary to make the expansion up to third order.
The two-point correlation of $\zeta$ is, schematically, 
\begin{figure}[t]
	\begin{center}
		\includegraphics[width=1\textwidth]{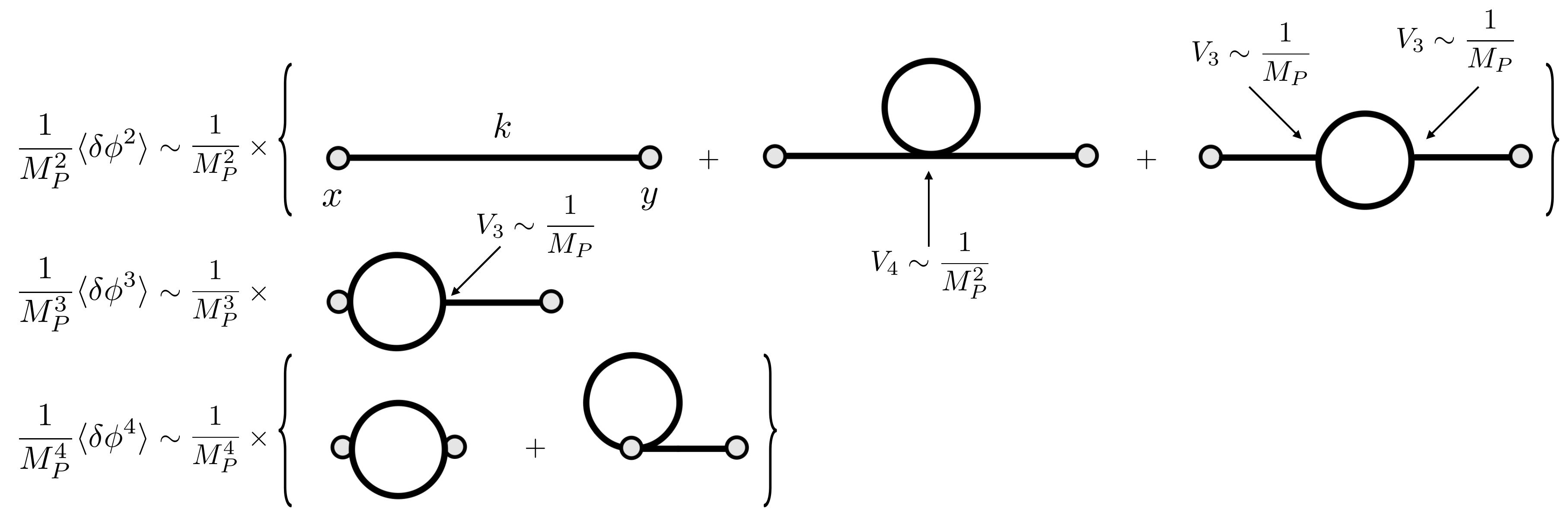}
		\caption{\label{fig:diagstotales}  \small \it Complete set of diagrams that affect the calculation at one-loop of $\expval{\zeta^2}$ in terms of correlations of $\delta\phi$. Bubble diagrams and terms proportional to tadpoles have been omitted.}
	\end{center}
\end{figure}
\begin{align}
	\expval{\zeta^2} \sim \dfrac{1}{M_P^2} \expval{\delta\phi^2} + \dfrac{1}{M_P^3 } \expval{\delta\phi^3} + \dfrac{1}{M_P^4 } \expval{\delta\phi^4} +\order{\dfrac{H}{M_P}}^5 \,,
\end{align}
where we introduce $H$ as it is the only energy scale of inflation.
Assuming the set of interactions we work with in this paper, the contributions to the two-point correlation of $\zeta$ at order $(H/M_P)^4$ are shown in Figure \ref{fig:diagstotales}. Since $V_3 \sim H ( H/M_P )$ and $V_4 \sim ( H/M_P )^2$, the one-loop diagrams that contribute to $\langle\delta\phi^2\rangle$ are of the same order in $( H/M_P )$ as the contribution to $\langle\zeta^2\rangle$ coming from $\langle\delta\phi^3\rangle$ and $\langle\delta\phi^4\rangle$. 

In the model we consider, $\eta = 0$ at late times and therefore only the first term of the expansion (\ref{eq:App_Expanion_Zeta}) survives. However, we stress that (\ref{eq:App_Expanion_Zeta}) is valid for any inflationary model as long as at late times $\dot\varphi\propto 1/(a^3 \epsilon) \to 0$.

\section{Tadpoles}\label{sec:Tadpoles}
\begin{figure}[t]
	\begin{center}
		\includegraphics[width=1\textwidth]{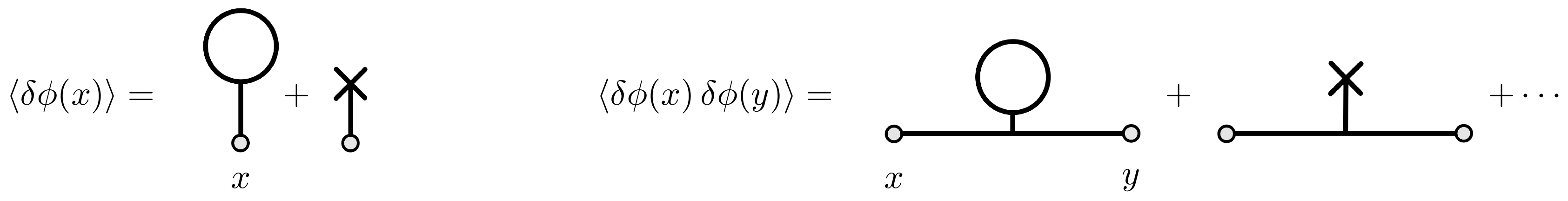}
		\caption{\label{tadfig}  \small \it Diagrams that contribute to the tadpole $\expval{\delta\phi(x)}$ (left) or are proportional to it and contribute to two-point correlation $\expval{\delta\phi(x) \delta\phi(y)}$ (right).
 }
	\end{center}
\end{figure}

In this appendix we discuss why we do not need to include tadpoles in our calculations. Since we work with a theory with cubic interactions, we inevitably get a non-zero one-loop contribution to the correlator $\expval{\delta\phi(x)}$, as shown in Figure \ref{tadfig}. In principle, this type of  
interaction implies the need of considering contributions to the power spectrum beyond those studied in the main body of this work. 
However, by using counterterms we can avoid having to include these effects, imposing that $\expval{\delta\phi(x)} = 0$. 

The interaction Hamiltonian we are going to use is
\begin{equation}
	H_I(\tau) = \int \dd^3 \vb{x} \left[ a^4 \dfrac{V_3 \delta\phi^3}{3!} + \delta_{\rm tp} \delta\phi \right] \,.
\end{equation}
The coupling of the counterterm, $\delta_{\rm tp}$, will be responsible for the renormalization of the tadpole. 
{Since this counterterm comes from renormalizing the potential (in particular, from making $V_1\to V_1 + \delta_{V_1}$), $\delta_{\rm tp}$ will have an arbitrary time dependence, as well as $\delta_{V}\equiv \delta_{V_2}$.}
The reason why we do not include the quartic interaction here is that it contributes to the next order in loops.
Using eq.\ (\ref{eq:In-In_General}), we see that if we impose
\begin{align}
	\nonumber
	& \expval{\delta\phi(x)} = 2 \Im{\int_{-\infty}^\tau \dd \tau' \bra{0} \delta\phi(x) \int \dd^3 \vb{x}' \left[ a^4(\tau'_-) \dfrac{V_3(\tau'_-) \delta\phi^3(x'_-)}{3!} + \delta_{\rm tp}(\tau'_-) \delta\phi(x'_-) \right] \ket{0}} \\
	 &= \int_{-\infty}^\tau \dd \tau' \Im{\delta\phi_0(\tau) \delta\phi_0^*(\tau'_-) \left[a^4(\tau'_-)V_3(\tau'_-) \int \dfrac{\dd^3 \vb{p}}{(2\pi)^3} \abs{\delta\phi_p(\tau'_-)}^2 + 2\delta_{\rm tp}(\tau'_-)\right] } = 0\,, \quad \forall \, \tau \,, \label{eq:Countertem_Tadpole}
\end{align}
in order for this equation to be satisfied, we need the counterterm to be 
\begin{equation}
	2\delta_{\rm tp}(\tau) = -a^4(\tau)V_3(\tau) \int \dfrac{\dd^3 \vb{p}}{(2\pi)^3} \abs{\delta\phi_p(\tau)}^2 \,.
\end{equation}
{Considering this functional form of $\delta_{\rm tp}$, it is easy to see that the tadpole contribution to the two-point correlation (diagram in Figure \ref{tadfig}) through the eq.\ (\ref{eq:In-In_General}) vanish.}

\section{Asymptotic limits} \label{sec:limits}

In this appendix we analyze in detail some aspects of the calculation of $\mathcal{P}_\zeta$ at one-loop. First, we consider the limit $k\to 0$, concluding that all the contributions to $\mathcal{P}_\zeta$ are scale invariant in this limit, but different from zero in general. For completeness, we also discuss briefly the limit $k\to \infty$. Finally, we study the parameter dependence of $\mathcal{P}_\zeta$ and use this dependence to estimate its magnitude.

\subsection{Limit $k\to 0$}
Looking at the equation of motion for $\delta\phi_k$,
\begin{equation}
	\delta\phi_k'' + 2Ha \delta\phi_k' + (k^2 + a^2V_2) \delta\phi_k = 0\,,
\end{equation}
and the Bunch-Davies initial conditions $\delta\phi_k(\tau\to-\infty) \to e^{-ik\tau}/\sqrt{2ka^2}$,
we note that $\delta\phi_k$ admits a series expansion around $k=0$
\begin{equation}
	\delta\phi_k(\tau) = \dfrac{1}{k^{3/2}} \sum_{n=0}^\infty k^n \delta\phi_n(\tau) \,.
\end{equation}
Each term $\delta\phi_n$ can be obtained by solving the equation of motion and imposing the initial (and boundary) conditions order by order in $k$. In particular, the solution of the modes in the first SR phase is
\begin{equation}
	\delta\phi_k (\tau < \tau_1) = \dfrac{iHe^{-ik\tau}}{\sqrt{2k^3}}(1+ik\tau) \,{.}
\end{equation}
Then, it is straightforward to obtain the evolution of $\delta\phi_n$ along inflation. Since $\delta\phi_k/\sqrt{\epsilon}$ is constant in the limit $k\to 0${,}
\begin{equation}
	\delta\phi_0(\tau) = \sqrt{\dfrac{\epsilon(\tau)}{\epsilon_{SR}}} \dfrac{iH}{\sqrt{2}}\,,
\end{equation}
where $\epsilon_{SR} \equiv \epsilon(\tau < \tau_1) $ is constant. 
We can now analyze the limit $k\to0$ of the loop integral $l(k;i,j)$, which is IR divergent.
Starting from the general expression, eq.\ \eq{eq:loop_integral},
\begin{align} \label{reploopint} \nonumber
	l(k;i,j) =\quad\quad\quad\quad\quad\quad\quad\quad\quad\quad\quad\quad\quad\quad\quad\quad\quad\quad\quad\quad\quad\quad\quad\quad\quad \quad\quad\quad\quad\quad\quad\quad\quad\quad\quad  \\ \dfrac{k^3}{16\pi^2} \int_{1+\Lambda_{\rm IR}/k}^\infty \dd s \int_0^1 \dd d \: (s^2 - d^2) \delta\phi_{k(s-d)/2}(\tau_i) \delta\phi^*_{k(s-d)/2}(\tau_{j-}) \delta\phi_{k(s+d)/2}(\tau_i) \delta\phi^*_{k(s+d)/2}(\tau_{j-}) \,,
\end{align}
we observe that the IR divergence comes from the region
$s \sim 1$ and $d\sim s$, given that
\begin{equation}
	\delta\phi_{k(s-d)/2}(\tau_i) = \dfrac{1}{(k(s-d)/2)^{3/2}}\left(\delta\phi_0(\tau_i) + \order{k^2(s-d)^2} \right) \,.
\end{equation}
The prescription $\tau_{j-}$ has an effect only on the UV, so we can omit it for the calculation of the IR divergence.
Then, the IR divergent part of $l(k;i,j)$ is
\begin{align}
	\nonumber
	l^{\rm IR}(k;i,j) &= \dfrac{\delta\phi_0(\tau_i) \delta\phi_0^*(\tau_j)}{\pi^2} \int_{1+\Lambda_{\rm IR}/k} \dd s \int_0^1 \dd d \: \dfrac{s}{(s-d)^2}\left(  \delta\phi_{ks}(\tau_i) \delta\phi_{ks}^*(\tau_j) + \order{s-d}^2\right) \\
	\nonumber
	& = \dfrac{\delta\phi_0(\tau_i) \delta\phi_0^*(\tau_j)}{\pi^2} \int_{1+\Lambda_{\rm IR}/k} \dd s  \dfrac{1}{s-1}\left(  \delta\phi_{ks}(\tau_i) \delta\phi_{ks}^*(\tau_j) + \textrm{finite}\right) \\
	&= -\dfrac{\log \Lambda_{\rm IR}/k}{\pi^2} \delta\phi_0(\tau_i) \delta\phi_0^*(\tau_j) \delta\phi_k(\tau_i) \delta\phi_k^*(\tau_j) + \textrm{finite} \,.
\end{align}
The IR divergence is $\sim \log k/\Lambda_{\rm IR}$, a standard result in this type of calculations. As already mentioned, it is not the aim of our work to deal with IR divergences, so we eliminate them by hand, by redefining $l(k;i,j)$ subtracting from it its IR divergent part, $l^{\rm IR}(k;i,j)$. With this redefinition there are no IR divergences (by construction), but neither UV divergences due to the $\tau_{j-}$ prescription.

Now, we compute the limit $k\to0$,
\begin{align}
	\nonumber
	\lim_{k\to0}l(k;i,j) &= \dfrac{4}{\pi^2k^3}\delta\phi_0^2(\tau_i) \delta\phi_0^{*2}(\tau_j) \int_{1+\Lambda_{\rm IR}/k}^\infty \dd s \int_0^1 \dfrac{\dd d}{\left(s^2 - d^2 \right)^2} +\dfrac{\log \Lambda_{\rm IR}/k}{\pi^2k^3} \delta\phi_0^2(\tau_i) \delta\phi_0^{*2}(\tau_j) \\
	& = \dfrac{\log2 -1}{\pi^2 k^3} \delta\phi_0^2(\tau_i) \delta\phi_0^{*2}(\tau_j) = \dfrac{\epsilon(\tau_i) \epsilon(\tau_j)H^4}{4\pi^2 \epsilon_{SR}^2  k^3}\left(\log2 -1 \right) + \order{1/k^2}\,,
\end{align}
{which is} real and its dependence on $k$ is $\Re{l(k)} \sim k^{-3}$. The first imaginary contribution 
is $\Im{l(k)} \sim k^{0}$. However, 
{calculating it}
is complicated because it receives a contribution from both the IR and the UV, so we cannot do an expansion around $k = 0$ as we have done to calculate the dominant real part. Nevertheless, the imaginary contribution plays an important role in the IR limit ($k\to 0$) of the power spectrum at one-loop.

We are now in a position to analyze each of the contributions to the power spectrum in this limit. Starting with the tree-level, 
\begin{equation}
	\mathcal{P}^{\rm tl}_\zeta (\tau,k) = \dfrac{k^3}{4\pi^2M_P^2\epsilon(\tau)} \abs{\delta\phi_k(\tau)}^2 \to \dfrac{k^3}{4\pi^2M_P^2\epsilon(\tau)} \dfrac{1}{k^3} \abs{\delta\phi_0(\tau)}^2 =\dfrac{H^2}{8\pi^2M_P^2\epsilon_{SR}} + \order{k^2}\,,
\end{equation}
we recover the usual scale invariant spectrum. The contribution of the counterterms is 
\begin{align}
	\nonumber
	\mathcal{P}^{\rm ct}_\zeta (\tau,k) &= \dfrac{k^3}{4\pi^2M_P^2\epsilon(\tau)}\left(  2\,\delta_\phi\, \abs{\delta\phi_k(\tau)}^2 +
	8 \Im{\delta\phi_k^2(\tau) \int_{-\infty}^\tau \dd \tau' a^2(\delta_\phi k^2 + \tilde{\delta}_V) \delta\phi_k^{*2}\eval_{\tau'}} \right) \\
	& \to 2\,\delta_\phi\, \mathcal{P}^{\rm tl}_\zeta (\tau,k) + \dfrac{2k^3}{\pi^2M_P^2\epsilon(\tau)} \Im{\delta\phi_k^2(\tau) \int_{-\infty}^\tau \dd \tau' a^2 \tilde{\delta}_V \delta\phi_k^{*2}\eval_{\tau'}}\,.
\end{align}
We have to analyze the term $\Im{\delta\phi_k^2(\tau) \delta\phi_k^{2*}(\tau')} = 2 \Im{\delta\phi_k(\tau) \delta\phi_k^{*}(\tau')} \Re{\delta\phi_k(\tau) \delta\phi_k^{*}(\tau')}$.
Making an expansion around $k\to 0$, we get
\begin{align}
	\dfrac{1}{\sqrt{\epsilon(\tau) \epsilon(\tau')}}\Re{\delta\phi_k(\tau) \delta\phi_k^{*}(\tau')} &\to \dfrac{H^2}{2k^3\epsilon_{SR}} + \order{1/k}\,,\\
	\dfrac{1}{\sqrt{\epsilon(\tau) \epsilon(\tau')}}\Im{\delta\phi_k(\tau) \delta\phi_k^{*}(\tau')} &\to \dfrac{1-\left(a(\tau')/ a(\tau)\right)^{3+\eta} }{2(3+\eta)a^3(\tau')H\epsilon(\tau')} + \order{k^2}\,,
\end{align}
where we have made a set of simplifications, taking $\eta$ as a constant and assuming that $\tau$ and $\tau'$ belong in the same phase. We note that the imaginary part is $\sim k^0$, because it satisfies the boundary conditions $\Im{\delta\phi_k(\tau) \delta\phi_k^{*}(\tau)} = 0$ and {$\Im{\delta\phi'_k(\tau) \delta\phi_k^{*}(\tau)} = -1/(2a^2(\tau))$}. We also note that this imaginary part has two solutions: one constant and one decaying (with the well-known exception occurring when $\eta <-3$).
When the simplifications we have made are no longer valid, the solution for the real part remains the same, but the imaginary part is more complicated. 
{However, its dependence with the momentum $k$ in the IR limit will still be $\Im{\delta\phi_k(\tau) \delta\phi_k^{*}(\tau')} \to \order{k^0}$.}
Keeping this in mind, we observe that the leading contribution of the counterterms to the power spectrum is also scale invariant, analogous to the tree-level one.
Finally, we analyze the one-loop contribution:
\begin{align} \nonumber
	\mathcal{P}^{\rm 1l}_{\zeta}(\tau,k) &= \dfrac{k^3}{4\pi^2M_P^2\epsilon(\tau)} \dfrac{2 H^2}{M_P^2} \sum_{i = 1}^4 \sum_{j=i + 1}^4 \dfrac{a^3\Delta\nu^2}{\sqrt{\epsilon}}\eval_{\tau_i} \dfrac{a^3\Delta\nu^2}{\sqrt{\epsilon}}\eval_{\tau_j} \times \\ \quad & \times \Im{\delta\phi_k(\tau) \delta\phi_{k}^*(\tau_j)} \Im{\delta\phi_k(\tau) \delta\phi_{k}^*(\tau_i)\, l(k;j,i)}\,{.}
\end{align}
As with the counterterms, we note that $\Im{\delta\phi_k(\tau) \delta\phi_{k}^*(\tau_j)} \Im{\delta\phi_k(\tau) \delta\phi_{k}^*(\tau_i)\, l(k;j,i)} \to \order{1/k^3}$, indicating that the one-loop contribution is scale invariant as well in this limit. 
Consequently, we deduce that all contributions to the total power spectrum (tree-level, one-loop and counterterms) are scale invariant in the limit $k\to 0$.

\subsection{Limit $k\to\infty$}
For completeness, we will analyze the behavior in the UV limit $k\to\infty$ of the one-loop contribution to the power spectrum. In this limit, the modes are sub-Hubble, they do not feel the effects of the background and then they are still in Bunch-Davies, $\delta\phi_{k\to\infty}(\tau) \to e^{-ik\tau}/\sqrt{2ka^2}$. Then, the loop integral is
\begin{align}
	\nonumber
	\lim_{k\to \infty} l(k;i,j)  &= \dfrac{k}{16\pi^2 a_i^2 a_j^2} \int_{1+\Lambda_{\rm IR}/k}^\infty \dd s \int_0^1 \dd d \: e^{-ik s(\tau_i - \tau_{j-})} + \dfrac{\log \Lambda_{\rm IR}/k}{\pi^2} \dfrac{\sqrt{\epsilon_i\epsilon_j}H^2 e^{-ik(\tau_i-\tau_j)}}{4\,\epsilon_{SR}\,k\,a_i\,a_j} \\
	& = \dfrac{i e^{-ik(\tau_i- \tau_j)}}{16\pi^2 a_i^2 a_j^2 (\tau_j- \tau_i)} + \order{1/k} \,.
\end{align}
Here we see the importance of the prescription $\tau_{j-}$ in the convergence of the integral. If it were missing, we would obtain a result that oscillates in the UV. 
We also note that if we had not subtracted the IR divergence, it would not play any role since it is suppressed by a factor $\sim \order{1/k}$. Because in the UV limit $l(k;i,j) \sim k^0$, the one-loop contribution to the power spectrum in this limit is
\begin{align}
	\nonumber
	\mathcal{P}^{\rm 1l}_{\zeta}(\tau,k) &= \dfrac{k^3}{4\pi^2M_P^2\epsilon(\tau)} \dfrac{2 H^2}{M_P^2} \sum_{i = 1}^4 \sum_{j=i + 1}^4 \dfrac{a^3\Delta\nu^2}{\sqrt{\epsilon}}\eval_{\tau_i} \dfrac{a^3\Delta\nu^2}{\sqrt{\epsilon}}\eval_{\tau_j} \times \\ \quad & \times  \Im{\delta\phi_k(\tau) \delta\phi_{k}^*(\tau_j)} \Im{\delta\phi_k(\tau) \delta\phi_{k}^*(\tau_i)\, l(k;j,i)}\\
	& \sim k^2 \delta\phi^2_k(\tau) \sim \dfrac{1}{k} \mathcal{P}^{\rm tl}_{\zeta}(\tau,k)
	\,.
\end{align}
That is, the contribution of the one-loop is suppressed with respect to that of the tree-level. 

\section{Dimensional analysis and a quick estimate of $\mathcal{P}_\zeta$}\label{sec:Estimation 1l}
Let us define a dimensionless comoving wave number $\kappa \equiv k/k_0 $. In terms of $k_0$, $k_0 \tau = -\dfrac{\tau}{\tau_0} = -e^{N_0 - N}$, where 
{$k_0 \tau_0 \equiv -1$}. We use the freedom we have to set $N_0 = 0$.
Examining the equations of motion for $\delta\phi_k(\tau)$, we observe that its solution will always be a function of $x \equiv k\tau = - \kappa e^{-N}$. Therefore, $\delta\phi_k(\tau) = \mathcal{C} \tilde{\delta\phi}_\kappa(x)$, where $\tilde{\delta\phi}_\kappa(x)$ is a dimensionless function and $\mathcal{C}$ is a constant that may depend on $k$ and captures all the dimensions of $\delta\phi_k(\tau)$. To determine $\mathcal{C}$, we analyze the initial conditions of $\delta\phi_k(\tau)$, $\delta\phi_k(\tau \to -\infty) \propto H \tau e^{-i x}/\sqrt{k} $, concluding that 
\begin{equation}
	\delta\phi_k(\tau) \equiv \dfrac{H}{k_0^{3/2}} \tilde{\delta\phi}_\kappa(x) \,.
\end{equation}
Now, the loop integral is $l(k;i,j) \sim k^3 \int \dd s \dd d \: \delta \phi^4 \equiv \dfrac{H^4}{k_0^3} \tilde{l}(\kappa;i,j)$, where again $\tilde{l}(\kappa;i,j)$ is dimensionless. We will also need $a(\tau) = -\dfrac{1}{H\tau} = \dfrac{k_0 \tau_0}{H\tau} = \dfrac{k_0}{H} \tilde{a}(x/\kappa)$. 
With all that, we can move on to analyze the contribution to the power spectrum of the tree-level 
\begin{equation}
	\mathcal{P}^{\rm tl}_\zeta (\tau,k) = \dfrac{k^3}{4\pi^2M_P^2\epsilon(\tau)} \abs{\delta\phi_k(\tau)}^2 =  \dfrac{H^2}{M_P^2} \dfrac{\kappa^3}{4\pi^2\epsilon(\tau)} \abs{\tilde{\delta\phi}_\kappa(x)}^2 \,,
\end{equation}
and the one-loop
\begin{align} 
	\nonumber
	\mathcal{P}^{\rm 1l}_{\zeta}(\tau,k) &= \dfrac{k^3}{4\pi^2M_P^2\epsilon(\tau)} \dfrac{2 H^2}{M_P^2} \sum_{i = 1}^4 \sum_{j=i + 1}^4 \dfrac{a^3\Delta\nu^2}{\sqrt{\epsilon}}\eval_{\tau_i} \dfrac{a^3\Delta\nu^2}{\sqrt{\epsilon}}\eval_{\tau_j} \times \\\quad & \times  \Im{\delta\phi_k(\tau) \delta\phi_{k}^*(\tau_j)} \Im{\delta\phi_k(\tau) \delta\phi_{k}^*(\tau_i)\, l(k;j,i)}\\
	&= \dfrac{H^4}{M_P^4} \dfrac{\kappa^3}{2\pi^2\epsilon(\tau)} \sum_{i = 1}^4 \sum_{j=i + 1}^4 \dfrac{\tilde{a}^3\Delta\nu^2}{\sqrt{\epsilon}}\eval_{\tau_i} \dfrac{\tilde{a}^3\Delta\nu^2}{\sqrt{\epsilon}}\eval_{\tau_j}\times \nonumber  \\ \quad & \times  \Im{\tilde{\delta\phi}_\kappa(x) \tilde{\delta\phi}^*_\kappa(x_j)} \Im{\tilde{\delta\phi}_\kappa(x) \tilde{\delta\phi}^*_\kappa(x_i)\, \tilde{l} (\kappa;j,i)}\,.
\end{align}
We obtain, as it should be, that both contributions to the power spectrum are dimensionless. The factor that controls (naively) the perturbative expansion is $H^2/M_P^2$.
Moreover, we find that the dependence on $k_0$ disappears. This was expected since $k_0$ is only a reference value without physical meaning. 

\begin{figure}[t]
	\begin{center}
		\includegraphics[width=0.90\textwidth]{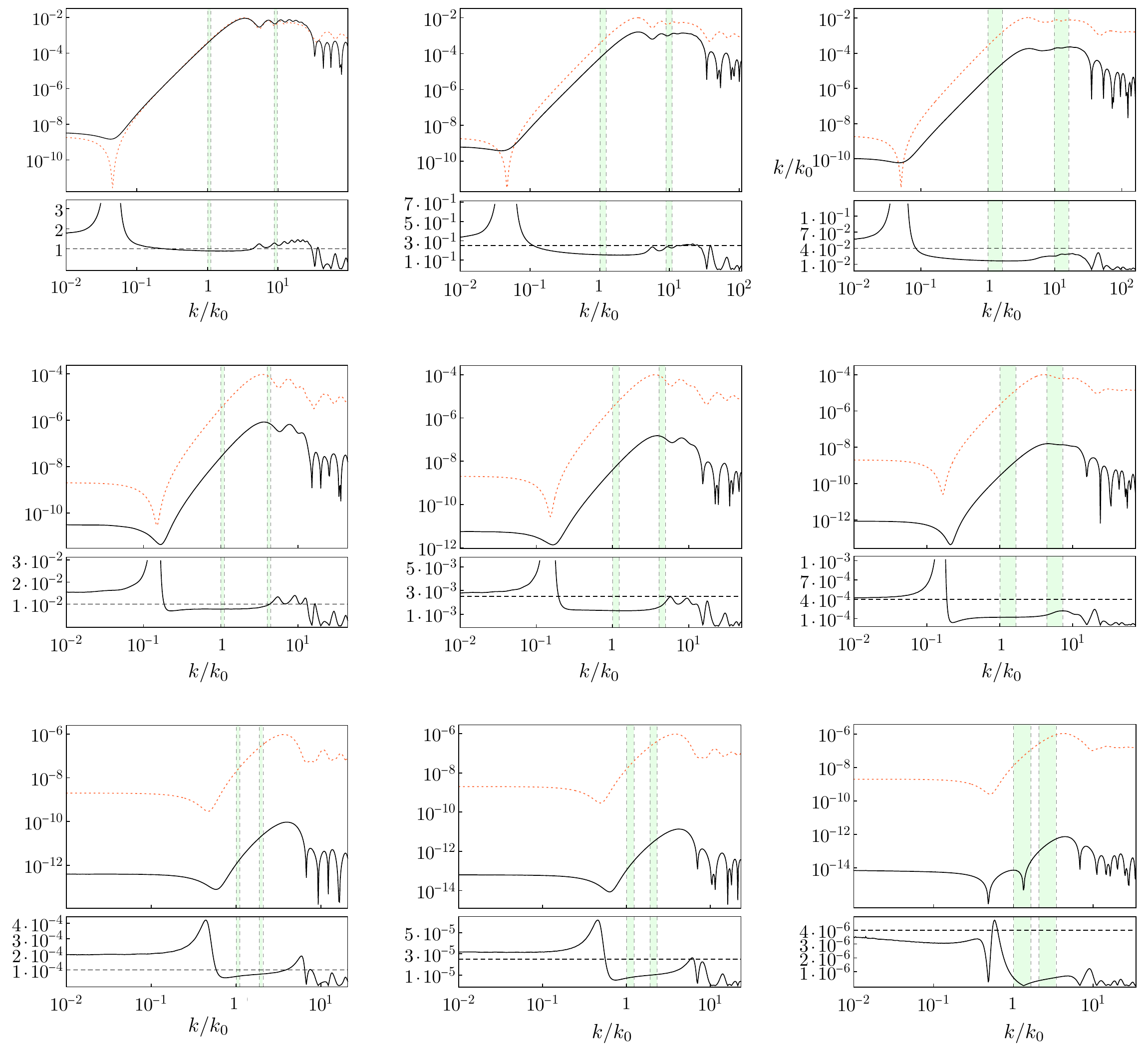}
		\caption{\label{fig:PTbreak3x3}  \small \it Each graphic shows, for nine different choices of $\delta N$ and $\mathcal{P}^{\rm tl}_\zeta(k_{\rm peak})$: At the top, tree-level (red dashed) and absolute value of renormalized one-loop (black continuous); at the bottom, absolute value of renormalized one-loop divided by tree-level (black continuous) and $\mathcal{P}^{\rm tl}_\zeta(k_{\rm peak}) / \delta N^2$ (black dashed). The rows, from left to right, represent $\delta N = \{0.1,\,0.2,\,0.5\}$, while the columns, from top to bottom, represent $\mathcal{P}^{\rm tl}_\zeta(k_{\rm peak}) = \{10^{-2},\,10^{-4},\,10^{-6}\}$.}
	\end{center}
\end{figure}

Finally, we will make an estimate of the value of $\mathcal{P}^{\rm 1l}_\zeta / \mathcal{P}^{\rm tl}_\zeta $ in a general scenario, i.e.\ for any value of $\delta N$ and $\Delta N$ (always being $\delta N \ll \Delta N$). 
From the previous analysis $\mathcal{P}^{\rm 1l}_\zeta / \mathcal{P}^{\rm tl}_\zeta \sim H^2/M_P^2 \sim \mathcal{P}^{\rm tl}_\zeta$. Moreover, since $\mathcal{P}^{\rm 1l}_\zeta$ includes two factors of $\Delta \nu^2 \sim 1/\delta N$, we expect
\begin{equation}
	\dfrac{\mathcal{P}^{\rm 1l}_\zeta(k)}{\mathcal{P}^{\rm tl}_\zeta(k)} \propto \dfrac{1}{\delta N^2} \mathcal{P}^{\rm tl}_\zeta(k)\,.
\end{equation}
However, it is challenging to determine the $k$-dependence because
it requires
an analytical estimate of the loop integral $l(k;i,j)$.
Hence, to estimate the magnitude of the ratio $\mathcal{P}^{\rm 1l}_\zeta / \mathcal{P}^{\rm tl}_\zeta$ we resort to numerical calculations. In {Figure \ref{fig:PTbreak3x3}} we illustrate its variation with different values of $\delta N$ and $\mathcal{P} ^{\rm tl}_\zeta(k_{\rm peak})$. We see that 
\begin{equation}
	\dfrac{\mathcal{P}^{\rm 1l}_\zeta(k)}{\mathcal{P}^{\rm tl}_\zeta(k)} = \order{1} \dfrac{\mathcal{P}^{\rm tl}_\zeta(k_{\rm peak})}{\delta N^2} 
\end{equation}
is a good estimate for all different values of $\delta N$ and $\Delta N$ and all the scales $k$.

\subsection{The limit $\delta N \to 0$}

As we have seen, by analysing the dimensions of $\mathcal{P}^{\rm 1l}_\zeta$ and through a numerical estimation, we obtain that $\mathcal{P}^{\rm 1l}_\zeta \propto 1/\delta N^2$. In the limit $\delta N \to 0$, $\mathcal{P}^{\rm 1l}_\zeta$ is therefore divergent, a difference with respect to previous results (see e.g.\ \cite{Kristiano:2022maq,Franciolini:2023lgy}). It is natural to ask what is the origin of this difference. 

In order to compare with the literature, we abandon in this Appendix the cutoff regularization that we have been using so far. This means that we now have to include the points along the symmetrical (red) diagonal of Figure \ref{fig:ttplott}, which up until now we have left out due to the regulator. Doing so, $\mathcal{P}^{\rm 1l}_\zeta$ becomes UV divergent, and its therefore necessary to introduce another regulator that does not eliminate the diagonal contributions, e.g.\ dimensional regularization or Pauli-Villars.
We will left the specific nature of the regulator unspecified and simply assume that we can indeed regularize the integrals in the UV. We will work directly in the limit $\delta N \to 0$, and for simplicity we will focus exclusively on the last transition from USR to SR, i.e.\ for times $\tau \in \left[\tau_3\,,\tau_4\right]$. As we shall see, this is enough to understand the behavior of $\mathcal{P}^{\rm 1l}_\zeta$ in the limit $\delta N \to 0$.
With these considerations, we are going to calculate the two-point correlation of $\zeta$. As a consistency check, we will first use the $\delta \phi$-gauge and then the $\zeta$-gauge. We will find that both gauges lead to the same result. 

Starting with the calculation via the $\delta \phi$-gauge, we have
\begin{align}
	\nonumber
	\expval{\zeta(x)\zeta(y)}^{\rm 1l}_{\delta \phi} &= \dfrac{1}{2\epsilon(\tau) M_p^2} \Bigg( 2 \Im{\int_{-\infty}^\tau \dd \tau' \expval{\delta \phi(x)\delta \phi(y) H_I(\tau'_-)}} \\
	&+ 2 \Re{ \int_{-\infty}^\tau \dd \tau' \int_{\tau'}^\tau \dd \tau'' \expval{\left( H_I(\tau''_+)\delta \phi(x)\delta \phi(y) -\delta \phi(x)\delta \phi(y)H_I(\tau''_-) \right)  H_I(\tau'_-)}} \Bigg) \,,
\end{align}
where
\begin{equation}
	H_I(\tau) = \int\dd^3\vb{x}\, a^4 \left(\dfrac{V_3}{3!}\delta\phi^3 + \dfrac{V_4}{4!}\delta\phi^4 \right) \,.
\end{equation}
Since in the limit $\delta N\to 0$ we have $\Delta \nu^2(\tau_3) = - \Delta \nu^2(\tau_4) = 3/ \delta N$, we see that the time dependence of $V_4$ relevant for the calculation is of the form $V_4 \sim \frac{1}{\delta N} \delta(\tau - \tau_i)$.\footnote{In fact, $V_4$ also has contributions $\sim \delta(\tau - \tau_i)'$. However, because the contribution from $V_4$ to $\mathcal{P}^{1l}_\zeta$ is integrated over all time, we can integrate by parts to move the time derivative from the Dirac delta to the fields. Therefore, in practice the argument by which we neglect the quartic contribution applies.} The contribution to the power spectrum of $\delta\phi$ coming from $V_4$ (and computed in the $\delta\phi$-gauge) turns out to scale as  
\begin{equation}
	\mathcal{P}^{\rm 1l, V_4}_{\delta \phi} \sim \dfrac{1}{\delta N} \left(f(\tau,\tau_3) - f(\tau,\tau_4) \right)\,,
\end{equation}
where $f$ is a continuous function whose explicit form we do not need to write. In the limit $\tau_4\to \tau_3$, there is a cancellation in the numerator which makes $\mathcal{P}^{\rm 1l, V_4}_{\delta \phi} \sim \order{\delta N^0}$ and so does not diverge as $\delta N \to 0$. As we shall see, the contribution coming from $V_3$ does instead diverge in that limit, and therefore $\mathcal{P}^{\rm 1l, V_4}_{\delta \phi}$ is a subdominant contribution that we can neglect. Thus, we have
\begin{align} \nonumber
	\expval{\zeta(x)\zeta(y)}^{\rm 1l}_{\delta \phi} = \dfrac{1}{\epsilon(\tau) M_p^2}\, \text{Re}\, \bigg\{ \int_{-\infty}^\tau \dd x' \dfrac{a^4V_3}{3!} \int_{\tau'}^\tau \dd x'' \dfrac{a^4V_3}{3!}  \bigg\langle \delta \phi^3(x'_-) \times \\ \times \left( \delta \phi^3(x''_+)\delta \phi(x)\delta \phi(y)  -\delta \phi(x)\delta \phi(y)\delta \phi^3(x''_-)\right)  \bigg\rangle\bigg\}\,.
\end{align}
This correlation induces the topology of diagrams depicted in the third row of Figure \ref{fig:one-loops}. Including all the diagrams that affect the power spectrum (the first and the second topologies),\footnote{As shown in {Appendix} \ref{sec:Tadpoles}, by imposing that $\expval{\zeta(x)} = 0$ the effect of the diagram proportional to the tadpole (second of the third row of Fig.\ \ref{fig:one-loops}) vanishes. However, we note that here we have not yet imposed any conditions on the counterterms, and therefore all contributions to $\expval{\zeta(x)\zeta(y)}$ must be included. Moreover, our aim is to compare the result with that obtained in another gauge, so the only way to do this consistently is to blindly include all diagrams, without the need to talk about counterterms.} we arrive at
\begin{align} \label{divUVsin}
	\nonumber
	\expval{\zeta(x)\zeta(y)}^{\rm 1l}_{\delta \phi} & = \int \dfrac{\dd^3\vb{k}}{(2\pi)^3} e^{i \vb{k}(\vb{x} - \vb{y})} \int \dfrac{\dd^3\vb{p}}{(2\pi)^3} \dfrac{36 M_P^4 H^2}{\delta N^2} \times \\ \nonumber & \times	\Re \Big\lbrace a_3^6 \epsilon_3^2 \zeta_k(\tau) \zeta_k^*(\tau_3) 2i \Im{\zeta_k^*(\tau) \zeta_k(\tau_3)} \abs{\zeta_p(\tau_3)}^2 \abs{\zeta_{\abs{\vb{k} - \vb{p}}}(\tau_3)}^2 + (3\to 4)\\
	\nonumber
	 & -2a_3^3\epsilon_3 a_4^3 \epsilon_4 \zeta_k(\tau) \zeta_k^*(\tau_3) 2i \Im{\zeta_k^*(\tau) \zeta_k(\tau_4)} \zeta_p(\tau_4) \zeta_p^*(\tau_{3-}) \zeta_{\abs{\vb{k} - \vb{p}}}(\tau_4) \zeta_{\abs{\vb{k} - \vb{p}}}^*(\tau_{3-})\\
	 & -2a_3^3\epsilon_3 a_4^3 \epsilon_4 i\Im{\zeta_k^2(\tau_4) \zeta_k^{*2}(\tau)} \abs{\zeta_p(\tau_3)}^2 \lim_{l\to 0} \zeta_l(\tau_4) \zeta_l^*(\tau_3)  \Big\rbrace \,,
\end{align} 
where $a_i = a(\tau_i)$ and $\epsilon_i = \epsilon(\tau_i)$.
The first two lines correspond to the first diagram, while the third line corresponds to the second one. Because the fields are in the interaction picture, i.e.\ they are determined by the free action, we have used the linear relation between the classical fields: $\delta \phi_k(\tau) = -\sqrt{2\epsilon(\tau)} M_p \zeta_k(\tau)$. We have also omitted the prescription $i\omega$ wherever it plays no role. Of course, eq.\ \eq{divUVsin} is divergent in the UV, and we stress that we would need to regularize it in order to obtain a finite result.
We will now take the limit $\delta N\to 0$. To do so, we note that $a^3_4\epsilon_4 = a^3_3 \epsilon_3 + \order{\delta N^2}$ and $\tau_4 - \tau_3 = \delta N/(a_3H)$. Moreover, since $\zeta$ is smooth in this limit, we can expand $\zeta_k(\tau_4)$ in series around $\tau_3$. Then, we arrive at
\begin{align}\label{eq:P_z from dp - delta N limit}
	\expval{\zeta(x)\zeta(y)}^{\rm 1l}_{\delta \phi} = \int \dfrac{\dd^3\vb{k}}{(2\pi)^3} e^{i \vb{k}(\vb{x} - \vb{y})} \int \dfrac{\dd^3\vb{p}}{(2\pi)^3} \dfrac{54 M_P^2 H}{\delta N} a_3^3\epsilon_3 \Im{\zeta_k^2(\tau) \zeta_k^{*2}(\tau_3)}\abs{\zeta_p(\tau_3)}^2\,,
\end{align}
where we have used $\Im{\zeta_k(\tau) \zeta_k^{*'}(\tau)} =1/(4\epsilon(\tau)M_P^2 a^2(\tau)) $, which is a general relation derived from the quantization condition.
The two-point correlation obtained is valid for the whole range of external comoving momentum $k$ in the limit $\delta N\to 0$.
We conclude that, using a regulator that does not forbid the points along the (symmetric) diagonal of Figure \ref{fig:ttplott}, 
the naive estimate of the divergence of the two-point correlation in the limit $\delta N\to 0$ relaxes from $1/\delta N^2$ to $1/\delta N$.
This, however, still contradicts the results in earlier works, see e.g.\ \cite{Kristiano:2022maq,Franciolini:2023lgy} (obtained in the $\zeta$-gauge) where no such divergence was identified. We will now try to address this apparent inconsistency by calculating the two-point correlation directly in the $\zeta$-gauge. As we will see, this calculation agrees with eq.\ (\ref{eq:P_z from dp - delta N limit}).

In the $\zeta$-gauge, the quadratic action together with the dominant term of the cubic action in the inflationary model under consideration is  \cite{Maldacena:2002vr}
\begin{equation}
	S=\int \dd \tau \dd^3\vb{x} \, M_P^2 a^2\epsilon\left(\left( \zeta'\right) ^2 - \left(\partial\zeta \right)^2 + \dfrac{\eta'}{2} \zeta' \zeta^2  \right)\,.
\end{equation}
We are leaving aside boundary terms which, of course, may have a relevant contribution to the correlation functions and, as a rule, always have to be included \cite{Arroja:2011yj,Burrage:2011hd,Braglia:2024zsl}. The reason why we do not include these boundary terms is because none of them has a coupling that presents a divergence in the limit $\delta N\to0$. In other words, they are subdominant terms in this limit.
In order to compute the two-point correlation using the in-in formalism, we need to calculate the interaction Hamiltonian in the interaction picture. The canonical conjugate momentum of $\zeta$ is 
\begin{equation}
	\pi_\zeta = \dfrac{\partial \mathcal{L}}{\partial \zeta'} =  2 M_P^2 a^2 \epsilon\left(\zeta' + \dfrac{\eta'}{4}\zeta^2 \right) \,, \quad \zeta' = \dfrac{\pi_\zeta}{2 M_P^2 a^2 \epsilon} - \dfrac{\eta'}{4}\zeta^2 \,,
\end{equation}
and therefore the Hamiltonian is 
\begin{equation}
	\mathcal{H} = \pi_\zeta \zeta' - \mathcal{L} = M_P^2 a^2 \epsilon\left( \dfrac{\pi_\zeta^2}{4M_P^4a^4\epsilon^2} + \left(\partial \zeta \right)^2 - \dfrac{\eta'}{4M_P^2 a^2\epsilon} \pi_\zeta \zeta^2 + \dfrac{\left( \eta'\right) ^2}{16} \zeta^4  \right) \,.
\end{equation}
We identify the interaction Hamiltonian as 
\begin{equation}
	\mathcal{H}_{\rm i} =  M_P^2 a^2 \epsilon\left( - \dfrac{\eta'}{4M_P^2 a^2\epsilon} \pi_\zeta \zeta^2 + \dfrac{\left( \eta'\right) ^2}{16} \zeta^4  \right)\,.
\end{equation}
Now, we take into account an important detail (see e.g.\ \cite{Chen:2006dfn} as well): in the in-in formalism we work in the interaction picture, so the conjugate momentum in this picture only sees the free action, i.e.\ $\pi_\zeta^I = 2 M_P^2 a^2\epsilon \left( \zeta^{I}\right)' $. Thus, the interaction Hamiltonian in the interaction picture is 
\begin{equation}\label{eq:Int Hamiltonian z-gauge}
	H_I(\tau) = \int \dd^3\vb{x}\,  M_P^2 a^2 \epsilon\left( - \dfrac{\eta'}{2} \zeta' \zeta^2 + \dfrac{\left( \eta'\right) ^2}{16} \zeta^4  \right)\,.
\end{equation}
We stress that $H_I$, which has a cubic and a quartic interaction, is induced exclusively by the cubic term in the original action.
The quartic interaction is usually neglected in the literature and, as we shall see, it is precisely the one that gives rise to the divergence in the limit $\delta N \to 0$. Knowing $H_I$, we can calculate the two-point correlation 
\begin{align} \label{inter34}
	\nonumber
	\expval{\zeta(x)\zeta(y)}^{\rm 1l}_{\zeta} &=  2 \Im{\int_{-\infty}^\tau \dd \tau' \expval{\zeta(x)\zeta(y) H_I(\tau'_-)}} \\
	&+ 2 \Re{ \int_{-\infty}^\tau \dd \tau' \int_{\tau'}^\tau \dd \tau'' \expval{\left( H_I(\tau''_+)\zeta(x)\zeta(y) -\zeta(x)\zeta(y)H_I(\tau''_-) \right)  H_I(\tau'_-)}} \,.
\end{align} 
In the limit $\delta N \to 0$, we have that $\eta'(\tau) = -6 a_3 H / \delta N $ for $\tau\in \left[ \tau_3\,, \tau_4\right] $.
This coupling (squared) introduces a factor $1/\delta N^2$, but the time integrals (over an integrand that is smooth in this limit) cover an integration region with area $\sim \delta N^2$. Therefore, the (purely) cubic contribution does not diverge, as obtained in previous works; see in particular~\cite{Franciolini:2023lgy}. However, the coupling of the (induced) quartic contribution also introduces a factor $1/\delta N^2$ and, since it appears inside just one time integral (from the first line of \eq{inter34}, which gives a factor $\sim \delta N$), this leads to the $1/\delta N$ divergence that we also find in the $\delta\phi$-gauge. We make the calculation explicitly to show that the results in the $\zeta$- and $\delta \phi$-gauges are consistent.
\begin{align}\label{eq:P_z from z - delta N limit}
	\nonumber
	\expval{\zeta(x)\zeta(y)}^{\rm 1l}_{\zeta} &=  2 \Im{\int_{-\infty}^\tau \dd^4x'\, M_P^2a^2\epsilon \dfrac{\left( \eta'\right) ^2}{16} \expval{\zeta(x)\zeta(y) \zeta^4(x'_-)}} \\
	&= \int \dfrac{\dd^3\vb{k}}{(2\pi)^3} e^{i \vb{k}(\vb{x} - \vb{y})} \int \dfrac{\dd^3\vb{p}}{(2\pi)^3} \dfrac{54 M_P^2 H}{\delta N} a_3^3\epsilon_3 \Im{\zeta_k^2(\tau) \zeta_k^{*2}(\tau_3)} \abs{\zeta_p(\tau_3)}^2\,.
\end{align}
The equations (\ref{eq:P_z from dp - delta N limit}) and (\ref{eq:P_z from z - delta N limit}) coincide. It appears that the $\delta N \rightarrow 0$ divergence may have been missed in previous works due to an incorrect derivation of $H_I$.

To conclude, we discuss the discrepancy between this result and the estimate we obtained earlier regularizing with a cutoff, which was a scaling $\mathcal{P}_\zeta^{\rm 1l} \sim 1/\delta N^2$. As we have seen, the reason why the divergence relaxes to $1/\delta N$ after removing the cutoff is a cancellation involving the contributions coming from the interactions located along the (symmetric) diagonal of Figure \ref{fig:ttplott}. 
If instead we keep the cutoff --see eq.\ \eq{eq:reg_time_integrals}--, this cancellation does not occur and the divergence remains as $1/\delta N^2$.
We stress that in any case, in the limit $\delta N \to 0$ we obtain that $\mathcal{P}_\zeta^{\rm 1l}$ grows unbounded; i.e.\ perturbation theory is broken in that limit. Therefore, we cannot trust the estimate of $\mathcal{P}_\zeta^{\rm 1l}$ obtained through any regulator. 

\section{Bispectrum and consistency condition}\label{sec:Bispectrum}

\begin{figure}[t]
	\begin{center}
		\includegraphics[width=1\textwidth]{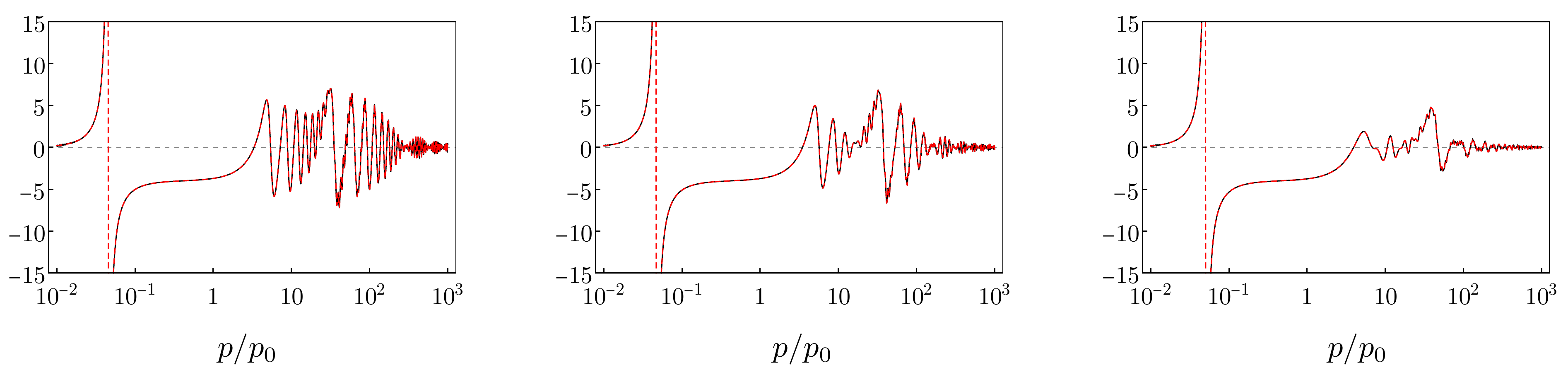}
		\caption{\label{fig:bispectrum} \it     \small The ratio $ {B^{\rm tl}_\zeta(\tau;\vb{k},\vb{p},-\vb{p})}/{(P^{\rm tl}_\zeta(\tau,p) P^{\rm tl}_\zeta(\tau,k))}$ in the squeezed limit (black continuous) and $- \dd \log \mathcal{P}_\zeta(\tau,p)/{\dd \log p}$ (red dashed) for the three models $\{\delta N, \Delta N\}$ of Figure \ref{fig:Set_Of_1loop}. Both quantities coincide for all values of $p/p_0$. The consistency relation is satisfied and the bispectrum is small for large $p/p_0$. }
	\end{center}
\end{figure}

The cubic interaction we consider in this work (see eq.\ (\ref{eq:Int_Hamiltonian})) implies a tree-level contribution to the three-point function of $\zeta$ that we can calculate using the in-in formalism {(eq.\ (\ref{eq:In-In_General}))}:
\begin{align} \nonumber
	\expval{\zeta(x) \zeta(y) \zeta(z)} &= -\dfrac{1}{(\sqrt{2\epsilon(\tau)}M_P)^3} \expval{\delta\phi(x) \delta\phi(y) \delta\phi(z)} \\
	&= -\dfrac{1}{(\sqrt{2\epsilon(\tau)}M_P)^3} 2\Im{\int_{-\infty}^\tau \dd^4 x' \dfrac{a^4(\tau')V_3(\tau')}{3!}\bra{0} \delta\phi(x) \delta\phi(y) \delta\phi(z) \delta\phi^3(x') \ket{0}}\,,
\end{align}
{where, as usual, $x' = \left(\tau',\, \vb{x}' \right) $.}
The prescription $i\omega$ plays no role due to the instantaneous interactions we consider, and so we have omitted it in the last expression. We also stress that the {linear} relation we have used between $\zeta$ and $\delta\phi$ {(eq.\ (\ref{eq:App_Expanion_Zeta}))} is only valid at late times. We obtain
\begin{equation}
	\expval{\zeta(x) \zeta(y) \zeta(z)} = \int \dfrac{\dd^3 \vb{p_1}}{(2\pi)^{3}} \dfrac{\dd^3 \vb{p_2} }{(2\pi)^{3}} \dfrac{\dd^3 \vb{p_3}}{(2\pi)^{3}} e^{i(\vb{p_1}\vb{x} + \vb{p_2}\vb{y} + \vb{p_3}\vb{z})}(2\pi)^3 \delta(\vb{p_1} + \vb{p_2} + \vb{p_3}) B_\zeta(\tau;\vb{p_1},\vb{p_2},\vb{p_3})
\end{equation}
where, at tree-level, the bispectrum is
\begin{equation}
	B^{\rm tl}_\zeta(\tau;\vb{p_1},\vb{p_2},\vb{p_3}) = \sum_{i=1}^4 \dfrac{H a^3_i \Delta \nu^2_i}{2M_P^4 \left( \epsilon(\tau) ^3 \epsilon_i \right)^{1/2}} \Im{\delta\phi_{p_1}(\tau) \delta\phi^*_{p_1}(\tau_i) \delta\phi_{p_2}(\tau) \delta\phi^*_{p_2}(\tau_i) \delta\phi_{p_3}(\tau) \delta\phi^*_{p_3}(\tau_i)} \,.
	\label{eq:bispectrum_general_dp-gauge}
\end{equation}
In the squeezed limit, $p_1 \equiv k \ll p_2 \simeq p_3 \equiv p$,
\begin{equation}
	B^{\rm tl}_\zeta(\tau;\vb{k},\vb{p},-\vb{p}) = \sum_{i=1}^4 \dfrac{H a^3_i \Delta \nu^2_i}{2M_P^4 \epsilon^2(\tau) } \Im{ \left( \delta\phi_{p}(\tau) \delta\phi^*_{p}(\tau_i) \right) ^2} \abs{\delta\phi_{k}(\tau) }^2 \,,
\end{equation}
where we have used that $\delta\phi_k(\tau_i) = \sqrt{\epsilon_i / \epsilon_k(\tau)} \delta\phi_k(\tau)\left(1+\order{k^{-2}} \right) $ in the limit $\abs{k\tau} \ll 1$. Defining 
\begin{equation}
	P^{\rm tl}_\zeta(\tau,k) \equiv \dfrac{2\pi^2}{k^3} \mathcal{P}^{\rm tl}_\zeta(\tau,k) = \dfrac{\abs{\delta\phi_k(\tau)}^2}{2\epsilon(\tau)M_P^2}\,,
\end{equation}
we finally arrive at the expression 
\begin{equation}
	B^{\rm tl}_\zeta(\tau;\vb{k},\vb{p},-\vb{p}) = 2\, H\,P^{\rm tl}_\zeta(\tau,p)\, P^{\rm tl}_\zeta(\tau,k) \sum_{i=1}^4 a^3_i \Delta \nu^2_i \Im{ \dfrac{\left( \delta\phi_{p}(\tau) \delta\phi^*_{p}(\tau_i) \right) ^2}{\abs{\delta\phi_p(\tau)}^2}}\,.
	\label{eq:Bispectrum-Squeezed_limit}
\end{equation}
Maldacena's consistency relation \cite{Maldacena:2002vr} 
imposes a constraint on the bispectrum in this squeezed limit,\footnote{See \cite{Bravo:2017wyw} for a generalization when $\zeta$ is not constant for super-Hubble modes.}
\begin{equation}
	B_\zeta(\tau;\vb{k},\vb{p},-\vb{p}) = - \dfrac{\dd \log \mathcal{P}_\zeta(\tau,p)}{\dd \log p} P_\zeta(\tau,p) P_\zeta(\tau,k) \,.
	\label{eq:Consistency-cond}
\end{equation}
Restricting the calculation to tree-level, the RHS of eq.\ (\ref{eq:Consistency-cond}) only depends on the quadratic action through the dynamics of the free fields. The LHS, computed in the in-in formalism and explicitly written in eq.\ (\ref{eq:Bispectrum-Squeezed_limit}), depends on the coupling of the cubic action. Therefore, this is a non-trivial check of the relation between the quadratic and cubic actions, which holds in our model, as shown in Figure \ref{fig:bispectrum}. 

{As in Appendix \ref{sec:Estimation 1l}, it is important to consider the limit $\delta N\to0$, since the model with instantaneous transitions originally considered in \cite{Kristiano:2022maq} is recovered.}
It was found in \cite{Kristiano:2023scm,Tada:2023rgp}, working on the $\zeta$-gauge, that 
in an analogous model defined in terms of $\eta$ the consistency relation is {also} satisfied and, moreover, the bispectrum is finite. 
The limit of instantaneous transitions must be treated with care to compute the bispectrum of $\zeta$ using the $\delta\phi$-gauge, since there is a subtle cancellation that removes the naive divergence (as $\delta N \rightarrow 0$) that might be expected to be introduced by the transitions. Being specific, in this limit
$\Delta \nu^2(\tau_1) = \Delta \nu^2(\tau_4) = - \Delta \nu^2(\tau_2) = - \Delta \nu^2(\tau_3) = -3/\delta N+\mathcal{O}(1)$, and eq.\ (\ref{eq:bispectrum_general_dp-gauge}) simplifies to
\begin{align}
	B_\zeta^{\rm tl}(\tau;\vb{p}_1,\vb{p}_2,\vb{p}_3) &= -\dfrac{12 H M_P^2 }{\delta N} \Im\Bigg\lbrace \dfrac{\delta\phi_{p_1}(\tau)}{\sqrt{2\epsilon(\tau)}M_P} \dfrac{\delta\phi_{p_2}(\tau)}{\sqrt{2\epsilon(\tau)}M_P} \dfrac{\delta\phi_{p_3}(\tau)}{\sqrt{2\epsilon(\tau)}M_P}\times\\ \nonumber
	&\times\left(a_1^3\epsilon_1 \dfrac{\delta\phi^*_{p_1}(\tau_1)}{\sqrt{2\epsilon_1}M_P} \dfrac{\delta\phi^*_{p_2}(\tau_1)}{\sqrt{2\epsilon_1}M_P} \dfrac{\delta\phi^*_{p_3}(\tau_1)}{\sqrt{2\epsilon_1}M_P}+ (\tau_1\rightarrow \tau_4)  - (\tau_1\rightarrow \tau_2)- (\tau_1\rightarrow \tau_3)\right) \Bigg\rbrace\,.
\end{align}
In order to facilitate a comparison with the earlier literature mentioned above, we define now $\tau_1 \equiv  \tau_s$ and  $\tau_3 \equiv \tau_e$ so that $\tau_2 - \tau_1 = \delta N / (a_s H)+\mathcal{O}(\delta N^2)$ and $\tau_4 - \tau_3 = \delta N / (a_e H)+\mathcal{O}(\delta N)^2$.
Care must be taken when taking the limit $\delta N \to 0$, since the time derivative of $\delta\phi_k(\tau)$ is not continuous.
However, we stress that $\zeta \propto \delta\phi / \sqrt{\epsilon}$ has a smooth derivative, and hence $\zeta_k(\tau_2) = \zeta_k(\tau_s) + (\tau_2 - \tau_1) \zeta'_k(\tau_s)$ and $\zeta_k(\tau_4) = \zeta_k(\tau_e) + (\tau_4 - \tau_3) \zeta'_k(\tau_e)$. Again, $\epsilon$ must be treated with care, since its derivative is not continuous in this limit, and we obtain $a^3_2\epsilon_2 = a^3_s\epsilon_s(1+\order{\delta N^2})$ and $a^3_4\epsilon_4 = a^3_e\epsilon_e(1+\order{\delta N^2})$. In this way, we can rewrite the above expression in a more compact form, 
\begin{align} \nonumber
	B_\zeta^{\rm tl}(\tau;\vb{p}_1,\vb{p}_2,\vb{p}_3) = \quad \quad\quad \quad\quad \quad \quad\quad\quad \quad\quad \quad \quad \quad\quad \quad\quad \quad\quad \quad\quad \quad\quad \quad\quad \quad\quad \quad \\ 12 M_P^2 \Im\Bigg\lbrace \zeta_{p_1}(\tau) \zeta_{p_2}(\tau) \zeta_{p_2}(\tau)
	a_s^2\epsilon_s \dfrac{\dd}{\dd \tau'}\left(  \zeta^*_{p_1}(\tau') \zeta^*_{p_2}(\tau') \zeta^*_{p_3}(\tau')\right)\eval_{\tau'=\tau_s} \Bigg\rbrace - (s\leftrightarrow e)\,,
	\label{eq:bispectrum_limit_dp-gauge}
\end{align}
where we have used the first-order relation between $\zeta$ and $\delta\phi$. 
{As with the two-point correlation in Appendix \ref{sec:Estimation 1l}, a cancellation at the times of the transitions makes that, in this case, the divergence of the bispectrum disappears in the limit $\delta N\to0$.}
We can now compute the bispectrum of $\zeta$ directly in the $\zeta$-gauge using the in-in formalism with the interaction 
\begin{align} \label{revint}
		S \supset \frac{M_P^2}{2}\int \dd^4x \, a^2 \, \epsilon \, \eta'\,\zeta'\,\zeta^2\,,
\end{align} 
which is the most relevant interaction in this type of models, as discussed in Appendix \ref{sec:Estimation 1l}.
We obtain the same expression for the tree-level bispectrum as the eq.\ (\ref{eq:bispectrum_limit_dp-gauge}), i.e.\ the bispectrum obtained from the $\delta\phi$-gauge route; as it should be. This is a further confirmation that employing the $\delta\phi$-gauge to compute correlations of $\zeta$ in our model can be convenient. 

\bibliographystyle{JHEP}

\bibliography{draft}

\providecommand{\href}[2]{#2}\begingroup\raggedright\begin{thebibliography}{10}

\bibitem{Carr:2009jm}
B.J.~Carr, K.~Kohri, Y.~Sendouda and J.~Yokoyama, \emph{{New cosmological
  constraints on primordial black holes}},
  \href{https://doi.org/10.1103/PhysRevD.81.104019}{\emph{Phys. Rev. D}
  {\bfseries 81} (2010) 104019}
  [\href{https://arxiv.org/abs/0912.5297}{{\ttfamily 0912.5297}}].

\bibitem{Niikura:2017zjd}
H.~Niikura et~al., \emph{{Microlensing constraints on primordial black holes
  with Subaru/HSC Andromeda observations}},
  \href{https://doi.org/10.1038/s41550-019-0723-1}{\emph{Nature Astron.}
  {\bfseries 3} (2019) 524} [\href{https://arxiv.org/abs/1701.02151}{{\ttfamily
  1701.02151}}].

\bibitem{Katz:2018zrn}
A.~Katz, J.~Kopp, S.~Sibiryakov and W.~Xue, \emph{{Femtolensing by Dark Matter
  Revisited}}, \href{https://doi.org/10.1088/1475-7516/2018/12/005}{\emph{JCAP}
  {\bfseries 12} (2018) 005}
  [\href{https://arxiv.org/abs/1807.11495}{{\ttfamily 1807.11495}}].

\bibitem{Montero-Camacho:2019jte}
P.~Montero-Camacho, X.~Fang, G.~Vasquez, M.~Silva and C.M.~Hirata,
  \emph{{Revisiting constraints on asteroid-mass primordial black holes as dark
  matter candidates}},
  \href{https://doi.org/10.1088/1475-7516/2019/08/031}{\emph{JCAP} {\bfseries
  08} (2019) 031} [\href{https://arxiv.org/abs/1906.05950}{{\ttfamily
  1906.05950}}].

\bibitem{Ballesteros:2019exr}
G.~Ballesteros, J.~Coronado-Bl\'azquez and D.~Gaggero, \emph{{X-ray and
  gamma-ray limits on the primordial black hole abundance from Hawking
  radiation}},
  \href{https://doi.org/10.1016/j.physletb.2020.135624}{\emph{Phys. Lett. B}
  {\bfseries 808} (2020) 135624}
  [\href{https://arxiv.org/abs/1906.10113}{{\ttfamily 1906.10113}}].

\bibitem{Carr:2020gox}
B.~Carr, K.~Kohri, Y.~Sendouda and J.~Yokoyama, \emph{{Constraints on
  primordial black holes}},
  \href{https://doi.org/10.1088/1361-6633/ac1e31}{\emph{Rept. Prog. Phys.}
  {\bfseries 84} (2021) 116902}
  [\href{https://arxiv.org/abs/2002.12778}{{\ttfamily 2002.12778}}].

\bibitem{Green:2020jor}
A.M.~Green and B.J.~Kavanagh, \emph{{Primordial Black Holes as a dark matter
  candidate}}, \href{https://doi.org/10.1088/1361-6471/abc534}{\emph{J. Phys.
  G} {\bfseries 48} (2021) 043001}
  [\href{https://arxiv.org/abs/2007.10722}{{\ttfamily 2007.10722}}].

\bibitem{Carr:1974nx}
B.J.~Carr and S.W.~Hawking, \emph{{Black holes in the early Universe}},
  \href{https://doi.org/10.1093/mnras/168.2.399}{\emph{Mon. Not. Roy. Astron.
  Soc.} {\bfseries 168} (1974) 399}.

\bibitem{Nakama:2013ica}
T.~Nakama, T.~Harada, A.G.~Polnarev and J.~Yokoyama, \emph{{Identifying the
  most crucial parameters of the initial curvature profile for primordial black
  hole formation}},
  \href{https://doi.org/10.1088/1475-7516/2014/01/037}{\emph{JCAP} {\bfseries
  01} (2014) 037} [\href{https://arxiv.org/abs/1310.3007}{{\ttfamily
  1310.3007}}].

\bibitem{Musco:2018rwt}
I.~Musco, \emph{{Threshold for primordial black holes: Dependence on the shape
  of the cosmological perturbations}},
  \href{https://doi.org/10.1103/PhysRevD.100.123524}{\emph{Phys. Rev. D}
  {\bfseries 100} (2019) 123524}
  [\href{https://arxiv.org/abs/1809.02127}{{\ttfamily 1809.02127}}].

\bibitem{Tsamis:2003px}
N.C.~Tsamis and R.P.~Woodard, \emph{{Improved estimates of cosmological
  perturbations}},
  \href{https://doi.org/10.1103/PhysRevD.69.084005}{\emph{Phys. Rev. D}
  {\bfseries 69} (2004) 084005}
  [\href{https://arxiv.org/abs/astro-ph/0307463}{{\ttfamily
  astro-ph/0307463}}].

\bibitem{Kinney:2005vj}
W.H.~Kinney, \emph{{Horizon crossing and inflation with large eta}},
  \href{https://doi.org/10.1103/PhysRevD.72.023515}{\emph{Phys. Rev. D}
  {\bfseries 72} (2005) 023515}
  [\href{https://arxiv.org/abs/gr-qc/0503017}{{\ttfamily gr-qc/0503017}}].

\bibitem{Fixsen:1996nj}
D.J.~Fixsen, E.S.~Cheng, J.M.~Gales, J.C.~Mather, R.A.~Shafer and E.L.~Wright,
  \emph{{The Cosmic Microwave Background spectrum from the full COBE FIRAS data
  set}}, \href{https://doi.org/10.1086/178173}{\emph{Astrophys. J.} {\bfseries
  473} (1996) 576} [\href{https://arxiv.org/abs/astro-ph/9605054}{{\ttfamily
  astro-ph/9605054}}].

\bibitem{Kristiano:2022maq}
J.~Kristiano and J.~Yokoyama, \emph{{Ruling Out Primordial Black Hole Formation
  From Single-Field Inflation}},
  \href{https://arxiv.org/abs/2211.03395}{{\ttfamily 2211.03395}}.

\bibitem{Schwinger:1960qe}
J.S.~Schwinger, \emph{{Brownian motion of a quantum oscillator}},
  \href{https://doi.org/10.1063/1.1703727}{\emph{J. Math. Phys.} {\bfseries 2}
  (1961) 407}.

\bibitem{Keldysh:1964ud}
L.V.~Keldysh, \emph{{Diagram technique for nonequilibrium processes}},
  {\emph{Zh. Eksp. Teor. Fiz.} {\bfseries 47} (1964) 1515}.

\bibitem{Weinberg:2005vy}
S.~Weinberg, \emph{{Quantum contributions to cosmological correlations}},
  \href{https://doi.org/10.1103/PhysRevD.72.043514}{\emph{Phys. Rev. D}
  {\bfseries 72} (2005) 043514}
  [\href{https://arxiv.org/abs/hep-th/0506236}{{\ttfamily hep-th/0506236}}].

\bibitem{Riotto:2023hoz}
A.~Riotto, \emph{{The Primordial Black Hole Formation from Single-Field
  Inflation is Not Ruled Out}},
  \href{https://arxiv.org/abs/2301.00599}{{\ttfamily 2301.00599}}.

\bibitem{Firouzjahi:2023bkt}
H.~Firouzjahi, \emph{{Revisiting Loop Corrections in Single Field USR
  Inflation}},  \href{https://arxiv.org/abs/2311.04080}{{\ttfamily
  2311.04080}}.

\bibitem{Franciolini:2023lgy}
G.~Franciolini, A.~Iovino, Junior., M.~Taoso and A.~Urbano, \emph{{One loop to
  rule them all: Perturbativity in the presence of ultra slow-roll dynamics}},
  \href{https://arxiv.org/abs/2305.03491}{{\ttfamily 2305.03491}}.

\bibitem{Tada:2023rgp}
Y.~Tada, T.~Terada and J.~Tokuda, \emph{{Cancellation of quantum corrections on
  the soft curvature perturbations}},
  \href{https://arxiv.org/abs/2308.04732}{{\ttfamily 2308.04732}}.

\bibitem{Iacconi:2023ggt}
L.~Iacconi, D.~Mulryne and D.~Seery, \emph{{Loop corrections in the separate
  universe picture}},  \href{https://arxiv.org/abs/2312.12424}{{\ttfamily
  2312.12424}}.

\bibitem{Ballesteros:2017fsr}
G.~Ballesteros and M.~Taoso, \emph{{Primordial black hole dark matter from
  single field inflation}},
  \href{https://doi.org/10.1103/PhysRevD.97.023501}{\emph{Phys. Rev. D}
  {\bfseries 97} (2018) 023501}
  [\href{https://arxiv.org/abs/1709.05565}{{\ttfamily 1709.05565}}].

\bibitem{Ballesteros:2020qam}
G.~Ballesteros, J.~Rey, M.~Taoso and A.~Urbano, \emph{{Primordial black holes
  as dark matter and gravitational waves from single-field polynomial
  inflation}}, \href{https://doi.org/10.1088/1475-7516/2020/07/025}{\emph{JCAP}
  {\bfseries 07} (2020) 025}
  [\href{https://arxiv.org/abs/2001.08220}{{\ttfamily 2001.08220}}].

\bibitem{Wands:1998yp}
D.~Wands, \emph{{Duality invariance of cosmological perturbation spectra}},
  \href{https://doi.org/10.1103/PhysRevD.60.023507}{\emph{Phys. Rev. D}
  {\bfseries 60} (1999) 023507}
  [\href{https://arxiv.org/abs/gr-qc/9809062}{{\ttfamily gr-qc/9809062}}].

\bibitem{Leach:2000yw}
S.M.~Leach and A.R.~Liddle, \emph{{Inflationary perturbations near horizon
  crossing}}, \href{https://doi.org/10.1103/PhysRevD.63.043508}{\emph{Phys.
  Rev. D} {\bfseries 63} (2001) 043508}
  [\href{https://arxiv.org/abs/astro-ph/0010082}{{\ttfamily
  astro-ph/0010082}}].

\bibitem{Leach:2001zf}
S.M.~Leach, M.~Sasaki, D.~Wands and A.R.~Liddle, \emph{{Enhancement of
  superhorizon scale inflationary curvature perturbations}},
  \href{https://doi.org/10.1103/PhysRevD.64.023512}{\emph{Phys. Rev. D}
  {\bfseries 64} (2001) 023512}
  [\href{https://arxiv.org/abs/astro-ph/0101406}{{\ttfamily
  astro-ph/0101406}}].

\bibitem{Chen:2010xka}
X.~Chen, \emph{{Primordial Non-Gaussianities from Inflation Models}},
  \href{https://doi.org/10.1155/2010/638979}{\emph{Adv. Astron.} {\bfseries
  2010} (2010) 638979} [\href{https://arxiv.org/abs/1002.1416}{{\ttfamily
  1002.1416}}].

\bibitem{Senatore:2009cf}
L.~Senatore and M.~Zaldarriaga, \emph{{On Loops in Inflation}},
  \href{https://doi.org/10.1007/JHEP12(2010)008}{\emph{JHEP} {\bfseries 12}
  (2010) 008} [\href{https://arxiv.org/abs/0912.2734}{{\ttfamily 0912.2734}}].

\bibitem{Weinberg:2008nf}
S.~Weinberg, \emph{{Non-Gaussian Correlations Outside the Horizon}},
  \href{https://doi.org/10.1103/PhysRevD.78.123521}{\emph{Phys. Rev. D}
  {\bfseries 78} (2008) 123521}
  [\href{https://arxiv.org/abs/0808.2909}{{\ttfamily 0808.2909}}].

\bibitem{Weinberg:2010wq}
S.~Weinberg, \emph{{Ultraviolet Divergences in Cosmological Correlations}},
  \href{https://doi.org/10.1103/PhysRevD.83.063508}{\emph{Phys. Rev. D}
  {\bfseries 83} (2011) 063508}
  [\href{https://arxiv.org/abs/1011.1630}{{\ttfamily 1011.1630}}].

\bibitem{Hollik:1988ii}
W.F.L.~Hollik, \emph{{Radiative Corrections in the Standard Model and their
  Role for Precision Tests of the Electroweak Theory}},
  \href{https://doi.org/10.1002/prop.2190380302}{\emph{Fortsch. Phys.}
  {\bfseries 38} (1990) 165}.

\bibitem{Bunch:1978yq}
T.S.~Bunch and P.C.W.~Davies, \emph{{Quantum Field Theory in de Sitter Space:
  Renormalization by Point Splitting}},
  \href{https://doi.org/10.1098/rspa.1978.0060}{\emph{Proc. Roy. Soc. Lond. A}
  {\bfseries 360} (1978) 117}.

\bibitem{Espinosa:2018eve}
J.R.~Espinosa, D.~Racco and A.~Riotto, \emph{{A Cosmological Signature of the
  SM Higgs Instability: Gravitational Waves}},
  \href{https://doi.org/10.1088/1475-7516/2018/09/012}{\emph{JCAP} {\bfseries
  09} (2018) 012} [\href{https://arxiv.org/abs/1804.07732}{{\ttfamily
  1804.07732}}].

\bibitem{Gerstenlauer:2011ti}
M.~Gerstenlauer, A.~Hebecker and G.~Tasinato, \emph{{Inflationary Correlation
  Functions without Infrared Divergences}},
  \href{https://doi.org/10.1088/1475-7516/2011/06/021}{\emph{JCAP} {\bfseries
  06} (2011) 021} [\href{https://arxiv.org/abs/1102.0560}{{\ttfamily
  1102.0560}}].

\bibitem{Giddings:2011zd}
S.B.~Giddings and M.S.~Sloth, \emph{{Cosmological observables, IR growth of
  fluctuations, and scale-dependent anisotropies}},
  \href{https://doi.org/10.1103/PhysRevD.84.063528}{\emph{Phys. Rev. D}
  {\bfseries 84} (2011) 063528}
  [\href{https://arxiv.org/abs/1104.0002}{{\ttfamily 1104.0002}}].

\bibitem{Senatore:2012nq}
L.~Senatore and M.~Zaldarriaga, \emph{{On Loops in Inflation II: IR Effects in
  Single Clock Inflation}},
  \href{https://doi.org/10.1007/JHEP01(2013)109}{\emph{JHEP} {\bfseries 01}
  (2013) 109} [\href{https://arxiv.org/abs/1203.6354}{{\ttfamily 1203.6354}}].

\bibitem{Gorbenko:2019rza}
V.~Gorbenko and L.~Senatore, \emph{{$\lambda \phi^4$ in dS}},
  \href{https://arxiv.org/abs/1911.00022}{{\ttfamily 1911.00022}}.

\bibitem{Taoso:2021uvl}
M.~Taoso and A.~Urbano, \emph{{Non-gaussianities for primordial black hole
  formation}}, \href{https://doi.org/10.1088/1475-7516/2021/08/016}{\emph{JCAP}
  {\bfseries 08} (2021) 016}
  [\href{https://arxiv.org/abs/2102.03610}{{\ttfamily 2102.03610}}].

\bibitem{Ezquiaga:2019ftu}
J.M.~Ezquiaga, J.~Garc\'\i{}a-Bellido and V.~Vennin, \emph{{The exponential
  tail of inflationary fluctuations: consequences for primordial black holes}},
  \href{https://doi.org/10.1088/1475-7516/2020/03/029}{\emph{JCAP} {\bfseries
  03} (2020) 029} [\href{https://arxiv.org/abs/1912.05399}{{\ttfamily
  1912.05399}}].

\bibitem{Figueroa:2020jkf}
D.G.~Figueroa, S.~Raatikainen, S.~Rasanen and E.~Tomberg, \emph{{Non-Gaussian
  Tail of the Curvature Perturbation in Stochastic Ultraslow-Roll Inflation:
  Implications for Primordial Black Hole Production}},
  \href{https://doi.org/10.1103/PhysRevLett.127.101302}{\emph{Phys. Rev. Lett.}
  {\bfseries 127} (2021) 101302}
  [\href{https://arxiv.org/abs/2012.06551}{{\ttfamily 2012.06551}}].

\bibitem{Firouzjahi:2023aum}
H.~Firouzjahi, \emph{{One-loop corrections in power spectrum in single field
  inflation}}, \href{https://doi.org/10.1088/1475-7516/2023/10/006}{\emph{JCAP}
  {\bfseries 10} (2023) 006}
  [\href{https://arxiv.org/abs/2303.12025}{{\ttfamily 2303.12025}}].

\bibitem{Fumagalli:2023hpa}
J.~Fumagalli, \emph{{Absence of one-loop effects on large scales from small
  scales in non-slow-roll dynamics}},
  \href{https://arxiv.org/abs/2305.19263}{{\ttfamily 2305.19263}}.

\bibitem{Kristiano:2023scm}
J.~Kristiano and J.~Yokoyama, \emph{{Response to criticism on ''Ruling Out
  Primordial Black Hole Formation From Single-Field Inflation'': A note on
  bispectrum and one-loop correction in single-field inflation with primordial
  black hole formation}},  \href{https://arxiv.org/abs/2303.00341}{{\ttfamily
  2303.00341}}.

\bibitem{Maldacena:2002vr}
J.M.~Maldacena, \emph{{Non-Gaussian features of primordial fluctuations in
  single field inflationary models}},
  \href{https://doi.org/10.1088/1126-6708/2003/05/013}{\emph{JHEP} {\bfseries
  05} (2003) 013} [\href{https://arxiv.org/abs/astro-ph/0210603}{{\ttfamily
  astro-ph/0210603}}].

\bibitem{Wang:2013zva}
Y.~Wang, \emph{{Inflation, Cosmic Perturbations and Non-Gaussianities}},
  \href{https://doi.org/10.1088/0253-6102/62/1/19}{\emph{Commun. Theor. Phys.}
  {\bfseries 62} (2014) 109} [\href{https://arxiv.org/abs/1303.1523}{{\ttfamily
  1303.1523}}].

\bibitem{Burrage:2011hd}
C.~Burrage, R.H.~Ribeiro and D.~Seery, \emph{{Large slow-roll corrections to
  the bispectrum of noncanonical inflation}},
  \href{https://doi.org/10.1088/1475-7516/2011/07/032}{\emph{JCAP} {\bfseries
  07} (2011) 032} [\href{https://arxiv.org/abs/1103.4126}{{\ttfamily
  1103.4126}}].

\bibitem{Braglia:2024zsl}
M.~Braglia and L.~Pinol, \emph{{No time to derive: unraveling total time
  derivatives in in-in perturbation theory}},
  \href{https://arxiv.org/abs/2403.14558}{{\ttfamily 2403.14558}}.

\bibitem{Firouzjahi:2023ahg}
H.~Firouzjahi and A.~Riotto, \emph{{Primordial Black Holes and loops in
  single-field inflation}},
  \href{https://doi.org/10.1088/1475-7516/2024/02/021}{\emph{JCAP} {\bfseries
  02} (2024) 021} [\href{https://arxiv.org/abs/2304.07801}{{\ttfamily
  2304.07801}}].

\bibitem{Inomata:2024lud}
K.~Inomata, \emph{{Curvature Perturbations Protected Against One Loop}},
  \href{https://arxiv.org/abs/2403.04682}{{\ttfamily 2403.04682}}.

\bibitem{Senatore:2016aui}
L.~Senatore, \emph{{Lectures on Inflation}},  in \emph{{Theoretical Advanced
  Study Institute in Elementary Particle Physics}: {New Frontiers in Fields and
  Strings}}, pp.~447--543, 2017,
  \href{https://doi.org/10.1142/9789813149441_0008}{DOI}
  [\href{https://arxiv.org/abs/1609.00716}{{\ttfamily 1609.00716}}].

\bibitem{Arnowitt:1962hi}
R.L.~Arnowitt, S.~Deser and C.W.~Misner, \emph{{The Dynamics of general
  relativity}}, \href{https://doi.org/10.1007/s10714-008-0661-1}{\emph{Gen.
  Rel. Grav.} {\bfseries 40} (2008) 1997}
  [\href{https://arxiv.org/abs/gr-qc/0405109}{{\ttfamily gr-qc/0405109}}].

\bibitem{Weinberg:2003sw}
S.~Weinberg, \emph{{Adiabatic modes in cosmology}},
  \href{https://doi.org/10.1103/PhysRevD.67.123504}{\emph{Phys. Rev. D}
  {\bfseries 67} (2003) 123504}
  [\href{https://arxiv.org/abs/astro-ph/0302326}{{\ttfamily
  astro-ph/0302326}}].

\bibitem{Baumann:2022mni}
D.~Baumann, \emph{{Cosmology}}, Cambridge University Press (7, 2022),
  \href{https://doi.org/10.1017/9781108937092}{10.1017/9781108937092}.

\bibitem{Arroja:2011yj}
F.~Arroja and T.~Tanaka, \emph{{A note on the role of the boundary terms for
  the non-Gaussianity in general k-inflation}},
  \href{https://doi.org/10.1088/1475-7516/2011/05/005}{\emph{JCAP} {\bfseries
  05} (2011) 005} [\href{https://arxiv.org/abs/1103.1102}{{\ttfamily
  1103.1102}}].

\bibitem{Gibbons:1976ue}
G.W.~Gibbons and S.W.~Hawking, \emph{{Action Integrals and Partition Functions
  in Quantum Gravity}},
  \href{https://doi.org/10.1103/PhysRevD.15.2752}{\emph{Phys. Rev. D}
  {\bfseries 15} (1977) 2752}.

\bibitem{York:1972sj}
J.W.~York, Jr., \emph{{Role of conformal three geometry in the dynamics of
  gravitation}}, \href{https://doi.org/10.1103/PhysRevLett.28.1082}{\emph{Phys.
  Rev. Lett.} {\bfseries 28} (1972) 1082}.

\bibitem{Dyer:2008hb}
E.~Dyer and K.~Hinterbichler, \emph{{Boundary Terms, Variational Principles and
  Higher Derivative Modified Gravity}},
  \href{https://doi.org/10.1103/PhysRevD.79.024028}{\emph{Phys. Rev. D}
  {\bfseries 79} (2009) 024028}
  [\href{https://arxiv.org/abs/0809.4033}{{\ttfamily 0809.4033}}].

\bibitem{Malik_2008im}
K.A.~Malik and D.~Wands, \emph{{Cosmological perturbations}},
  \href{https://doi.org/10.1016/j.physrep.2009.03.001}{\emph{Phys. Rept.}
  {\bfseries 475} (2009) 1} [\href{https://arxiv.org/abs/0809.4944}{{\ttfamily
  0809.4944}}].

\bibitem{Malik_2009fp}
A.J.~Christopherson and K.A.~Malik, \emph{{Practical tools for third order
  cosmological perturbations}},
  \href{https://doi.org/10.1088/1475-7516/2009/11/012}{\emph{JCAP} {\bfseries
  11} (2009) 012} [\href{https://arxiv.org/abs/0909.0942}{{\ttfamily
  0909.0942}}].

\bibitem{Sugiyama:2012tj}
N.S.~Sugiyama, E.~Komatsu and T.~Futamase, \emph{{$\delta$N formalism}},
  \href{https://doi.org/10.1103/PhysRevD.87.023530}{\emph{Phys. Rev. D}
  {\bfseries 87} (2013) 023530}
  [\href{https://arxiv.org/abs/1208.1073}{{\ttfamily 1208.1073}}].

\bibitem{Chen:2006dfn}
X.~Chen, M.-x.~Huang and G.~Shiu, \emph{{The Inflationary Trispectrum for
  Models with Large Non-Gaussianities}},
  \href{https://doi.org/10.1103/PhysRevD.74.121301}{\emph{Phys. Rev. D}
  {\bfseries 74} (2006) 121301}
  [\href{https://arxiv.org/abs/hep-th/0610235}{{\ttfamily hep-th/0610235}}].

\bibitem{Bravo:2017wyw}
R.~Bravo, S.~Mooij, G.A.~Palma and B.~Pradenas, \emph{{A generalized
  non-Gaussian consistency relation for single field inflation}},
  \href{https://doi.org/10.1088/1475-7516/2018/05/024}{\emph{JCAP} {\bfseries
  05} (2018) 024} [\href{https://arxiv.org/abs/1711.02680}{{\ttfamily
  1711.02680}}].

\end{thebibliography}\endgroup

\end{document}